\documentclass[onecolumn,pra,floatfix]{revtex4}
\usepackage{amsmath}
\usepackage{amssymb}
\usepackage{graphicx}
\usepackage{subfigure}
\usepackage{epsfig}
\usepackage{amsfonts}
\usepackage[german, english]{babel}
\usepackage{ifthen}

\DeclareGraphicsExtensions{.eps, .ps}

\begin{document}

\title{Spontaneous symmetry breaking of binary fields in a nonlinear
double-well structure}
\author{Arturas Acus}
\affiliation{Institute of Theoretical Physics and Astronomy, Vilnius University, Vilnius
LT-01108, Lithuania\\
email: {Arturas.Acus@tfai.vu.lt}, phone: +370 219 32 67}
\author{Boris A. Malomed}
\affiliation{Department of Physical Electronics, School of Electrical Engineering,
Faculty of Engineering, Tel Aviv University, Tel Aviv 69978, Israel\\
email: {malomed@post.tau.ac.il}, phone +972-3-640-6413}
\author{Yakov Shnir}
\affiliation{Institute of Physics, Carl von Ossietzky University Oldenburg, Germany;\\
Department of Theoretical Physics and Astrophysics, BSU, Minsk
Belarus\\
email: shnir@maths.tcd.ie; phone +44 191 33 43110}

\begin{abstract}
We introduce a one-dimensional two-component system with the self-focusing
cubic nonlinearity concentrated at a symmetric set of two spots. Effects of
the spontaneous symmetry breaking (SSB) of localized modes were previously
studied in the single-component version of this system. In this work, we
study the evolution (in the configuration space of the system) and SSB
scenarios for two-component modes of three generic types, as concerns the
spatial symmetry of each component: symmetric-symmetric (Sm-Sm),
antisymmetric-antisymmetric (AS-AS), and symmetric-antisymmetric (S-AS)
ones. In the limit case of the nonlinear potential represented by two $%
\delta $-functions, solutions are obtained in a semi-analytical form. They
feature novel properties, in comparison with the previously studied
single-component model. In particular, the SSB of antisymmetric modes is
possible solely in the two-component system, and, obviously, S-AS states
exist only in the two-component system too. In the general case of the
symmetric pair of finite-width nonlinear potential wells, evolution
scenarios are very complex. In this case, new results are reported, first,
for the single-component model. These are pairs of broken-antisymmetry
modes, and of twin-peak symmetric ones, which are generated by saddle-mode
bifurcations separated from the transformations previously studied in the
the single-component setting. With regard to these findings, complex
scenarios of the evolution of the two-component solution families are
realized in terms of links connecting pairs of modes of three simplest
types: (A) two-component ones with unbroken symmetries; (B) single-component
modes featuring density peaks in both potential wells; (C) single-component
modes which are trapped, essentially, in a single well.
\end{abstract}

\maketitle


\newcounter{fig}

\section{Introduction and the model}

A fundamental effect caused by the interplay of nonlinearity with symmetric
potentials is spontaneous symmetry breaking (SSB). The simplest setting
where this effect occurs is represented by double-well potentials. A
commonly known property of one-dimensional quantum mechanics is that the
ground state shares the symmetry of the underlying double-well potential
\cite{LL}. If the self-attractive cubic nonlinearity is added to the
consideration, the corresponding Schr\"{o}dinger equation is transformed
into the Gross-Pitaevskii equation for a Bose-Einstein condensate (BEC)\
loaded into the double-well potential \cite{BEC}, or the nonlinear Schr\"{o}%
dinger equation for photonic counterparts of the system \cite{NLS}. The
nonlinear term may break the symmetry of the ground state, replacing it by
an asymmetric one which minimizes the energy of the system, provided that
the strength of the nonlinearity exceeds a certain critical value (see Ref.
\cite{misc} for the general analysis, and Ref. \cite{misc2} for the
consideration of the SSB of self-trapped states in BEC). The experimental
realization of the SSB in double-well potentials was reported in BEC \cite%
{Markus} (this was done in the condensate with the self-repulsive
nonlinearity, which implies the spontaneous breaking of the \emph{%
antisymmetry} of the lowest-energy antisymmetric mode) and in nonlinear
optics \cite{photo}.

The \textit{symmetry-breaking bifurcation}, which destabilizes the original
symmetric ground state and gives rise to the SSB in the nonlinear systems,
was originally predicted in a discrete model of self-trapping \cite{Chris}.
In nonlinear optics, a similar bifurcation was analyzed in Ref. \cite{Snyder}
for continuous-wave (spatially uniform) light signals in dual-core fibers.
The allied bifurcation for solitons in nonlinear dual-core fibers was
studied in detail in Ref. \cite{dual-core}. Later, the SSB was studied for
gap solitons in the model of the dual-core fiber Bragg gratings with the
same cubic nonlinearity as in the ordinary fibers \cite{Mak}. The SSB\
effects were also predicted for matter-wave solitons in the self-attractive
BEC loaded into a double-channel potential trap \cite{Arik}-\cite{Luca}.

Typically, the self-attractive cubic nonlinearity gives rise to the soliton
bifurcations of the subcritical (\textit{backward}) type (in other words,
these are phase transitions of the first kind). In that case, branches of
asymmetric modes emerge as unstable ones, going backward (i.e., in the
direction of the decrease of the soliton's norm (energy)). They get
stabilized after switching the evolution direction forward at turning points
of the bifurcation diagram \cite{bif}. It was also demonstrated that the
combination of the self-focusing nonlinearity with a periodic potential
applied in the free direction (perpendicular to the direction of the action
of the double-well potential) tends to change the character of the soliton
bifurcation from sub- to supercritical (i.e., replace the phase transition
of the first kind by a transition of the second kind). In the supercritical
case, the asymmetric branches emerge as stable ones, going in the forward
direction, i.e., in the direction of the increase of the soliton's norm \cite%
{Arik,Warsaw2}. The bifurcation for gap solitons in the dual-core fiber
Bragg grating is of the forward type too \cite{Mak}.

A challenging alternative to the use of the ordinary linear double-well
potential is the setting based on an effective nonlinear potential, alias
\textit{pseudopotential} (as it is called in solid-state physics \cite%
{Harrison}), which is induced by the double-peak spatial modulation of the
local strength of the self-focusing nonlinearity. Pseudopotentials featuring
the double-well shape can be realized in optics and matter waves \cite%
{Pu,Barcelona}. The ultimate form of such a setting is the one with the
nonlinearity concentrated at two points, represented by a symmetric pair of
delta-functions or narrow Gaussians \cite{Dong,Nir} (the simplest model of
that type, with the self-attractive nonlinearity represented by a single
delta-function, was first introduced in Ref. \cite{Mark}). Two-dimensional
counterparts of the system were considered too, in the form of two parallel
channels \cite{Hung}, or two symmetric circles \cite{new}, in which the
self-attractive nonlinearity is localized . The SSB of solitons in the
symmetric nonlinear double-well potentials was studied in detail, featuring
the subcritical type of the symmetry breaking \cite{Dong,Hung,Nir}.

It is relevant to mention that SSB effects were also analyzed for discrete
solitons in dual-core nonlinear chains, with the uniform linear coupling
between them \cite{Herring}, as well as with the linear coupling which links
a single pair of sites in the parallel chains \cite{Ljupco}. The form of the
linear coupling determines the type of the respective SSB bifurcation, which
is subcritical in the former case, and supercritical in the latter
situation. The action of the nonlinearity concentrated at a pair of sites
embedded into a single linear chain was studied too \cite{Molina,Valera}.
The SSB effects in this setting (for both straight and circular host linear
chains) were analyzed in recent work \cite{Valera}. The SSB in a similar
system, with a pair of symmetric nonlinear sites side-coupled to the
infinite linear chain, was recently studied in Ref. \cite{Almas}.

An obviously relevant generalization of the work outlined above is extension
to two-component systems. In BEC, this implies a mixture of two different
hyperfine states of the same atomic species \cite{BEC}, while in optics it
implies the co-propagation of light beams mixing two different polarizations
of light, or different carrier wavelengths. Recently, the SSB, along with
the related dynamical effect of Josephson oscillations between the wave
functions trapped in adjacent potential wells, have been analyzed in diverse
models of binary systems embedded into ordinary (linear) potentials of the
double-well type \cite{binarySSB}. An experiment was performed too, for the
corresponding BEC mixture \cite{Markus2}. However, SSB effects have not yet
been studied in two-component systems trapped in double-well \emph{nonlinear}
(pseudo)potentials. A straightforward possibility to implement the latter
setting can be found in nonlinear optics, where the spatially nonuniform
nonlinearity, which forms the double-well structure, will act similarly on
both polarization components, in the form of the SPM (self-phase
modulation), as well as on the XPM\ (cross-phase-modulation) nonlinear
interaction between them \cite{NLS}.

In this work, we aim to perform the analysis of the two-component system in
the nonlinear double-well potential, with emphasis on manifestations of the
SSB in the two-component mixture. To this end, we aim to consider the basic
setting with the symmetric pair of strongly localized nonlinear spots
embedded into the one-dimensional linear host medium. Thus, our model is the
generalization, for wave functions $\phi (x,t)$ and $\psi (x,t)$ of the two
components, of the single-component model introduced in Ref. \cite{Dong}:
\begin{eqnarray}
i\phi _{t} &=&-\frac{1}{2}\phi _{xx}+g(x)\left( |\phi |^{2}+G\left\vert \psi
\right\vert ^{2}\right) \phi ,  \label{phi1} \\
i\psi _{t} &=&-\frac{1}{2}\psi _{xx}+g(x)\left( |\psi |^{2}+G\left\vert \phi
\right\vert ^{2}\right) \psi ,  \label{psi1}
\end{eqnarray}%
where $G>0$ is the relative strength of the XPM nonlinearity, while the SPM
coefficient is normalized to be $1$. In the optical model corresponding to
these equations, the evolutional variable $t$ is actually the propagation
distance (usually denoted as $z$), while $x$ is the transverse coordinate
(in the BEC model, $t$ is simply time). As for the XPM coefficient, its
typical values in optics are $G=2/3$ for the coupled linear polarizations,
and $G=2$ for the pair of orthogonal polarizations or different carrier
wavelengths \cite{NLS}. Other values of $G$ are also possible, for
elliptically polarized beams.

\begin{figure}[h]
\refstepcounter{fig} \centering\includegraphics[width=3.5in]{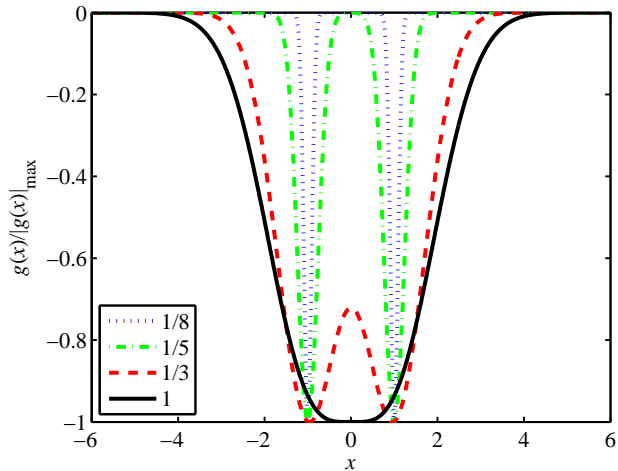}
\caption{(Color online) Shapes of the double-well nonlinearity-modulation
function (\protect\ref{g}), normalized to its maximum value, are shown (as
per Ref. \protect\cite{Dong}) for different values of scaled width $a$ of
the individual well.}
\label{Fig1}
\end{figure}

Following Ref. \cite{Dong}, the nonlinearity-modulation function in Eqs. (%
\ref{phi1}) and (\ref{psi1}), which corresponds to the symmetric set of two
strongly localized nonlinear spots, is adopted in the form of
\begin{equation}
g(x)=-\frac{1}{a\sqrt{\pi }}\left[ \exp \left( -\frac{\left( x+1\right) ^{2}%
}{a^{2}}\right) +\exp \left( -\frac{\left( x-1\right) ^{2}}{a^{2}}\right) %
\right] \,,  \label{g}
\end{equation}%
which is subject to the normalization condition,
\begin{equation}
\int_{-\infty }^{+\infty }g(x)dx~\equiv -2.  \label{scaling}
\end{equation}%
Profiles of modulation function (\ref{g}), which keeps the double-well
structure at $a<\sqrt{2}$, are shown, for different values of $a$, in Fig. %
\ref{Fig1}.

Equations (\ref{phi1}) and (\ref{psi1}) conserve three dynamical invariants,
\textit{viz}., two norms and the energy (Hamiltonian),
\begin{eqnarray}
M &\equiv &\int_{-\infty }^{+\infty }|\phi (x)|^{2}dx,~N\equiv \int_{-\infty
}^{+\infty }|\psi (x)|^{2}dx,~  \label{N} \\
H &=&\frac{1}{2}\int_{-\infty }^{\infty }\left[ \left\vert \phi
_{x}\right\vert ^{2}+\left\vert \psi _{x}\right\vert ^{2}+g(x)\left(
\left\vert \phi \right\vert ^{4}+\left\vert \psi \right\vert
^{4}+2G\left\vert \phi \right\vert ^{2}\left\vert \psi \right\vert
^{2}\right) \right] dx.  \label{H}
\end{eqnarray}

Stationary localized solutions to Eqs. (\ref{phi1}) and (\ref{psi1}) are
sought for as $\phi =e^{-i\lambda t}u(x),~\psi =e^{-i\mu t}v(x)$, where the
chemical potentials of localized modes must be negative, $\lambda <0$, $\mu
<0$, and real functions $u(x)$ and $v(x)$ satisfy equations%
\begin{eqnarray}
\lambda u+\frac{1}{2}u^{\prime \prime }+\frac{1}{a\sqrt{\pi }}\left[ \exp
\left( -\frac{\left( x+1\right) ^{2}}{a^{2}}\right) +\exp \left( -\frac{%
\left( x-1\right) ^{2}}{a^{2}}\right) \right] \left( u^{2}+Gv^{2}\right) u
&=&0,  \label{u} \\
\mu v+\frac{1}{2}v^{\prime \prime }+\frac{1}{a\sqrt{\pi }}\left[ \exp \left(
-\frac{\left( x+1\right) ^{2}}{a^{2}}\right) +\exp \left( -\frac{\left(
x-1\right) ^{2}}{a^{2}}\right) \right] \left( v^{2}+Gu^{2}\right) v &=&0
\label{v}
\end{eqnarray}%
A special version of the model corresponds to the limit of $a\rightarrow 0$,
with Eqs. (\ref{u}) and (\ref{v}) going over into%
\begin{eqnarray}
\lambda u &=&-(1/2)u^{\prime \prime }-\left[ \delta (x+1)+\delta (x-1)\right]
\left( u^{2}+Gv^{2}\right) u,  \label{udelta} \\
\mu v &=&-(1/2)v^{\prime \prime }-\left[ \delta (x+1)+\delta (x-1)\right]
\left( v^{2}+Gu^{2}\right) v,  \label{vdelta}
\end{eqnarray}%
where $\delta (x)$ is the Dirac's delta-function.

To conclude the Introduction, it is relevant to mention that a two-field
model with a symmetric set of two strongly localized nonlinear spots was
recently introduced for the quadratic (second-harmonic-generating) optical
nonlinearity \cite{Asia}. In the limit case when the nonlinearity profile is
described by the pair of $\delta $-functions, cf. Eqs. (\ref{udelta}) and (%
\ref{vdelta}), the corresponding equations for the complex amplitudes of the
fundamental-frequency and second-harmonic waves, $U(x,z)$ and $V\left(
x,z\right) $, take the form of

\begin{eqnarray}
iU_{z}+(1/2)U_{xx}+\delta (x)U^{\ast }V &=&0,  \notag \\
&&  \label{UV} \\
2iV_{z}+(1/2)V_{xx}-QV+(1/2)\delta (x)U^{2} &=&0,  \notag
\end{eqnarray}%
where the asterisk stands for the complex conjugate, $z$ is the propagation
distance, and real constant $Q$ measures the mismatch frustrating the
parametric interaction between the two waves. Although involving the two
wave fields, this system (for which exact solutions for symmetric,
antisymmetric, and asymmetric localized modes have been found in Ref. \cite%
{Asia}) is actually closer to the single-component model in the case of the
cubic nonlinearity, due to the coherent character of the nonlinear coupling
between the waves $U$ and $V$. In particular, system (\ref{UV}) directly
reduces to its single-component cubic counterpart in the\textit{\ cascading
limit}, which corresponds to large positive values of $Q$.

The rest of the paper is organized as follows. The limit case of the
two-component system with the modulation profile based on the set of $\delta
$-functions, which corresponds to Eqs. (\ref{udelta}) and (\ref{vdelta}), is
considered in detail in Section 2 (in the beginning of the section, we also
give an exact solution for the two-component system with the single
nonlinear $\delta $-function, as this simple case was not considered
previously). The system with the pair of $\delta $-functions admits a
semi-analytical solution, which reduces to a system of four coupled cubic
algebraic equations for amplitudes of asymmetric modes, while symmetric (Sm)
and antisymmetric (AS) ones can be found in a fully analytical form.

In Section 3, we report results of the analysis of the two-component model
in the general case, corresponding to the nonlinearity-modulation function (%
\ref{g}) with finite $a$. In that case, the analysis is based on a
comprehensive numerical solution of ODEs (\ref{u}) and (\ref{v}) with $a>0$.
The consideration in Section 3 starts by revisiting the single-component
model, with the aim to report additional solutions of the Sm and \textit{%
broken}-AS types (the latter means modes with \textit{broken} antisymmetry).
These solutions were not reported in Ref. \cite{Dong}, as they represent
pairs of modes generated by saddle-node bifurcations which occur at
sufficiently large $a$, being detached from the branches which continuously
evolve from the limit case of $a\rightarrow 0$. The subsequent analysis of
the two-component system reported in Section 3 produces a full picture of
bifurcations for compound modes of the Sm-Sm (symmetric-symmetric), AS-AS
(antisymmetric-antisymmetric), and S-AS (symmetric-antisymmetric) types, and
their \textit{broken} counterparts, i.e., modes with spontaneously broken
(anti)symmetries. Because the full picture turns out to be extremely
complex, we report only parts of it, which may be presented in a relatively
simple form. Despite the complexity of the picture, some underlying
principles can be formulated in a general form: the fundamental role is
played by simplest solutions, namely the two-component \textit{unbroken}
modes (i.e., those with the \textit{unbroken} (anti)symmetries), and \emph{%
single-component} solutions of both the unbroken and broken types. As a
matter of fact, numerous families of two-component solutions, which feature
sophisticated evolution in the system's configuration space, may be
understood as branches linking various pairs of these simplest modes (the
complexity is often produced by the fact that the branches make one or
several loops, in the course of their evolution).

The paper is concluded by Section 4. In particular, we discuss the necessity
of the systematic analysis of the dynamical stability of the diverse
stationary modes found in this work. With a few exceptions, we do not
analyze the stability here, as the study of the stationary modes and their
bifurcations is by itself a sufficiently heavy topic for a single paper.

\section{The model with the delta-functions ($a\rightarrow 0$)}

\subsection{The single delta-function}

Before presenting results for the model with the symmetric pair of $\delta $%
-functions, it makes sense to briefly consider the case of the nonlinearity
represented by the single $\delta $-function, when Eqs. (\ref{phi1}) and (%
\ref{psi1}) are replaced by
\begin{eqnarray}
i\phi _{t} &=&-\frac{1}{2}\phi _{xx}+\delta (x)\left( |\phi
|^{2}+G\left\vert \psi \right\vert ^{2}\right) \phi ,  \label{ph1} \\
i\psi _{t} &=&-\frac{1}{2}\psi _{xx}+\delta (x)\left( |\psi
|^{2}+G\left\vert \phi \right\vert ^{2}\right) \psi ,  \label{ps1}
\end{eqnarray}%
and, accordingly, their stationary version, Eqs. (\ref{udelta}) and (\ref%
{vdelta}), is replaced by%
\begin{eqnarray}
\lambda u &=&-(1/2)u^{\prime \prime }-\delta (x)\left( u^{2}+Gv^{2}\right) u,
\label{u1} \\
\mu v &=&-(1/2)v^{\prime \prime }-\delta (x)\left( v^{2}+Gu^{2}\right) v.
\label{v1}
\end{eqnarray}%
It is straightforward to find solutions to Eqs. (\ref{u1}) and (\ref{v1})
with both fields different from zero:%
\begin{eqnarray}
\left\{ u(x),v(x)\right\} &=&\left\{ A,B\right\} \exp \left( -\sqrt{%
-2\left\{ \lambda ,\mu \right\} }|x|\right) ,  \label{exp} \\
\left\{ A^{2},B^{2}\right\} &=&\frac{\sqrt{2}}{G^{2}-1}\left( G\sqrt{%
-\left\{ \mu ,\lambda \right\} }-\sqrt{-\left\{ \lambda ,\mu \right\} }%
\right) .  \label{AB}
\end{eqnarray}%
These solutions exist for $G\neq 1$ ($G=1$ is the degenerate case
corresponding to the Manakov's nonlinearity \cite{NLS}), for negative
chemical potentials $\lambda $ and $\mu $, in the regions of
\begin{eqnarray}
\frac{1}{G} &<&\sqrt{\frac{\lambda }{\mu }}<G,~~\mathrm{if}~~G>1, \\
G &<&\sqrt{\frac{\lambda }{\mu }}<\frac{1}{G},~~\mathrm{if}~~G<1.
\end{eqnarray}

It is also instructive to cast Eq. (\ref{AB}) into the form of relations
between the chemical potentials and norms $M$ and $N$ of the two components
(see Eq. (\ref{N})). Obviously, $\left\{ M,N\right\} =\left\{
A^{2},B^{2}\right\} /\sqrt{-2\left\{ \lambda ,\mu \right\} }$ for solutions (%
\ref{exp}), hence%
\begin{equation}
\left\{ M,N\right\} =\frac{1}{G^{2}-1}\left( G\sqrt{\left\{ \frac{\mu }{%
\lambda },\frac{\lambda }{\mu }\right\} }-1\right) .  \label{MN}
\end{equation}%
Although we chiefly leave the issue of the dynamical stability aside in this
paper, Eq. (\ref{MN}) makes it possible to address the stability problem,
using the Vakhitov-Kolokolov (VK) criterion. It states that, for the
single-component model, with the soliton family characterized by dependence $%
M=M(\lambda )$, the necessary stability condition is $dM/d\mu \leq 0$ \cite%
{VK,Berge'}, while for the two-component model with separately conserved
norms and two independent chemical potentials the switch between the
stability and instability occurs at points where the corresponding Jacobian
determinant vanishes:%
\begin{equation}
J\left( \lambda ,\mu \right) \equiv \left\Vert \frac{\partial \left(
M,N\right) }{\partial \left( \lambda ,\mu \right) }\right\Vert =0.
\end{equation}

In the single-component model with the single nonlinear $\delta $-function,
the soliton family is \textit{degenerate}, in the sense that its norm does
not depend on the chemical potential: $M\equiv 1$, hence this family seems
as a neutrally-stable one, which actually does not guarantee the true
stability. In reality, the single-component family was found to be \emph{%
completely unstable} \cite{Nir}. In the present case, the calculation of
Jacobian $J\left( \lambda ,\mu \right) $ for the two-component solitons per
Eq. (\ref{MN}) demonstrates that this determinant identically vanishes, $%
J\left( \lambda ,\mu \right) \equiv 0$, suggesting that the total family of
solutions (\ref{exp}), (\ref{AB}) is completely unstable too. Indeed, one
can easily find the following set of \emph{exact nonstationary} solutions to
Eqs. (\ref{ph1}) and (\ref{ps1}), with arbitrary real constant $\alpha $ and
$\beta $:%
\begin{eqnarray}
\left\{ \phi \left( x,t\right) \psi (x,t)\right\} &=&\left\{ \alpha ,\beta
\right\} |t|^{-1/2}\exp \left[ \frac{i}{2t}\left( |x|-i\left\{ a,b\right\}
\right) ^{2}\right] ,  \label{exact} \\
\left\{ a,b\right\} &\equiv &-\left( \left\{ \alpha ^{2},\beta ^{2}\right\}
+G\left\{ \beta ^{2},\alpha ^{2}\right\} \right) \mathrm{sgn}(t),
\label{sgn}
\end{eqnarray}%
which generalize similar exact single-component solutions found in Ref. \cite%
{Nir}. At $t<0$, solutions (\ref{exact}), (\ref{sgn}) describe the collapse
of the perturbed soliton (the formation of a singularity at $x=0$), which
happens at $t\rightarrow -0$, while at $t>0$ the solution describes decay of
the soliton. Thus, the nonstationary solutions demonstrate that the solitons
belonging to the degenerate family (\ref{exp}), (\ref{AB}), supported by the
single $\delta $-function, suffer either the collapse or decay, being indeed
unstable, as conjectured above. On the other hand, we expect that the
replacement of the ideal $\delta $-functions by a finite-width profile may
stabilize the solutions, cf. Ref. \cite{KaiLi} where this was demonstrated
for the single-component model.

\subsection{Amplitude equations for the system with two delta-functions}

Proceeding to the model with the symmetric set of two $\delta $-functions,
based on Eqs. (\ref{udelta}) and (\ref{vdelta}), our first objective is to
obtain analytical results, following the pattern of Ref. \cite{Dong}, where
a full analytical solution was found for stationary states in the
corresponding single-component model. Off points $x=\pm 1$, equations (\ref%
{udelta}) and (\ref{vdelta}) are linear, a general solution to which,
decaying at $|x|\rightarrow \infty $, can be sought for as%
\begin{equation}
u(x)=\left\{
\begin{array}{c}
B_{1}e^{\sqrt{2|\lambda |}\left( x+1\right) },~\mathrm{at}~x<-1, \\
A_{0}e^{-\sqrt{2|\lambda |}\left( x-1\right) }+B_{0}e^{\sqrt{2|\lambda |}%
\left( x+1\right) },~\mathrm{at}~-1<x<+1, \\
A_{1}e^{-\sqrt{2|\lambda |}\left( x-1\right) },~\mathrm{at}~x>+1,%
\end{array}%
\right.  \label{ulinear}
\end{equation}%
\begin{equation}
v(x)=\left\{
\begin{array}{c}
D_{1}e^{\sqrt{2|\mu |}\left( x+1\right) },~\mathrm{at}~x<-1, \\
C_{0}e^{-\sqrt{2|\mu |}\left( x-1\right) }+D_{0}e^{\sqrt{2|\mu |}\left(
x+1\right) },~\mathrm{at}~-1<x<+1, \\
C_{1}e^{-\sqrt{2|\mu |}\left( x-1\right) },~\mathrm{at}~x>+1,%
\end{array}%
\right.  \label{vlinear}
\end{equation}%
with constant amplitudes $A_{0},A_{1}$, $B_{0},B_{1}$ and $%
C_{0},C_{1},~D_{0},D_{1}$. The continuity of the wave functions at $x=\pm 1$
imposes four relations on the amplitudes,%
\begin{eqnarray}
B_{1} &=&B_{0}+A_{0}e^{2\sqrt{2|\lambda |}},~A_{1}=A_{0}+B_{0}e^{2\sqrt{%
2|\lambda |}}, \\
D_{1} &=&D_{0}+C_{0}e^{2\sqrt{2|\mu |}},~C_{1}=C_{0}+D_{0}e^{2\sqrt{2|\mu |}%
}.
\end{eqnarray}%
Using these relations, one may eliminate ``inner" amplitudes $A_{0}$,$%
B_{0},C_{0},D_{0}$ in favor of their ``outer" counterparts $A_{1}$,$%
B_{1},C_{1},D_{1}$:
\begin{eqnarray}
A_{0} &=&\frac{e^{2\sqrt{2|\lambda |}}B_{1}-A_{1}}{e^{4\sqrt{2|\lambda |}}-1}%
,~B_{0}=\frac{e^{2\sqrt{2|\lambda |}}A_{1}-B_{1}}{e^{4\sqrt{2|\lambda |}}-1},
\label{A0B0} \\
C_{0} &=&\frac{e^{2\sqrt{2|\mu |}}D_{1}-B_{1}}{e^{4\sqrt{2|\mu |}}-1},~D_{0}=%
\frac{e^{2\sqrt{2|\mu |}}C_{1}-D_{1}}{e^{4\sqrt{2|\mu |}}-1}  \label{C0D0}
\end{eqnarray}%
Further, the integration of Eqs. (\ref{udelta}) and (\ref{vdelta}) in
infinitesimal vicinities of points $x=\pm 1$ yields expressions for jumps ($%
\Delta $) of the first derivative at these points,
\begin{subequations}
\label{Delta}
\begin{eqnarray}
\Delta \left( u^{\prime }\right) |_{x=\pm 1} &=&-2\left[ \left( u|_{x=\pm
1})^{2}+G(v|_{x=\pm 1}\right) ^{2}\right] , \\
\Delta \left( v^{\prime }\right) |_{x=\pm 1} &=&-2\left[ \left( v|_{x=\pm
1})^{2}+G(u|_{x=\pm 1}\right) ^{2}\right] .
\end{eqnarray}%
The substitution of solution (\ref{ulinear}) and (\ref{vlinear}) into
relations (\ref{Delta}) leads to a system of equations for the outer
amplitudes,
\end{subequations}
\begin{eqnarray}
\sqrt{|\lambda |/2}\left( B_{1}-B_{0}+A_{0}e^{2\sqrt{2|\lambda |}}\right)
&=&\left( B_{1}^{2}+GD_{1}^{2}\right) B_{1},  \label{cubicA} \\
\sqrt{|\lambda |/2}\left( A_{1}-A_{0}+B_{0}e^{2\sqrt{2|\lambda |}}\right)
&=&\left( A_{1}^{2}+GC_{1}^{2}\right) A_{1}.  \label{cubicB}
\end{eqnarray}%
\begin{eqnarray}
\sqrt{|\mu |/2}\left( D_{1}-D_{0}+C_{0}e^{2\sqrt{2|\mu |}}\right) &=&\left(
D_{1}^{2}+GB_{1}^{2}\right) D_{1},  \label{cubicC} \\
\sqrt{|\mu |/2}\left( C_{1}-C_{0}+D_{0}e^{2\sqrt{2|\mu |}}\right) &=&\left(
C_{1}^{2}+GA_{1}^{2}\right) C_{1}.  \label{cubicD}
\end{eqnarray}%
After the substitution of expressions (\ref{A0B0}) and (\ref{C0D0}) into
Eqs. (\ref{cubicB})-(\ref{cubicD}), we end up with a system of four coupled
cubic equations for $A_{1}$,$B_{1}$,$C_{1},D_{1}$:%
\begin{equation}
\sqrt{|\lambda |/2}\left( e^{2\sqrt{2|\lambda |}}B_{1}-A_{1}\right) =\sinh
\left( 2\sqrt{2|\lambda |}\right) \left( B_{1}^{2}+GD_{1}^{2}\right) B_{1},
\label{A}
\end{equation}%
\begin{equation}
\sqrt{|\lambda |/2}\left( e^{2\sqrt{2|\lambda |}}A_{1}-B_{1}\right) =\sinh
\left( 2\sqrt{2|\lambda |}\right) \left( A_{1}^{2}+GC_{1}^{2}\right) A_{1}.
\label{B}
\end{equation}%
\begin{equation}
\sqrt{|\mu |/2}\left( e^{2\sqrt{2|\mu |}}D_{1}-C_{1}\right) =\sinh \left( 2%
\sqrt{2|\mu |}\right) \left( D_{1}^{2}+GB_{1}^{2}\right) D_{1},  \label{C}
\end{equation}%
\begin{equation}
\sqrt{|\mu |/2}\left( e^{2\sqrt{2|\mu |}}C_{1}-D_{1}\right) =\sinh \left( 2%
\sqrt{2|\mu |}\right) \left( C_{1}^{2}+GA_{1}^{2}\right) C_{1}.  \label{D}
\end{equation}

Symmetric solutions to Eqs. (\ref{A})-(\ref{D}) (as said above, they will be
referred to as ``Sm-Sm" modes, i.e., those which are spatially symmetric in
each component) are looked for by setting $A_{1}=B_{1} $ and $C_{1}=D_{1}$,
which reduces the system to equations
\begin{subequations}
\label{symm}
\begin{eqnarray}
A_{1}^{2}+GC_{1}^{2} &=&\frac{\sqrt{2|\lambda |}}{1+e^{-2\sqrt{2|\lambda |}}}%
, \\
C_{1}^{2}+GA_{1}^{2} &=&\frac{\sqrt{2|\mu |}}{1+e^{-2\sqrt{2|\mu |}}}.
\end{eqnarray}%
Similarly, solutions to Eqs. (\ref{A})-(\ref{D}) which are antisymmetric in
each component (``AS-AS" states) are looked for with $A_{1}=-B_{1}$ and $%
C_{1}=-D_{1}$, reducing Eqs. (\ref{A})-(\ref{D}) to
\end{subequations}
\begin{subequations}
\label{antisymm}
\begin{eqnarray}
A_{1}^{2}+GC_{1}^{2} &=&\frac{\sqrt{2|\lambda |}}{1-e^{-2\sqrt{2|\lambda |}}}%
, \\
C_{1}^{2}+GA_{1}^{2} &=&\frac{\sqrt{2|\mu |}}{1-e^{-2\sqrt{2|\mu |}}}.
\end{eqnarray}

There also exist solutions to Eqs. (\ref{A})-(\ref{D}) with mixed symmetry,
e.g., antisymmetric in component $u$ and symmetric in $v$, which implies $%
A_{1}=-B_{1}$ and $C_{1}=D_{1}$ (``S-AS" states). This restriction reduces
the system of Eqs. (\ref{A})-(\ref{D}) to a simplified one,
\end{subequations}
\begin{subequations}
\label{mixed}
\begin{eqnarray}
A_{1}^{2}+GC_{1}^{2} &=&\frac{\sqrt{2|\lambda |}}{1-e^{-2\sqrt{2|\lambda |}}}%
, \\
C_{1}^{2}+GA_{1}^{2} &=&\frac{\sqrt{2|\mu |}}{1+e^{-2\sqrt{2|\mu |}}}.
\end{eqnarray}

\subsection{Analysis of the solutions}

Systems (\ref{symm}), (\ref{antisymm}) and (\ref{mixed}) are obviously
analytically solvable, as each one is tantamount to a system of two
inhomogeneous linear equations for $A_{1}^{2}$ and $C_{1}^{2}$, while the
general algebraic system (\ref{A})-(\ref{D}) can be only solved in a
numerical form. Norms (\ref{N}) and energy (\ref{H}) for the solutions were
also calculated numerically, because analytical expressions for these
integrals are very cumbersome (they are actually analytically intractable
even in the single-component model \cite{Dong}). The results are displayed
below for $G=2$. As said above, in optics models this value corresponds to
the interaction between orthogonal circular polarizations, or between waves
with different carrier wavelengths.

The numerical solutions reveals a plethora of different branches of the
localized modes, with different degrees of the asymmetry. Typical examples
of the corresponding spatial models of the different types are presented in
Fig. \ref{Fig2a}. In this paper, we aim to report only a part of the
results, which represent the most characteristic findings, that may be
presented in a sufficiently clear form, as the full description of the
(anti)symmetry breaking in the two-component model (both for $a\rightarrow 0$
and finite $a$) is extremely complex. Generally, the results may be
understood as those pertaining to modes generated by the spontaneous
(anti)symmetry breaking from the above-mentioned solutions of three special
types, i.e., Sm-Sm, AS-AS, and S-AS ones. This is the principal difference
from the single-component model, where asymmetric modes could be generated
solely from the symmetric ones (antisymmetric states were dynamically
unstable at small $a$ and stable at larger $a,$ but they never underwent a
symmetry-breaking bifurcation).

\begin{figure}[tbh]
\refstepcounter{fig}
\par
\begin{center}
(a)\hspace{-0.4cm}\includegraphics[height=.22%
\textheight,angle=0]{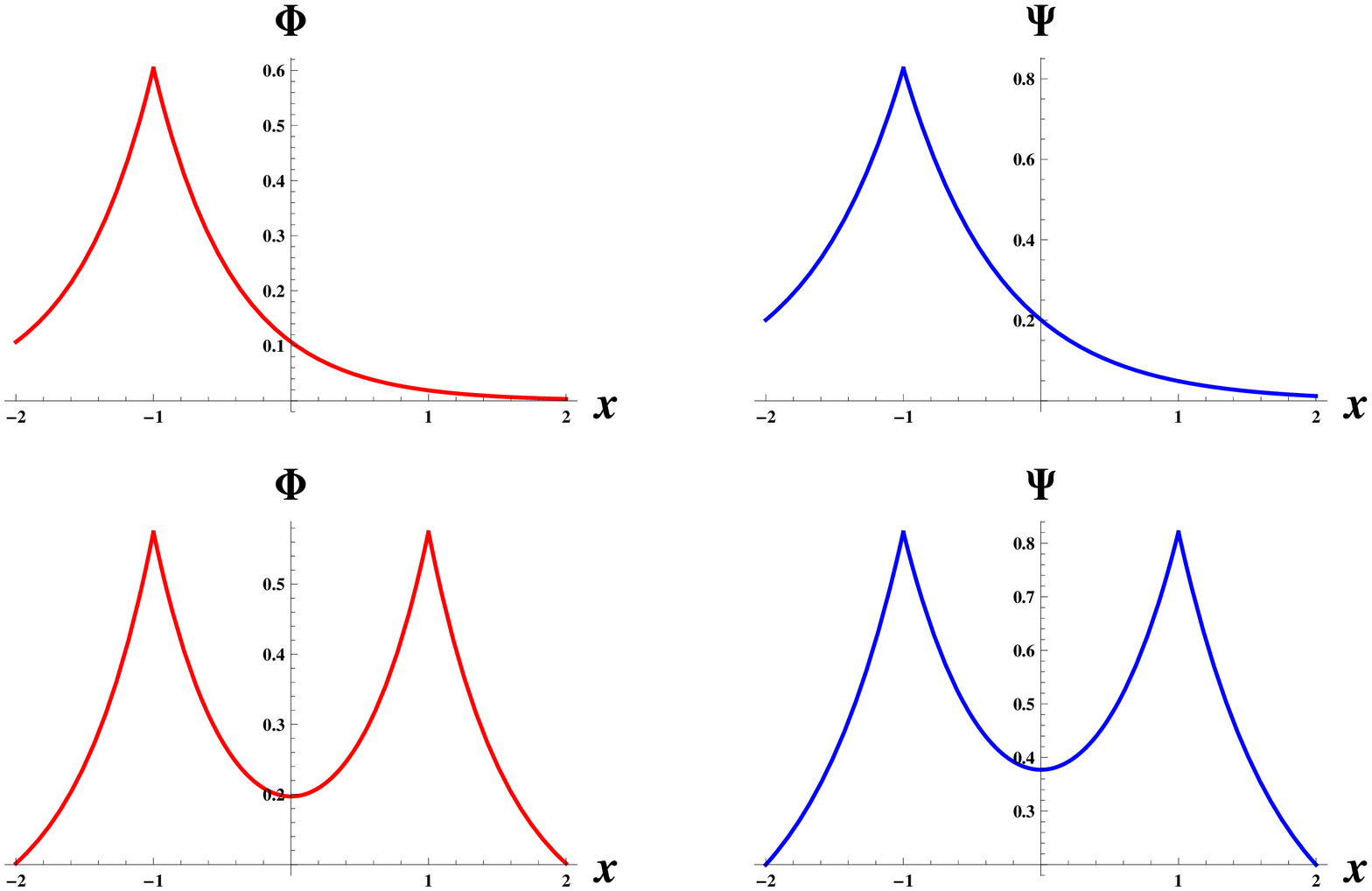} \hspace{0.5cm} \ (b)\hspace{-0.5cm} %
\includegraphics[height=.22\textheight, angle =0]{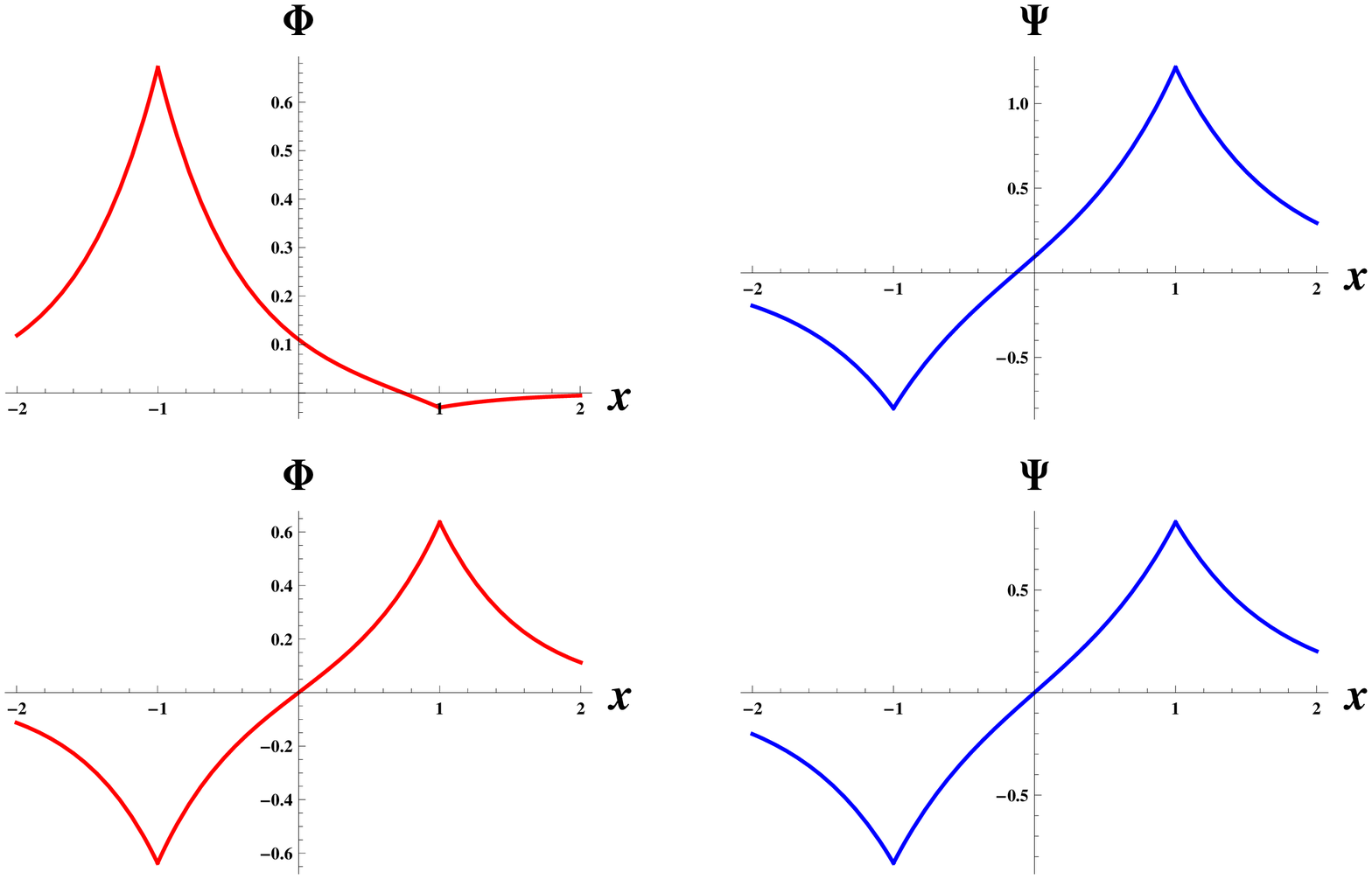}\hspace{%
0.0cm} (c)\hspace{-0.0cm}
\includegraphics[height=.22\textheight,angle
=-0]{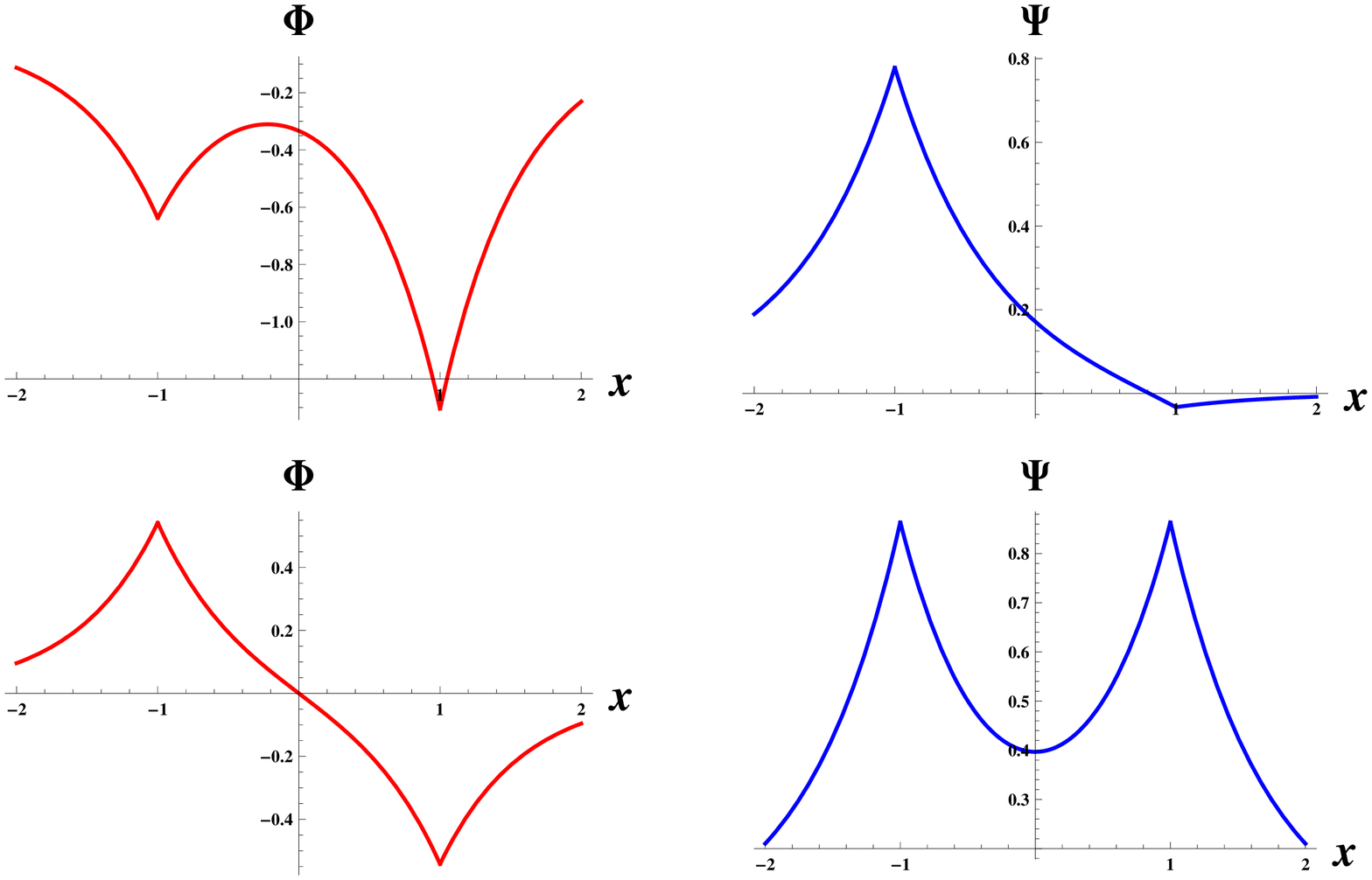}
\end{center}
\par
\vspace{-0.5cm}
\caption{(Color online) Typical examples of the exact modes in the model
with two $\protect\delta $-functions, produced by solution of algebraic
equations (\protect\ref{A})-(\protect\ref{D}) at $\protect\lambda =1.5$ and $%
\protect\mu =1.0$: Sm-Sm modes (a), AS-AS modes (b), and S-AS modes
(c), each with broken and unbroken (anti)symmetry.} \label{Fig2a}
\end{figure}

Each species of the modes may be characterized by their energy (\ref{H}) and
spatial asymmetries of the two fields, which are defined as follows:
\end{subequations}
\begin{equation}
\Delta \Theta _{\left\{ \phi ,\psi \right\} }\equiv \frac{1}{\left\{
M,N\right\} }\left( \int\limits_{0}^{+\infty }dx\left\vert \left\{ \phi
(x),\psi (x\right\} )\right\vert ^{2}-\int\limits_{-\infty }^{0}dx\left\vert
\left\{ \phi (x),\psi (x)\right\} \right\vert ^{2}\right) .  \label{Theta}
\end{equation}%
Exacerbating the complexity of the results, there is, in general, no
one-to-one relation between chemical potentials $\mu $,$\nu $ and norms $M,N$%
, as well as total energy $H$, see Eqs. (\ref{N}), (\ref{H}). However, the
asymmetries and energy of the solutions smoothly depend on $M$ and $N$,
allowing us to classify the solutions according to the type of their
(a)symmetry (Sm-Sm, AS-AS, S-AS, and their ``broken" versions).

\subsubsection{Solutions obtained from the symmetric-symmetric modes (Sm-Sm
type)}

\begin{figure}[tbh]
\refstepcounter{fig} 
\par
\begin{center}
(a)\hspace{-0.6cm}
\includegraphics[height=.35\textheight,
angle=-90]{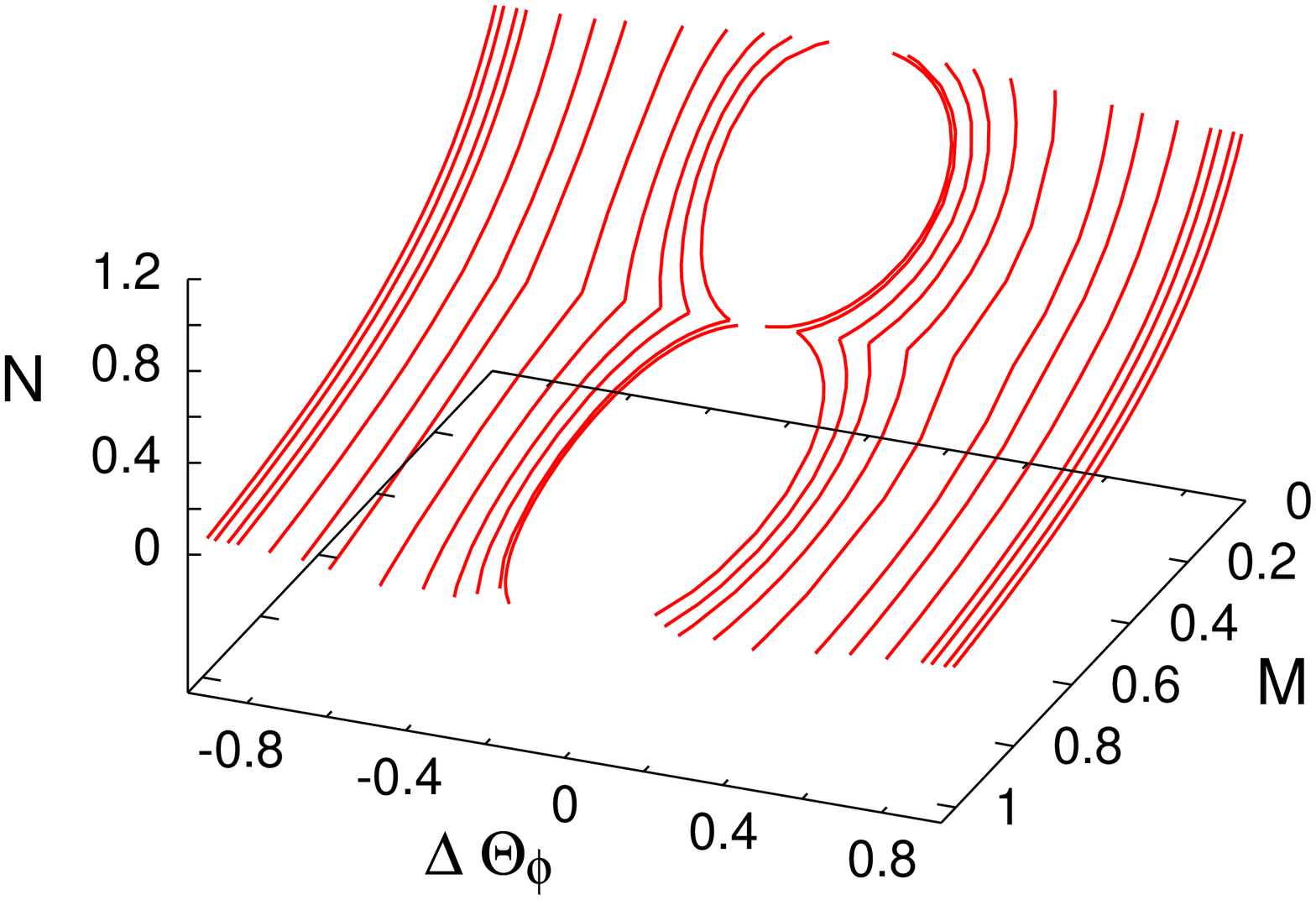} \hspace{0.5cm} (b)\hspace{-0.6cm} %
\includegraphics[height=.35\textheight, angle =-90]{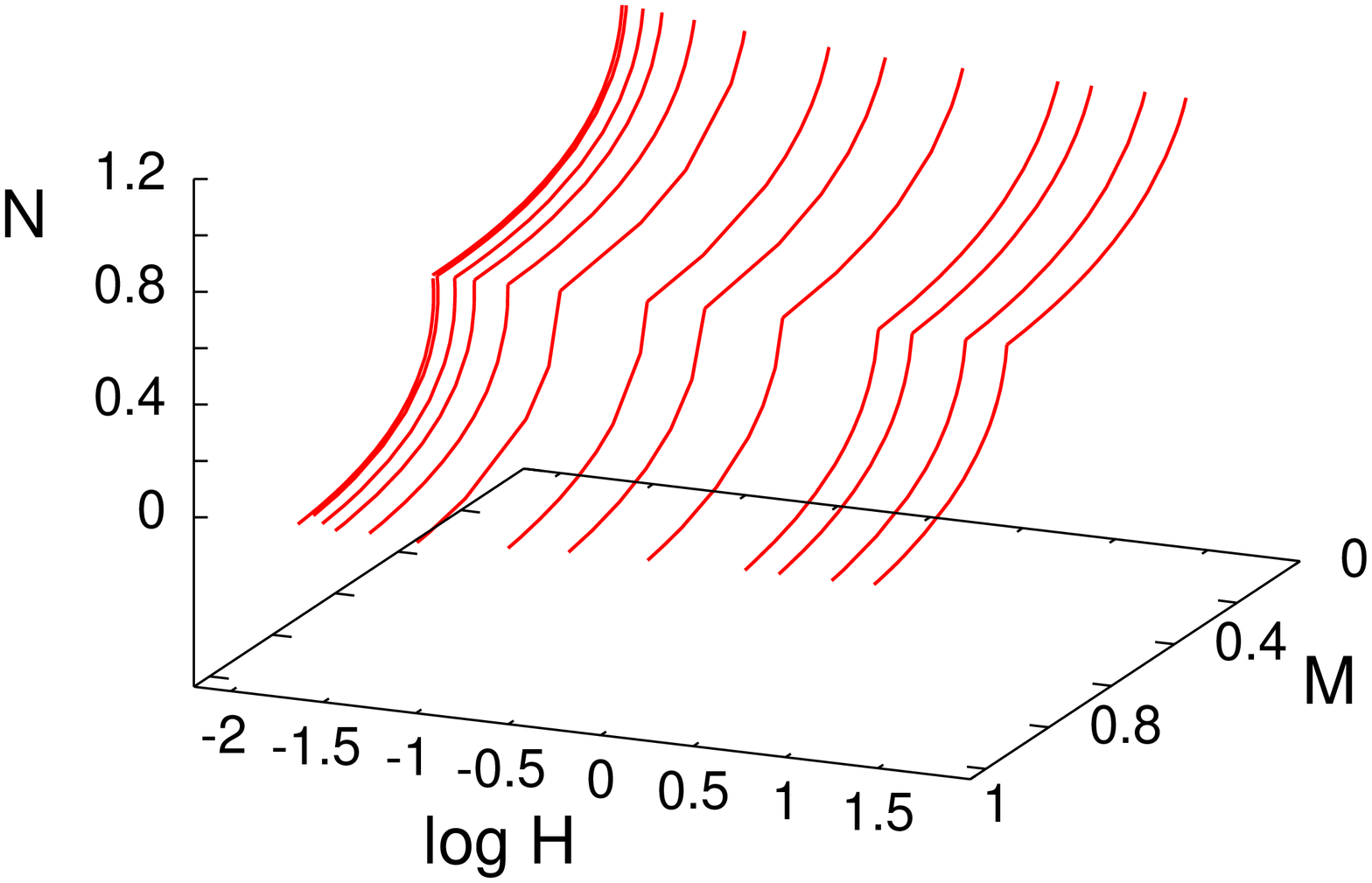}
\end{center}
\par
\vspace{-0.5cm}
\caption{(Color online) Symmetry-broken branches of the Sm-Sm type: the
asymmetry of the field $\protect\phi $ and the energy of the system in the
norms plane, $\left( M,N\right) $ . }
\label{Fig2}
\end{figure}

We start by the presentation of the results for modes generated by
the SSB of the states of the Sm-Sm type, as this case admits direct
comparison with the results obtained for the SSB of the symmetric
mode in the single-component model \cite{Dong}. Examples of the
symmetric modes with the broken and unbroken symmetry are shown in
Fig. \ref{Fig2a}(a), for $\lambda =1.5$ and $\mu =1.0$. In Fig.
\ref{Fig2}, the asymmetry of one component of the ``broken"
solutions of the Sm-Sm type is displayed, along with total energy
(\ref{H}), considered as functions of norms (\ref{N}). The asymmetry
of the second component actually follows the same pattern.

A characteristic feature of the solutions of the Sm-Sm type, both
``broken" and ``unbroken" ones, is that they allow to establish a
one-to-one correspondence between the norms and chemical potentials
of the two components. It can be checked that both the norms and
total energy (\ref{H}) of these modes grow monotonously with the
increase of $|\lambda |$ and $|\mu |$. It is worthy to note too that
the
energy of the symmetry-broken Sm-Sm modes (which is shown in Fig. \ref{Fig2}%
(b)) is always higher than the energy of their unbroken
counterparts, for the same values of $M$ and $N$, which suggests
that the ``unbroken" and ``broken" states may be stable and
unstable,
respectively, similar to the situation in the single-component model \cite%
{Dong}.

The cross section of the plots in Fig. \ref{Fig2} along $M=0$ and $N=0$
correspond to the single-component model. This fact explains why the
symmetry-broken modes exist only in a limited domain in the $\left(
M,N\right) $ plane, as in the single-component model with $a\rightarrow 0$
the solutions with the broken symmetry exist in a narrow interval of values
of the norm \cite{Dong},
\begin{equation}
1<M<2/3+(8/27)\left( 3/4+\ln 2\right) \approx 1.09.  \label{interval}
\end{equation}

The most salient feature observed in Fig. \ref{Fig2}(a) is the lacuna
(``hole") in the existence plot of the modes. In fact, the cross section of
the plot, at a fixed value of $M$ or $N$, which does not cut the lacuna
(including the sections running through $M=0$ and $N=0$), represents a
situation qualitatively similar to that observed in the single-component
model: the asymmetry parameter (\ref{Theta}) varies continuously from $-1$
to $+1$, passing the zero value, while the norm takes values in a narrow
interval, cf. Eq. (\ref{interval}). On the other hand, the cross section
which cuts through the lacuna reveals a qualitatively new situation, that
does not occur in the single-component model: the asymmetry parameter varies
in two disjoint \textit{outer} intervals (which are, naturally, mirror
images of each other), $\left( \Delta \Theta _{\phi }\right) _{\min
}<\left\vert \Delta \Theta _{\phi }\right\vert <1$ (the corresponding
interval of the variation of the norm remains narrow), while the \textit{%
inner} segment, $-\left( \Delta \Theta _{\phi }\right) _{\min }<\Delta
\Theta _{\phi }<+\left( \Delta \Theta _{\phi }\right) _{\min }$, remains
empty. Because there is no solution with $\Delta \Theta _{\phi }=0$ in the
latter case, this means that the corresponding bifurcation in the
single-component model (if such a bifurcation were possible) would be
completely different from the one observed in the actual single-component
equation: it would be a bifurcation loop disconnected from the line of the
solutions with the unbroken symmetry.

Nevertheless, the analysis of the full results for the two-component system
based on the set of two $\delta $-functions demonstrates that branches of
all the species of broken-symmetry solutions (those of the Sm-Sm, AS-AS, and
S-AS types, see results for the two latter species below), which are
parameterized by either chemical potential, $\lambda $ or $\mu $, while the
other one is fixed, actually originate, at bifurcation points, from the
families of allied solutions with unbroken (anti)symmetry, and eventually
merge back into the same families (see more details below). In other words,
the bifurcation curves are indeed shaped as loops, but they are not detached
from the respective (anti)symmetric solution families, which actually
constitute parts of the bifurcation loops.

\subsubsection{Solutions obtained from the antisymmetric-antisymmetric modes
(AS-AS type)}

\begin{figure}[tbh]
\refstepcounter{fig}
\par
\begin{center}
(a)\hspace{-0.6cm} \includegraphics[height=.35%
\textheight,angle=-90]{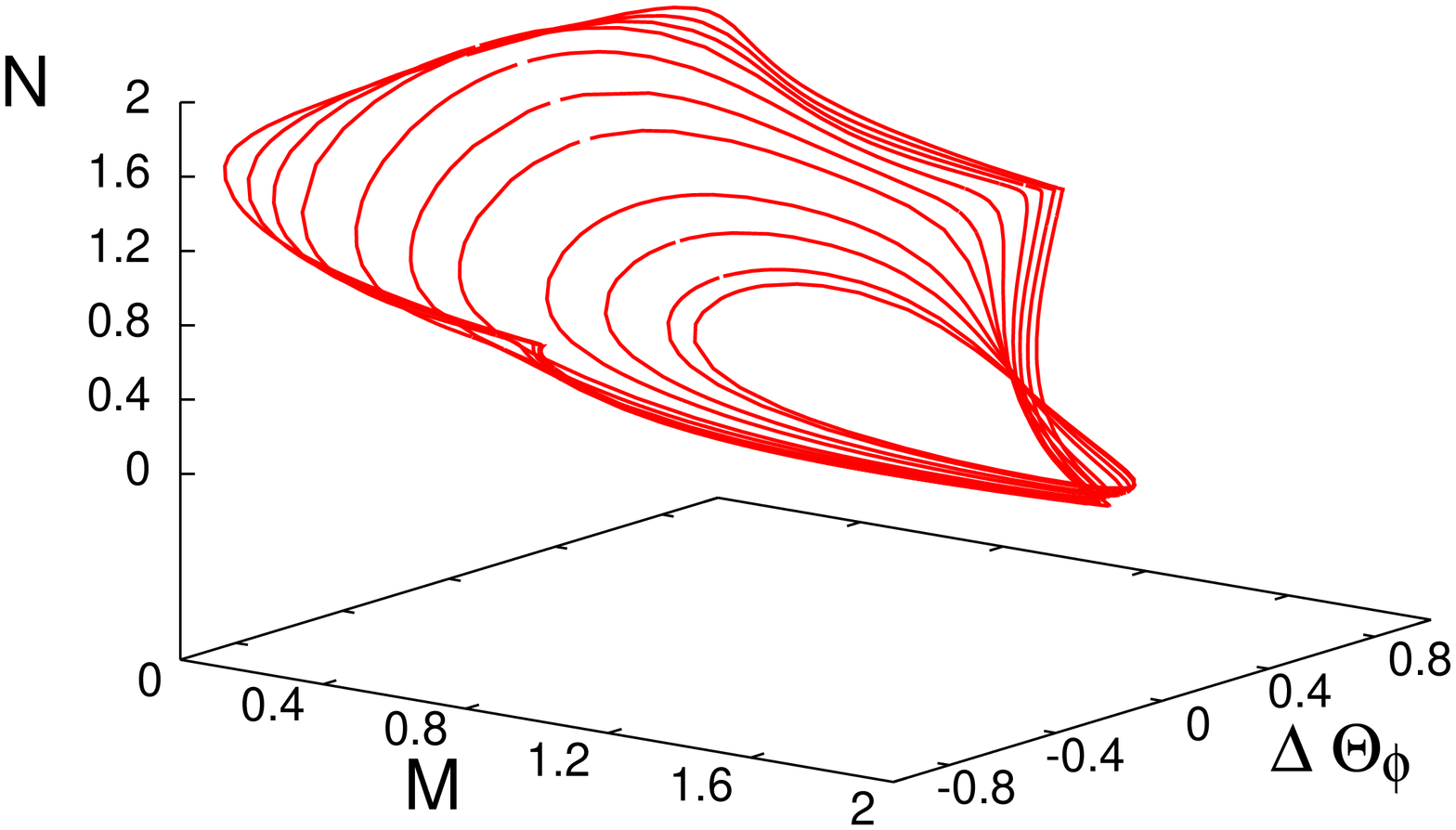} \hspace{0.5cm} (b)\hspace{-0.6cm%
} \includegraphics[height=.35\textheight, angle =-90]{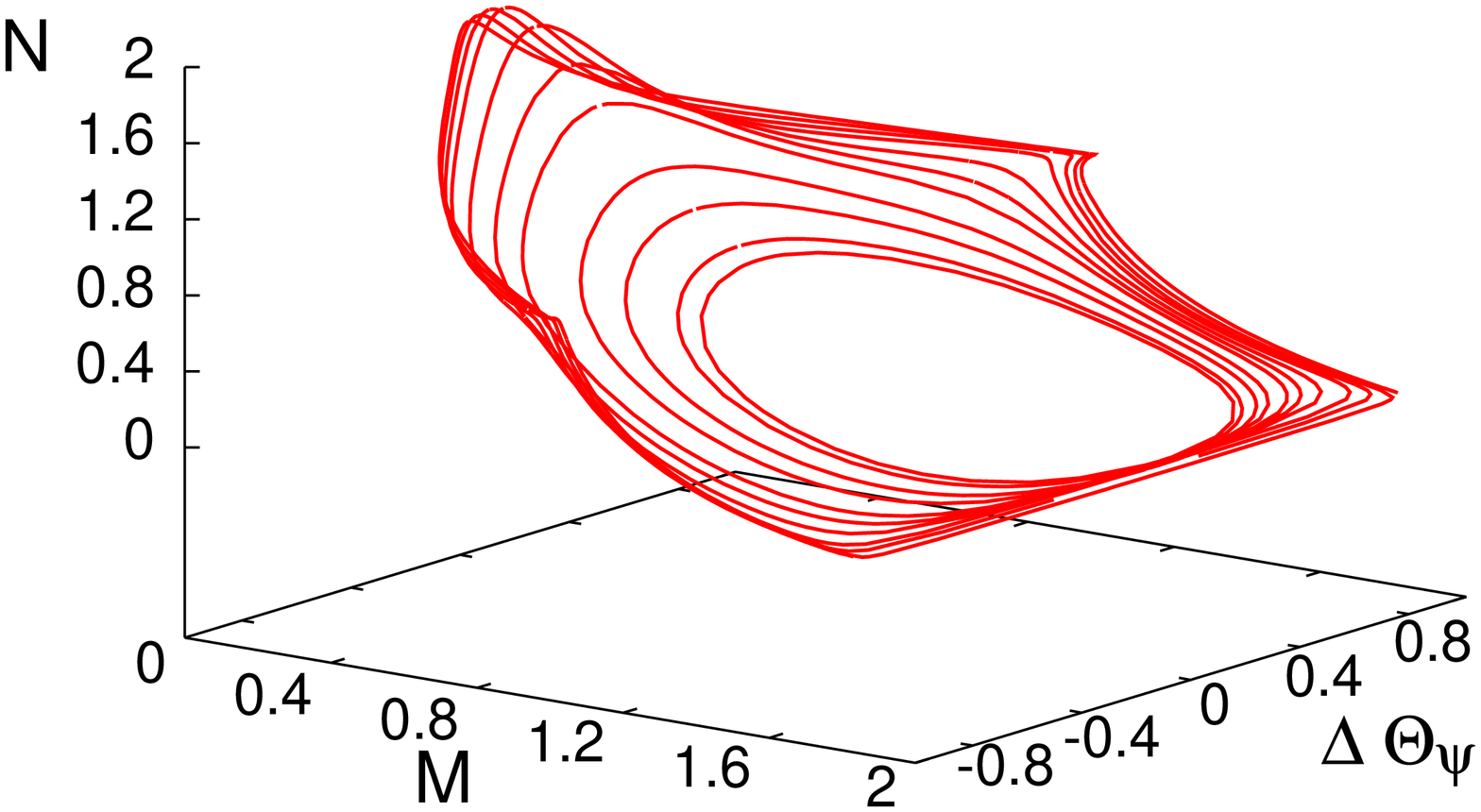}%
\hspace{0.5cm} (c)\hspace{-0.6cm} \includegraphics[height=.35%
\textheight,angle =-90]{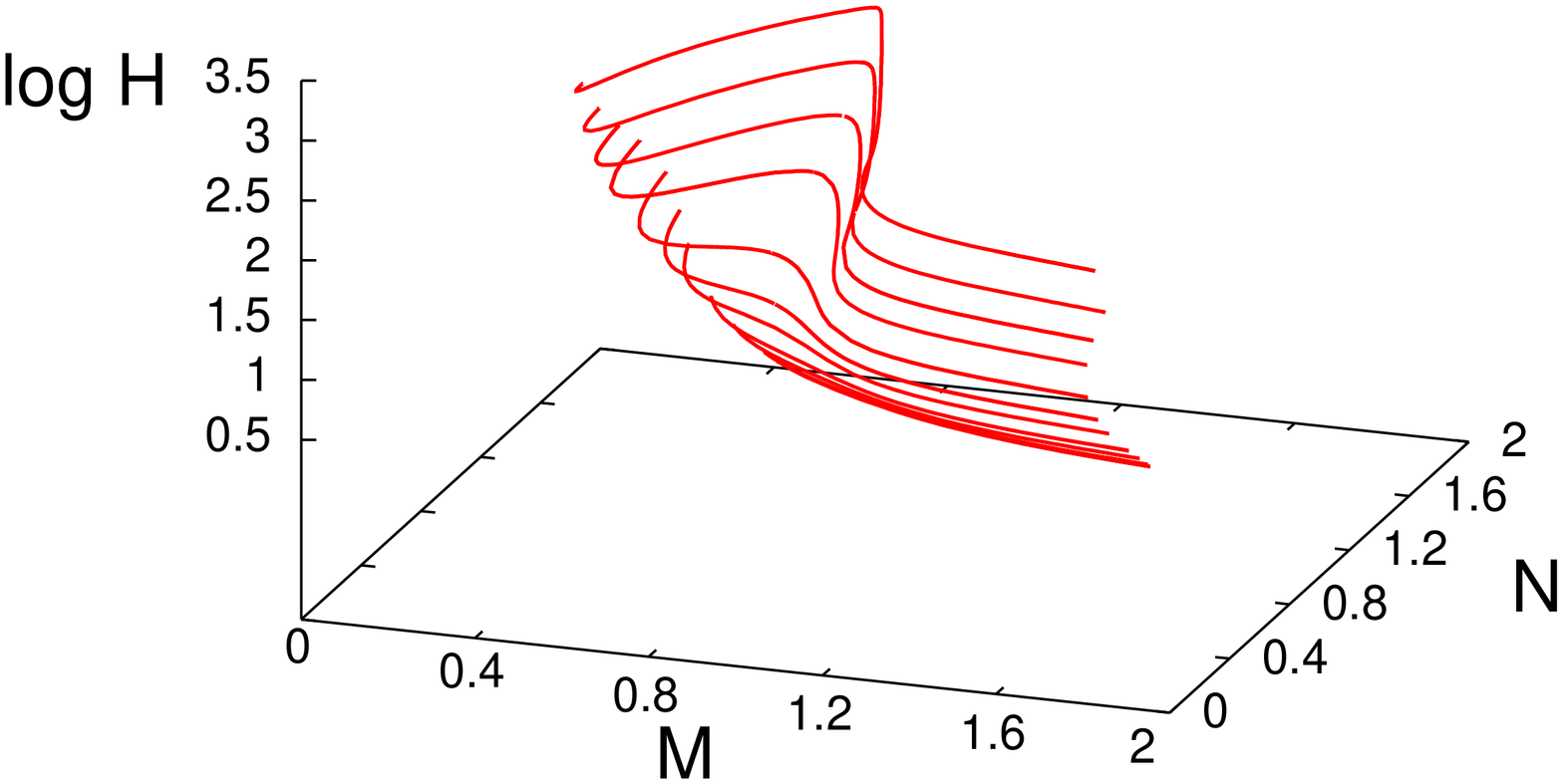}
\end{center}
\par
\vspace{-0.5cm}
\caption{(Color online) AS-AS modes with broken antisymmetry. Panels (a,b)
and (c) display, severally, the asymmetries of components $\protect\phi $
and $\protect\psi $, defined as per Eq. (\protect\ref{Theta}), and the total
energy vs. norms $M$ and $N$ of the two components. }
\label{Fig3}
\end{figure}

As mentioned above, the single-component model with $a\rightarrow 0$ does
not give rise to spontaneous antisymmetry breaking of antisymmetric modes
\cite{Dong}. In the present two-component system, double-antisymmetric modes
(the AS-AS type) do admit SSB, which is therefore a novel effect, in
comparison with the single-component model. In Fig. \ref{Fig3}, the global
picture of the AS-AS solution family with the broken antisymmetry is
displayed by means of two asymmetry measures (\ref{Theta}), $\Delta \Theta
_{\phi }$ and $\Delta \Theta _{\psi }$, along with total energy (\ref{H}),
shown as functions of norms $M$ and $N$. In fact, the plots presented in
panels (a) and (b) are tantamount to each other, but, being shown with
respect to the fixed frame of the axes ($M$ and $N$ ), they provide mutually
complementary views. To clarify qualitative aspects of the picture, Fig. \ref%
{Fig4} additionally displays the projection of the plots in Fig. \ref{Fig3}%
(a),(c) onto the planes of $\left( M,\Delta \Theta _{\phi }\right) $ and $%
\left( M,\ln H\right) $, respectively. Comparison of Fig. \ref{Fig3}(c) with
Fig. \ref{Fig2}(b) clearly demonstrates that the AS-AS solutions have much
higher energy than their counterparts of the Sm-Sm type, hence the AS-AS
modes may be more vulnerable to instabilities (as their AS counterpart in
the single-component model \cite{Dong}).

\begin{figure}[tbh]
\refstepcounter{fig}%
\includegraphics[height=.35\textheight,
angle=-90]{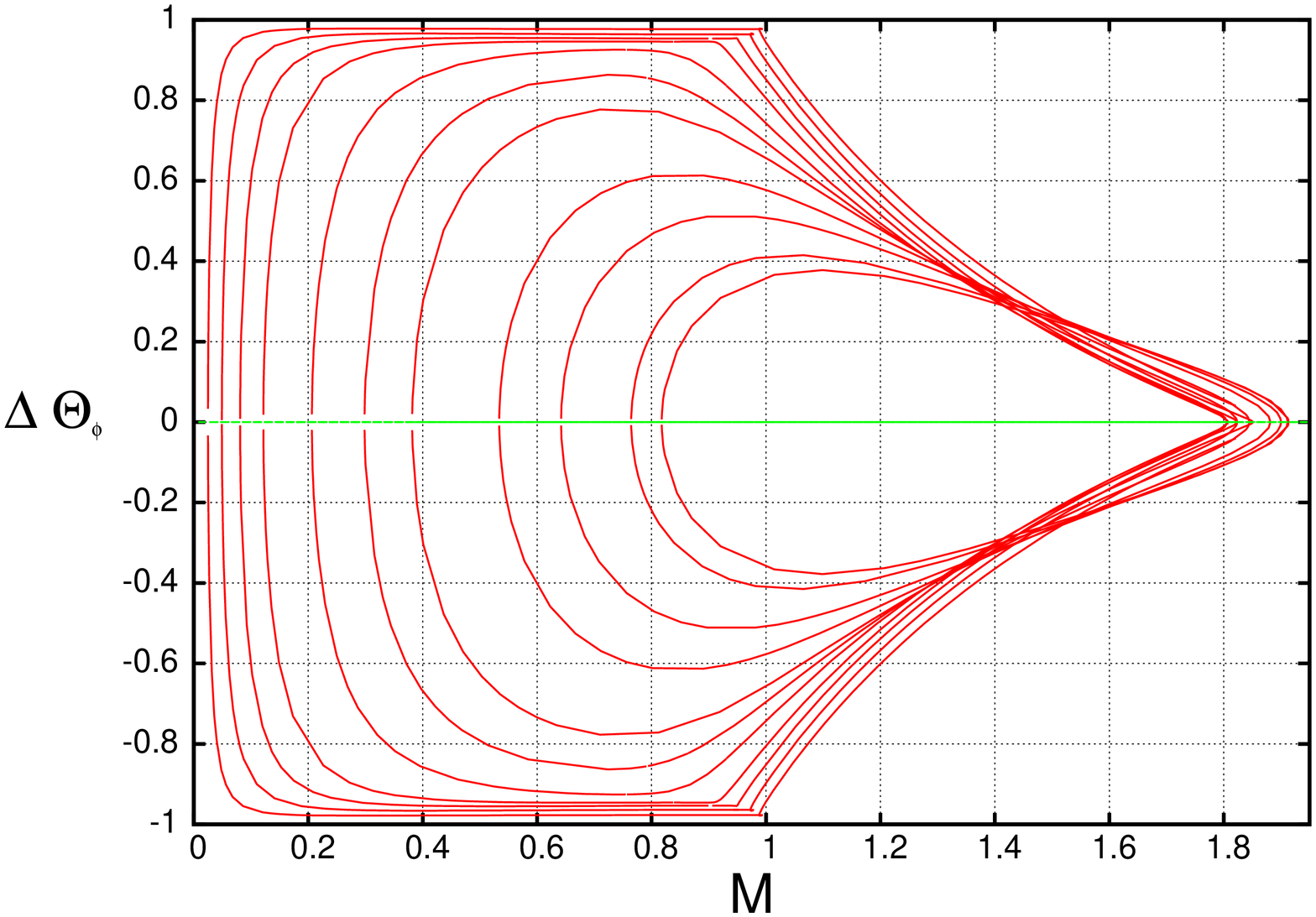}
\includegraphics[height=.35\textheight,
angle=-90]{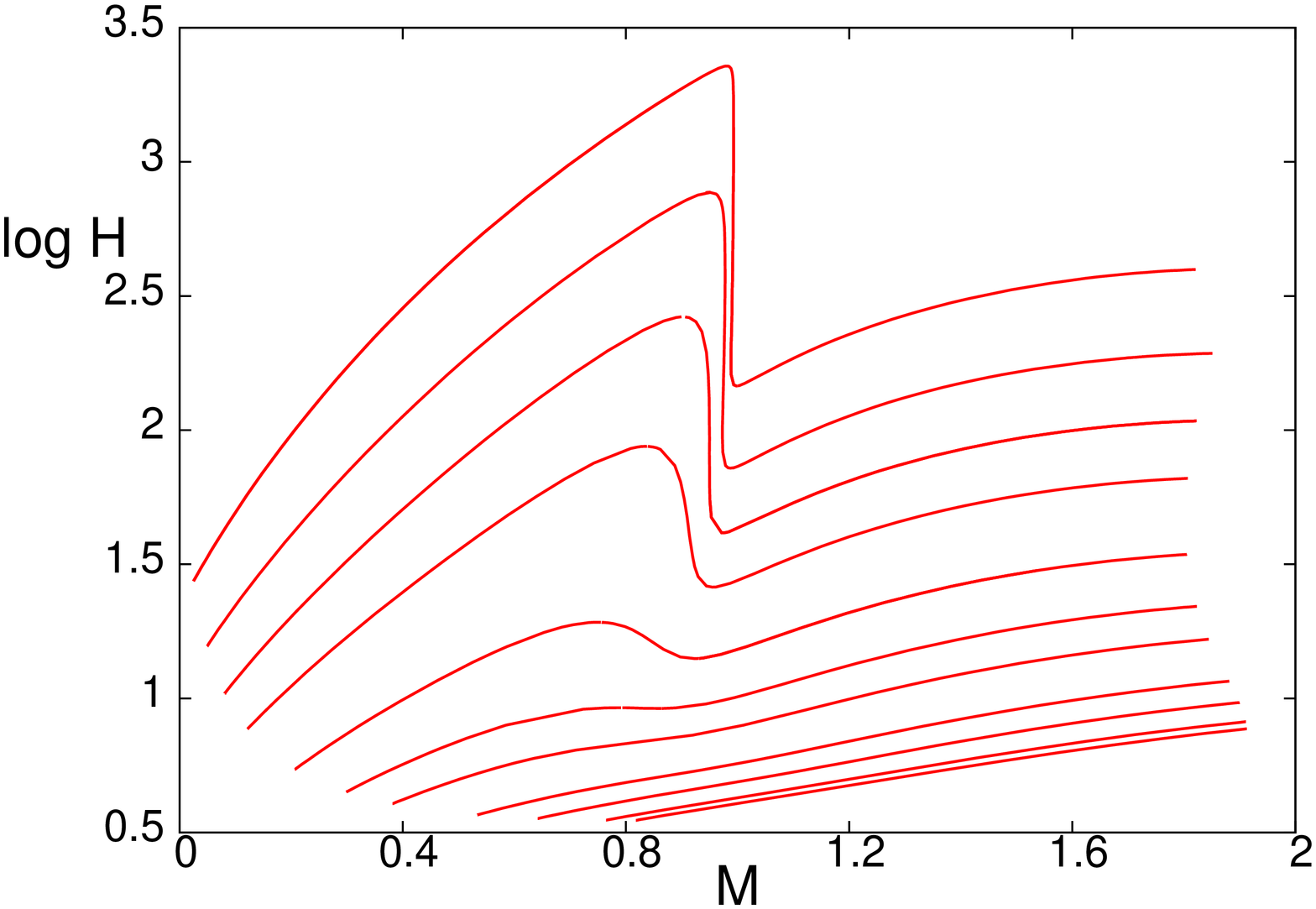}
\par
\caption{(Color online) The projections of the three-dimensional plots from
Fig. \protect\ref{Fig3}(a) (left panel) and Fig.~\protect\ref{Fig3}(c)
(right panel) onto the two-dimensional planes of $\left( M,\Delta \Theta _{%
\protect\phi }\right) $ and $\left( M,\ln H\right) $, respectively.}
\label{Fig4}
\end{figure}

In connection to the pictures displayed in Figs. \ref{Fig3} and \ref{Fig4},
it is relevant to mention that, in the single-component model, the mode with
the unbroken antisymmetry exists in an infinite domain limited from below, $%
1.84<M<\infty $. On the contrary, we here observe that the norms of the
``broken" AS-AS solutions take values in a limited domain. As seen in Fig. %
\ref{Fig4}, the corresponding existence plot also contains a lacuna, and, in
addition to that, it is confined to values $\left\vert \Delta \Theta _{\phi
}\right\vert <1$. A formally similar bifurcation in the single-component
model, should it be possible, might again seem as a loop detached from the
unbroken-antisymmetry modes, which does not reach the limit values of $%
\left\vert \Delta \Theta _{\phi ,\psi }\right\vert =1$.

To consider the bifurcation which actually accounts for emergence of the
``broken" AS-AS solutions from their counterparts with the exact
antisymmetry, it is convenient (as said above) to fix one of the chemical
potentials, $\lambda $ or $\mu $, varying the other one. The analysis of the
numerical results demonstrates that the antisymmetry-breaking bifurcation
occurs, with the variation of $\lambda $, when $|\mu |$ takes values
exceeding $\left( |\mu |\right) _{\min }=0.0601$ (indeed, for $|\mu
|\rightarrow 0$ one should actually return to the single-component model
parameterized by $\lambda $, where the AS mode does not undergo any
bifurcation). Emerging from the branch with the unbroken antisymmetry, the
solution of the broken AS-AS type eventually merges back into the original
branch, as $|\lambda |$ attains a certain maximal value (which increases
with the increase of fixed $|\mu |$). If $|\lambda |$ and/or $|\mu |$ remain
small enough, the numerical results demonstrate a one-to-one correspondence
between the chemical potentials and the corresponding norms, which grow
monotonously with the absolute values of the chemical potentials. However,
unlike the modes of the Sm-Sm type, for the AS-AS solutions this
correspondence breaks down at larger values of $|\lambda |$ and $|\mu |$.

If the fixed chemical potential exceeds another critical value, $\left( |\mu
|\right) _{\mathrm{cr}}=0.198$, an additional bifurcation occurs on the
broken-AS-AS branch, giving rise to two additional solution families. With
the further increase of the absolute value of the varying chemical
potential, $|\lambda |$, at some point $|\lambda |=\left( |\lambda |\right)
_{\max }$ one of the secondary branches merges with the original one from
which it had emerged. The other secondary branch extends to larger values of
$|\lambda |$, and eventually merges back into the underlying family of the
solutions with the unbroken antisymmetry. The range where the secondary
branches are found expands with the increase of fixed $|\mu |$. Another
generic feature of the bifurcation, which is responsible for the merger of
the broken-AS-AS branch back into its unbroken counterpart, is that this
happens when one norm is much larger than the other -- typically, at $%
M\simeq 0.05,$ while $N\simeq 2$.

\subsubsection{Solutions obtained from the symmetric-antisymmetric modes
(S-AS type)}

\begin{figure}[tbh]
\refstepcounter{fig} \label{f-10}
\par
\begin{center}
(a)\hspace{-0.6cm}
\includegraphics[height=.35\textheight,
angle=-90]{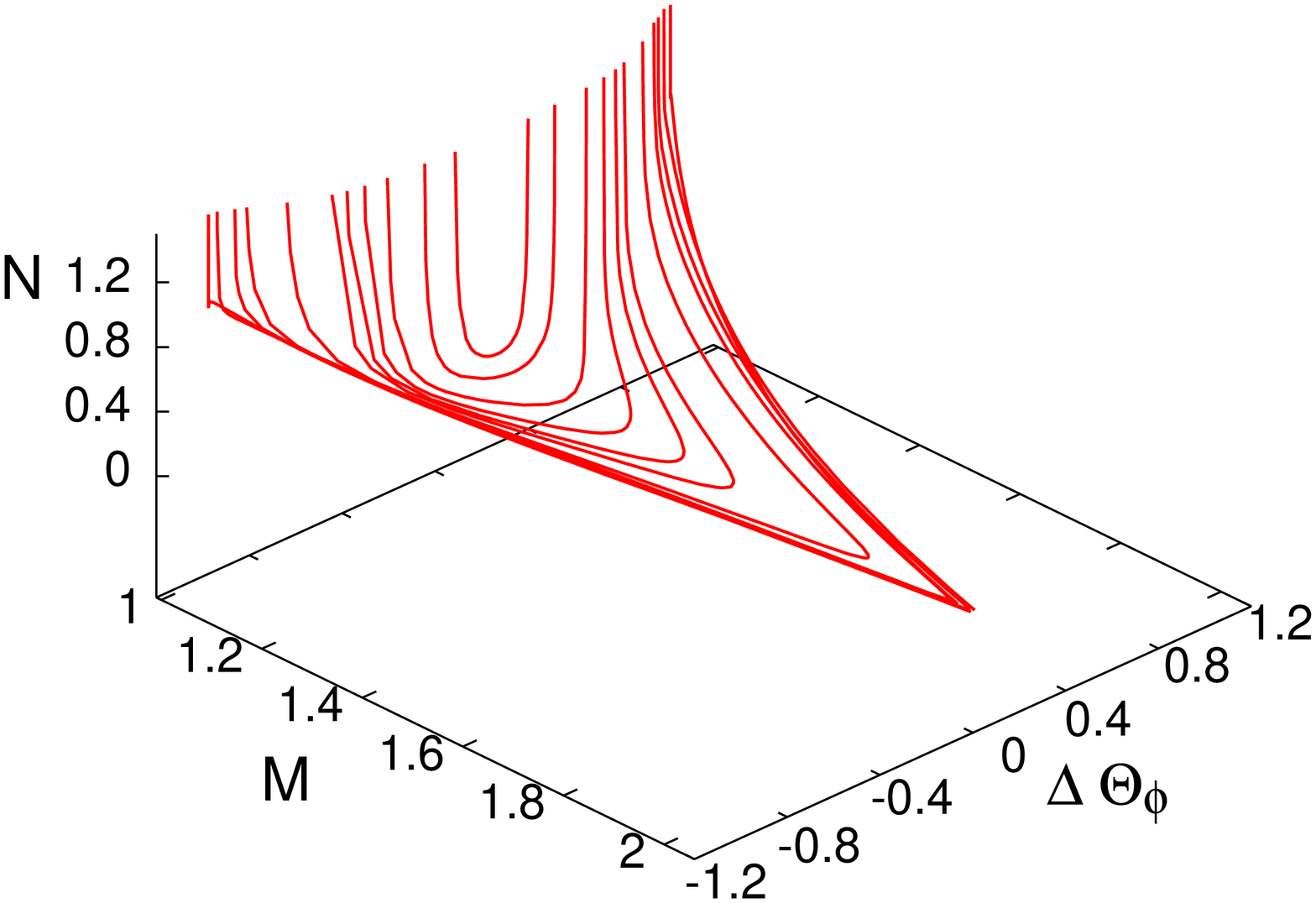} \hspace{0.5cm} (b) %
\includegraphics[height=.35\textheight, angle=-90]{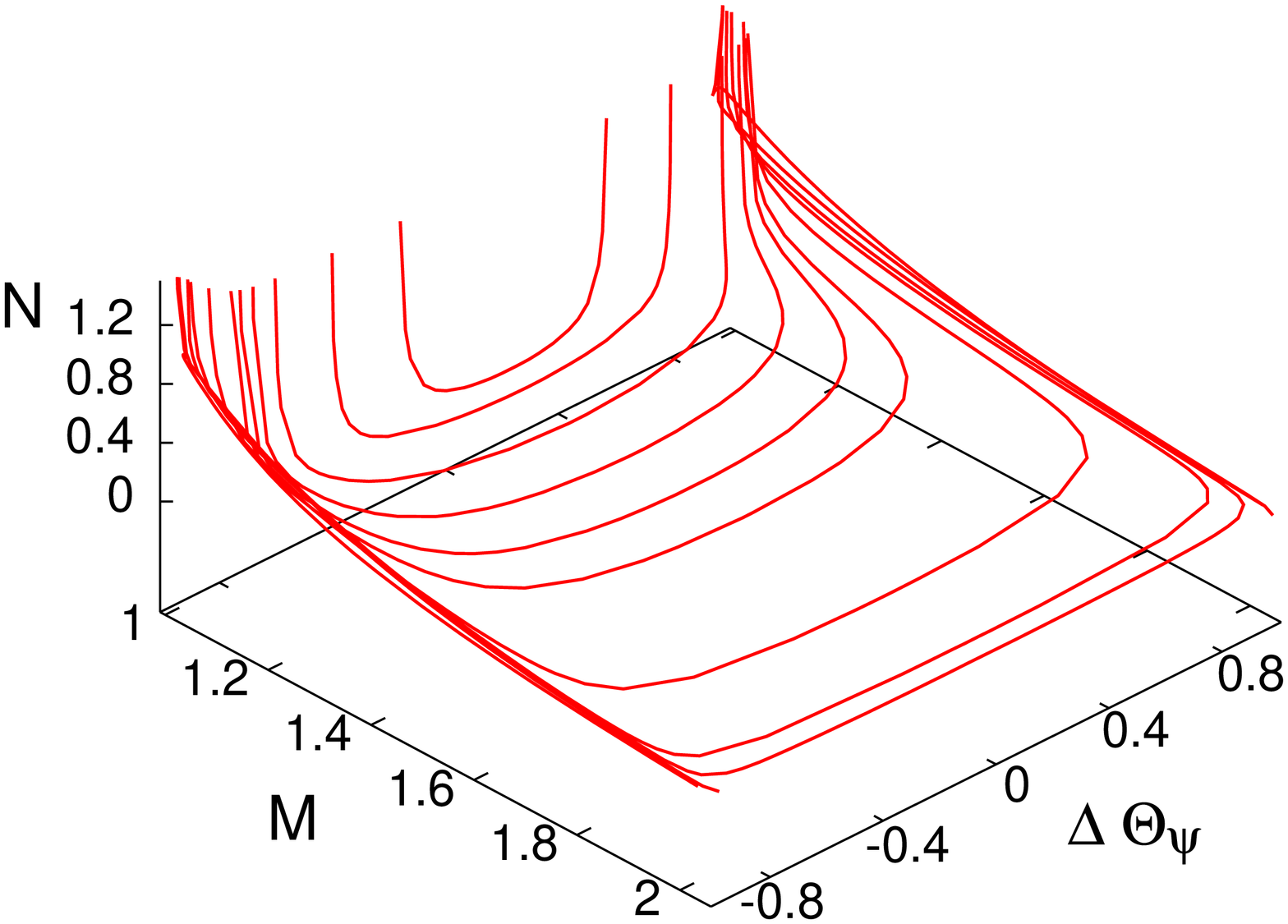} \
(c) \includegraphics[height=.35\textheight, angle =-90]{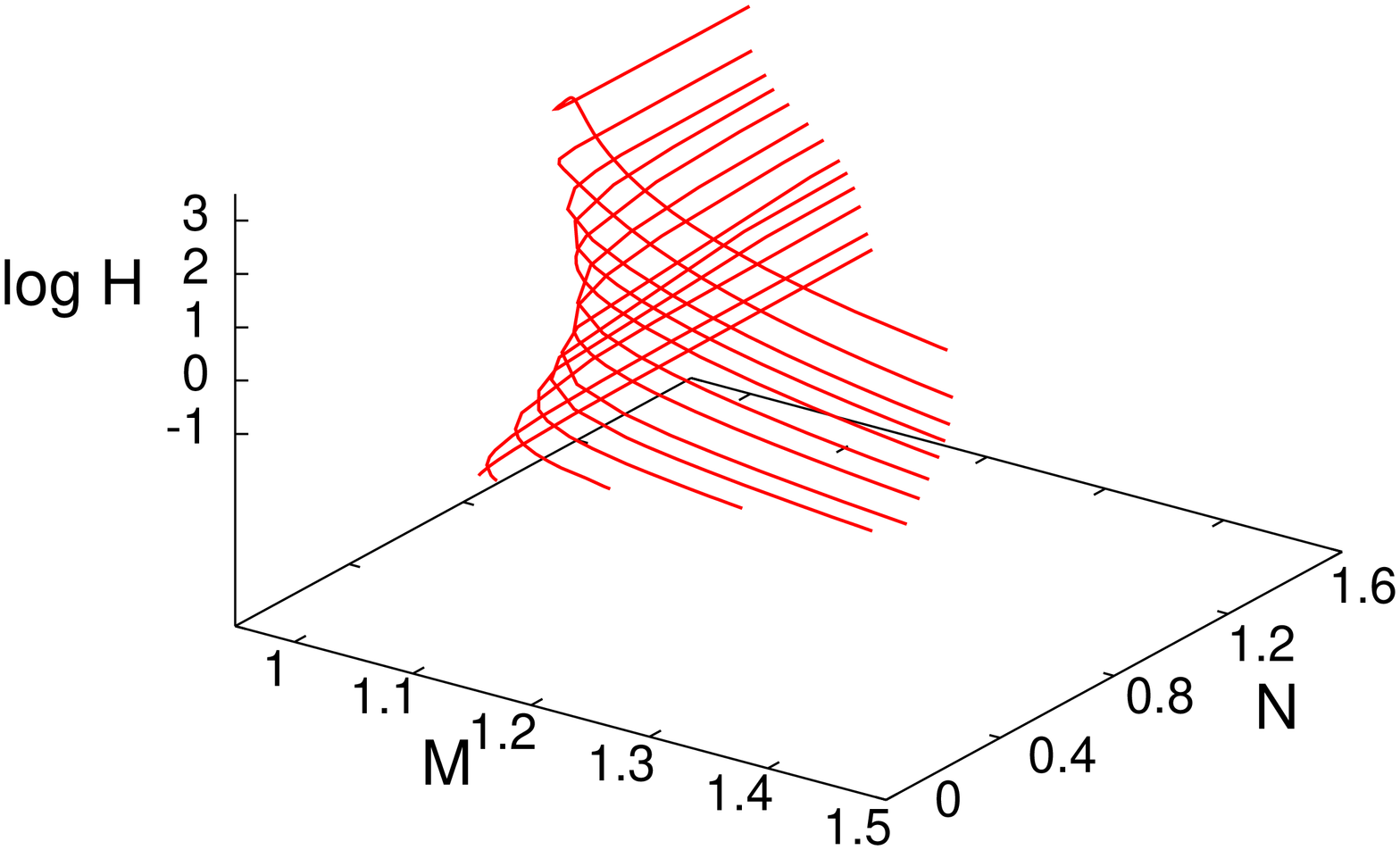}
\end{center}
\par
\vspace{-0.5cm}
\caption{(Color online) The family of mixed-symmetry S-AS modes, with the
broken (anti)symmetry. (a) and (b): The asymmetry measures of the symmetric
and antisymmetric fields, $\protect\phi $ and $\protect\psi $, respectively.
(c) The energy of the S-AS modes with the broken and unbroken
(anti)symmetries vs. norms $M$ and $N$.}
\label{Fig5}
\end{figure}

\begin{figure}[tbh]
\refstepcounter{fig}%
\includegraphics[height=.35\textheight,
angle=-90]{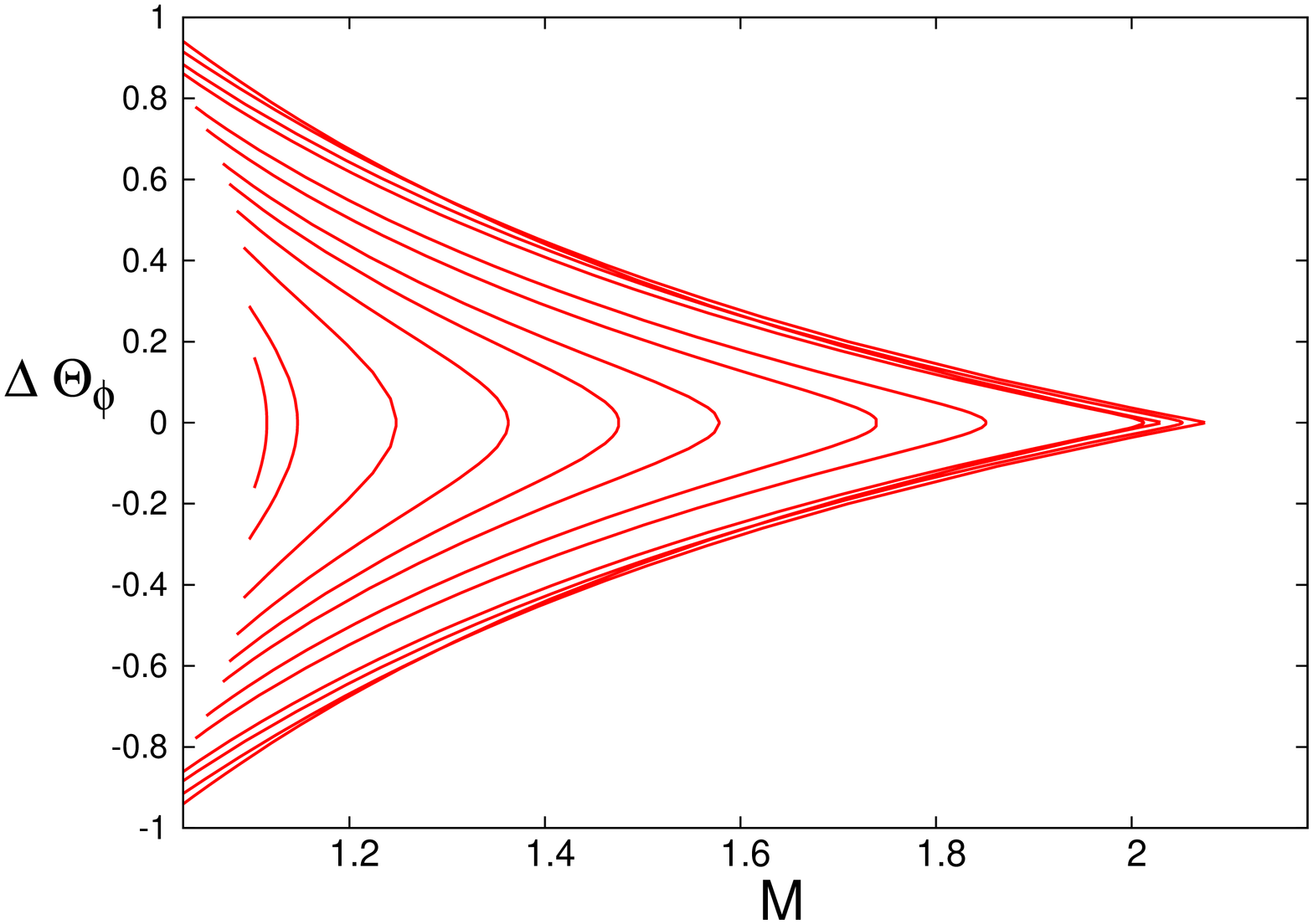}
\includegraphics[height=.35\textheight,
angle=-90]{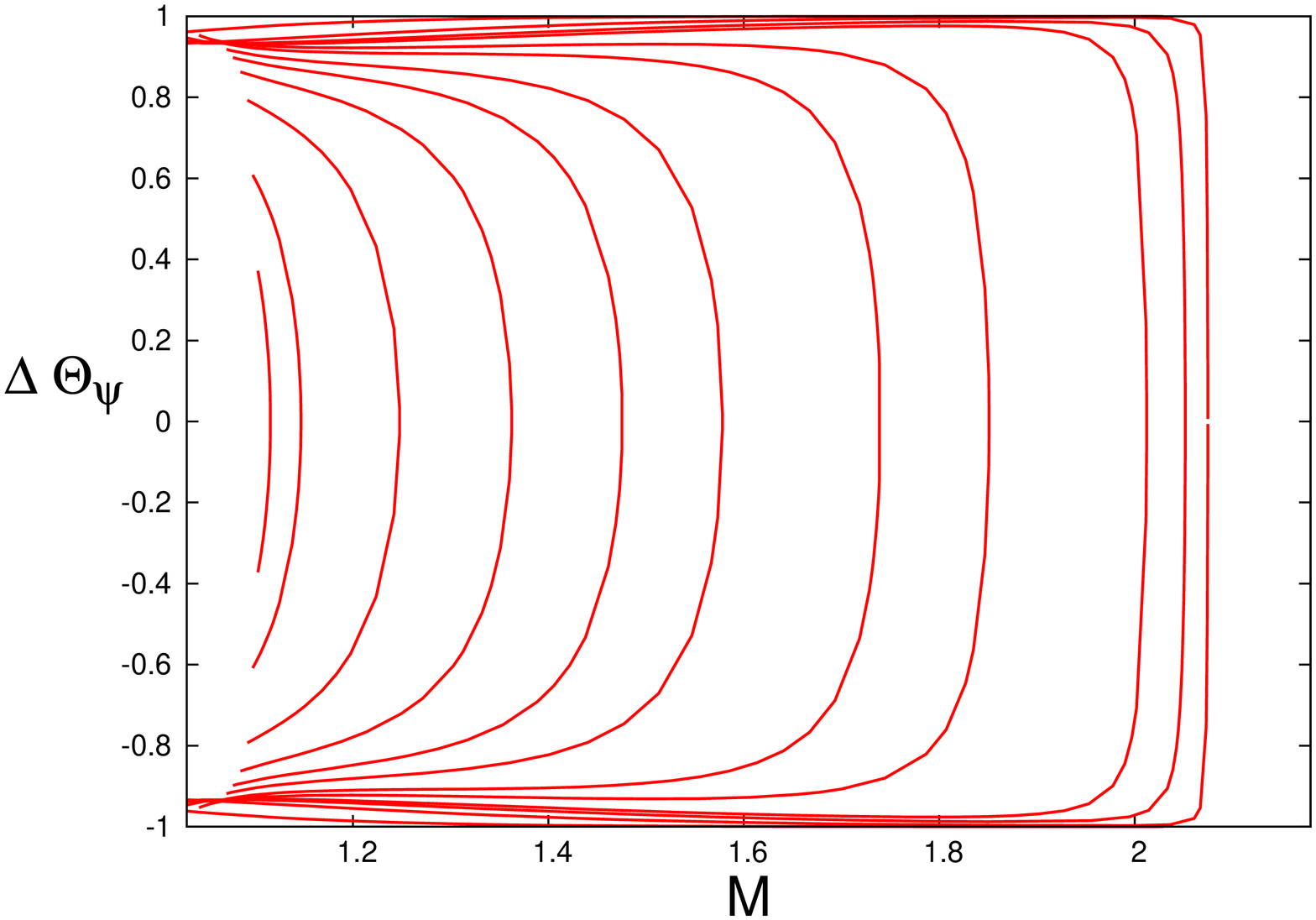}
\par
\caption{(Color online) Projections of the three-dimensional plots from Fig.
\ref{f-10}(a) (left panel) and Fig.~\protect\ref{f-10} (b) (right
panel) onto the two-dimensional planes of $\left( M,\Delta \Theta _{\protect%
\phi }\right) $ and $\left( M,\Delta \Theta _{\protect%
\psi }\right) $,
respectively.}
\label{Fig5a}
\end{figure}

The existence of modes with unbroken or broken mixed symmetry, of the S-AS
type, is another obvious difference of the two-component system from its
single-component predecessor. As might be expected, these solutions exist
only in the region of the $\left( M,N\right) $ plane where the Sm-Sm and
AS-AS states coexist. The global structure of the S-AS family with the
broken symmetry and antisymmetry is displayed in Fig. \ref{Fig5} by means of
plots showing the asymmetry measures of its initially symmetric ($\phi $)
and antisymmetric ($\psi $) components, as well as the total energy, vs. the
norms. Since the pattern of the solutions is rather complex, we complement
Fig. \ref{Fig5} by Fig.~\ref{Fig5a}, where the projection of the
three-dimensional plots of the asymmetry measure of components $\phi $ and $%
\psi $ from Figs.~\ref{f-10}(a),(b) onto the two-dimensional planes of $%
\left( M,\Delta \Theta _{\phi }\right) $ and $\left( M,\Delta \Theta _{\psi
}\right) $, respectively, is presented.

These plots are very different from those for the Sm-Sm and AS-AS\ modes,
cf. Figs. \ref{Fig2}-\ref{Fig4}. To consider the relation between the
mixed-symmetry modes with the their Sm-Sm and AS-AS counterparts, it is
relevant to stress, that, as may be naturally expected, the branch of S-AS
solutions with unbroken (anti)symmetries smoothly merges into the
corresponding solutions on the Sm-Sm type in the limit of the vanishing norm
$N$ in the AS component. On the other hand, when the latter norm ($N$)
increases, the energy of the system increases too, and the configuration
smoothly approaches the opposite limit of the AS-AS solution. If chemical
potential $|\lambda |$ is large enough, the latter limit corresponds to
vanishing norm $M$ of the Sm component. However, for $|\lambda |<0.71$, the $%
M$-norm does \emph{not} vanish in the limiting case; instead, it approaches
some finite value, whereas the $N$-norm rapidly diverges, as seen in Figs. %
\ref{Fig5}(a) and (b). In fact, this singular behavior is a peculiarity of
the model with the $\delta $-functions ($a\rightarrow 0$), while at $a>0$
(in the model with a finite width of the pseudopotential wells) the
situation is different, as shown in the next section.

As well as in the cases considered above, the states of the S-AS type with
the broken (anti)symmetries emerge from their ``unbroken" counterparts
through a bifurcation (i.e., they are connected). We discuss this
bifurcations in more detail below, considering the model with finite $a$.

It can also be checked, using detailed results of the numerical analysis,
that, similar to the case of the Sm-Sm modes, the norms and energy of the
S-AS states monotonously grow with the increase of $|\lambda |$ and $|\mu |$%
, which, in particular, means that the norms are in one-to-one
correspondence with the chemical potentials, also similar to the case of the
Sm-Sm modes, and in contrast to their AS-AS counterparts. Lastly, comparing
Fig. \ref{Fig5}(c) to Fig. \ref{Fig2}(b), we conclude that, for the same
values of norms $M$ and $N$, the energy of the S-AS modes is essentially
higher than the energy of the fundamental Sm-Sm states, which suggests that
the S-AS states may be more vulnerable to dynamical instabilities.

\section{The regular model ($a>0$)}

\subsection{Preliminaries}

The results reported in the previous section were based on the numerical
solution of coupled cubic equations (\ref{A})-(\ref{D}), which were derived
from the analysis of the singular version of the model, with the nonlinear
potential represented by the symmetric set of two $\delta $-functions, see
Eqs. (\ref{udelta}) and (\ref{vdelta}). It is known from the analysis of the
single-component model that this limit case, corresponding to $a\rightarrow
0 $ in Eq. (\ref{g}), is degenerate in several aspects (in particular, the
symmetric mode exists only up to a finite value of the norm, see Eq. (\ref%
{interval})). In the single-component model, the degeneracy was lifted by
the transition to the nonlinearity-modulation function in the general form (%
\ref{g}), with finite $a$ (recall that the modulation function keeps its
double-well shape up to $a=\sqrt{2}$). Another generic difference of the
situation with finite $a$ is that, unlike the case of the nonlinear
potential represented by the $\delta $-functions, smooth modulation function
(\ref{g}) may support localized patterns with local maxima shifted from the
bottom points, $x=\pm 1$.

We have expanded the analysis of the two-component system to the general
case of $g(x)$ taken in the form of (\ref{g}) with $a>0$. As demonstrated
below, the shape of various modes and bifurcations of the respective
branches strongly alter, against the limit of $a\rightarrow 0$, starting
from values $a\gtrsim 0.1$ (in the single-component model, the essential
change in the picture of stationary modes was observed in, roughly, the same
range \cite{Dong}). All the types of the two-component modes, i.e., Sm-Sm,
AS-AS, and S-AS, in their symmetry-broken and unbroken forms alike, feature
strong changes against the case of $a\rightarrow 0$ in this region. In fact,
these changes are strongly affected by additional modes with broken
(anti)symmetry, which appear, at $a$ large enough, still in the \emph{%
single-component} model. These modes were not found in Ref. \cite{Dong},
because they are generated ``from nothing" by saddle-node bifurcations, that
are disconnected from the symmetry-breaking bifurcation for the symmetric
mode, and the stabilization transition for the non-bifurcating antisymmetric
one, on which the analysis was focused in Ref. \cite{Dong}. Therefore, in
this section we first present the new results for the single-component
model, and then report, in a rather brief form, changes which occur to the
compound modes in the two-component system. A full description of the
bifurcations in the two-component system with $a\in \left( 0,\sqrt{2}\right)
$ turns out to be extremely complex, as there is a large number of
ramifications of the bifurcation scenarios.

Some comments are relevant too as concerns the numerical methods used for
the search of localized solutions to ODE system (\ref{u}), (\ref{v}) with $%
a>0$. In addition to the obvious fact that the numerical analysis had to
cover the three-dimensional parameter space, ($a$, $\lambda $, $\mu $), one
of difficulties in the numerical study is related to the fact that, for some
solutions that had to be obtained, components $u(x)$ and $v(x)$ feature very
different (up to a few orders of magnitude) variation rates. Another problem
is that, because for small $a$ the coupling between the equations is
confined to small neighborhoods of points $x=\pm 1$, it was necessary to use
a numerical scheme with a higher density of grid points in those areas.

One~numerical method that was used to solve the system of equations (\ref{u}%
) and (\ref{v}) was based on the shooting algorithm \cite{Khalil2002}, which
is, actually, better suited for producing the solution at smaller values of $%
a$. Another approach, using the centered-difference-approximation method,
turns out to be more appropriate for larger $a$. In particular, the
application of the shooting method was assisted by the fact that, for
smallest values of $a$, the solution has to be close to that available in
the analytical form given by Eqs. (\ref{ulinear}) and (\ref{vlinear}) for $%
a\rightarrow 0$. This fact provided a good initial guess for the shooting
method at $a\lesssim 0.4$. An important issue was also the choice of the
shooting point. The method did not work with these points taken at $\pm $%
"infinity", but it worked well when the point was set in the middle of
either nonlinear-potential well (see Eq. (\ref{g})). By means of this
technique, $5\times 10^{5}$ solutions for the localized modes had been
collected, which made it possible to formulate general conclusions about the
bifurcation scenarios in the two-component system.

\subsection{New results for the single-component model}

\begin{figure}[tbh]
\refstepcounter{fig} \label{f-21}
\par
\begin{center}
(a)\hspace{-0.6cm}%
\includegraphics[height=.28\textheight,
angle=-90]{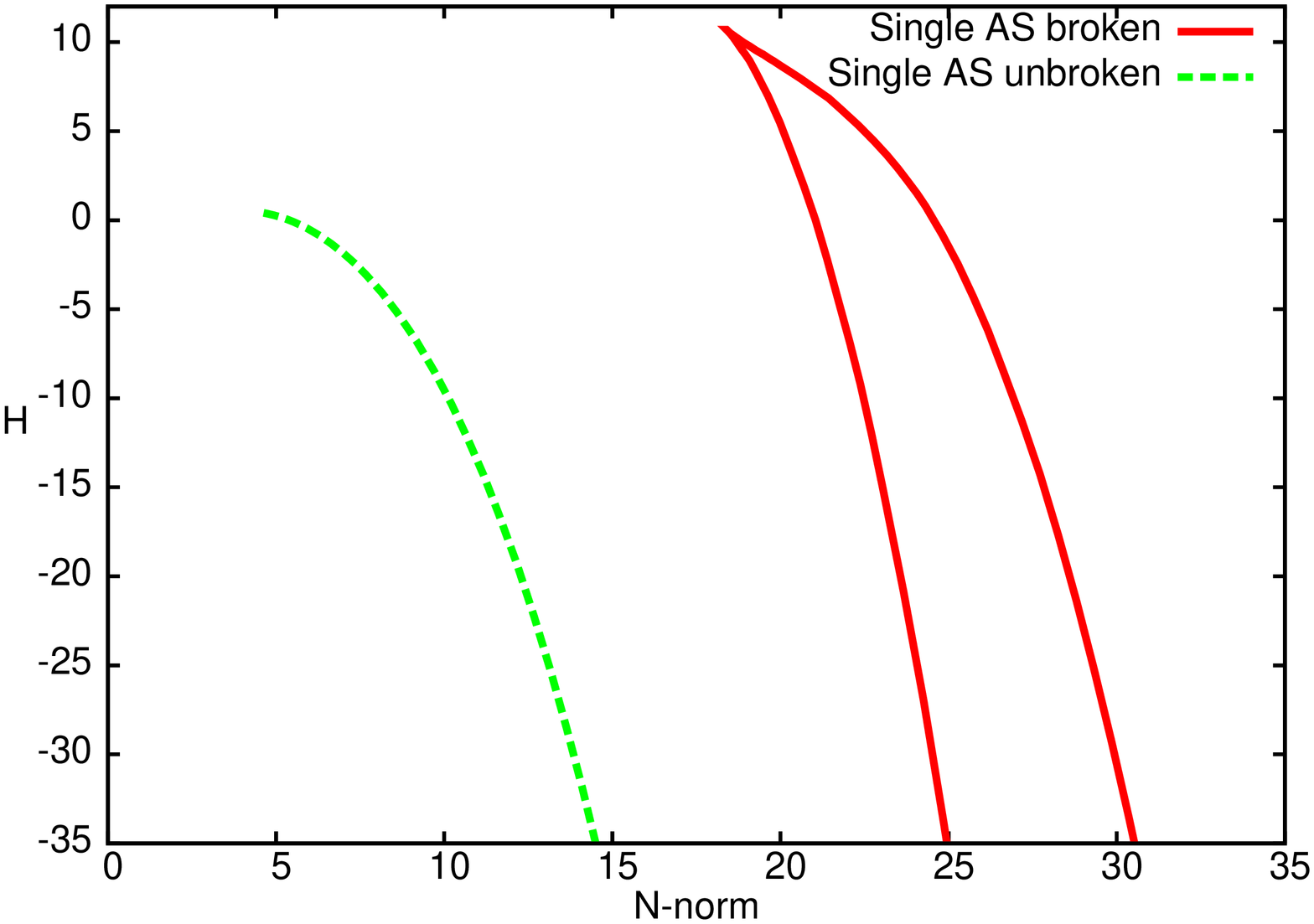} \hspace{0.5cm} (b)%
\includegraphics[height=.28\textheight, angle=-90]{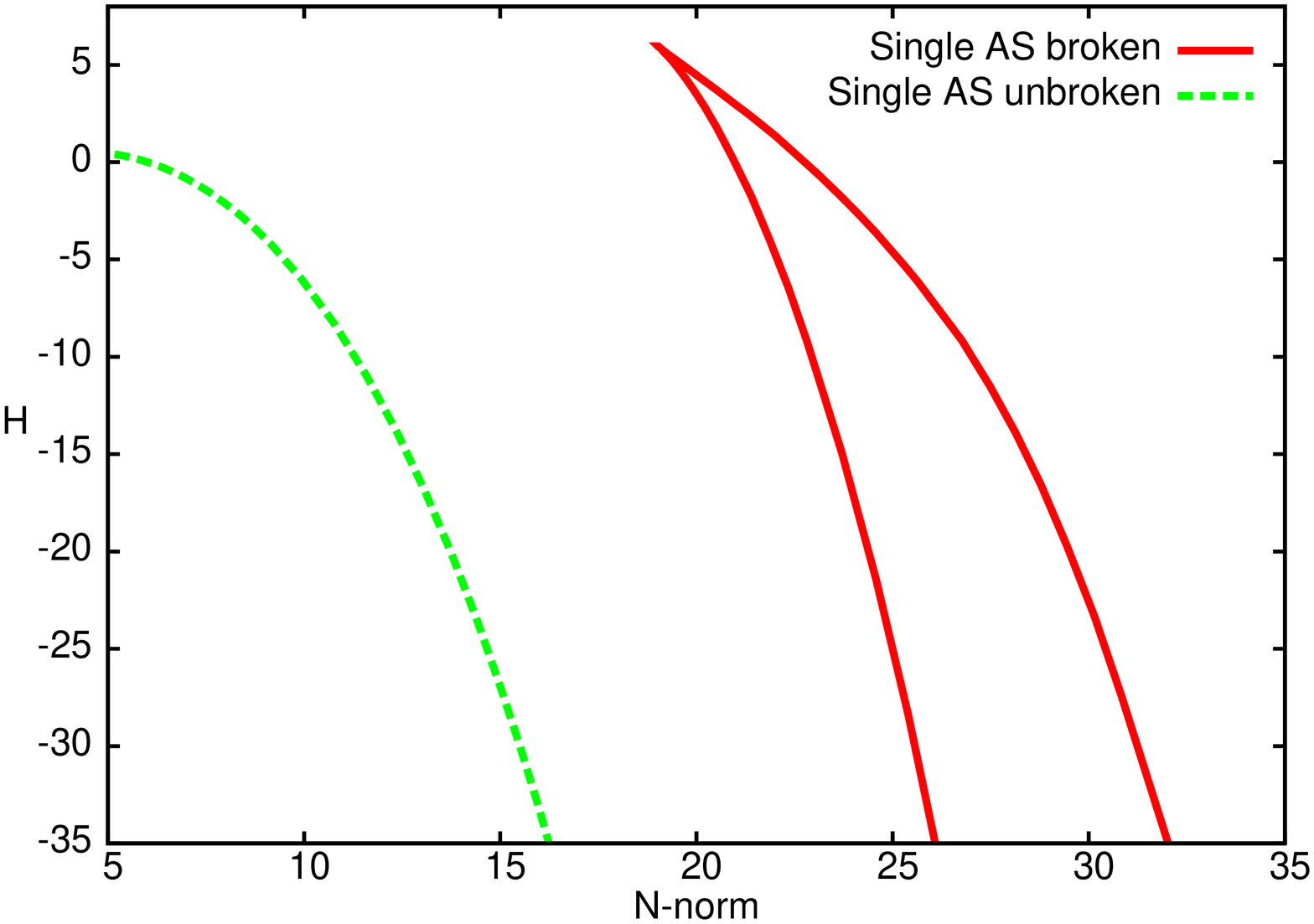}
\hspace{-0.6cm} (c)%
\includegraphics[height=.28\textheight, angle
=-90]{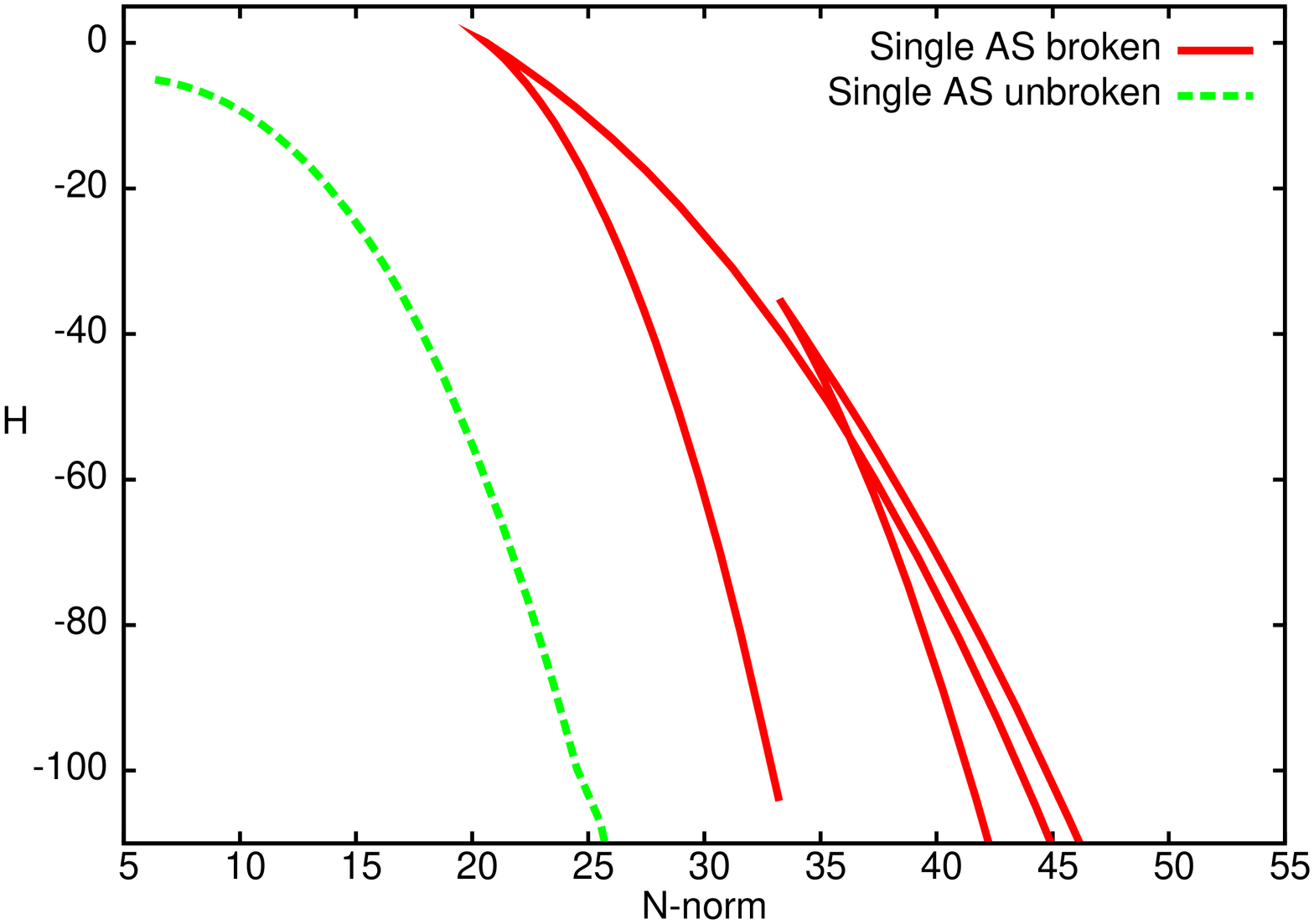} \hspace{0.5cm} (d)\hspace{-0.6cm}
\includegraphics[height=.28\textheight,
angle =-90]{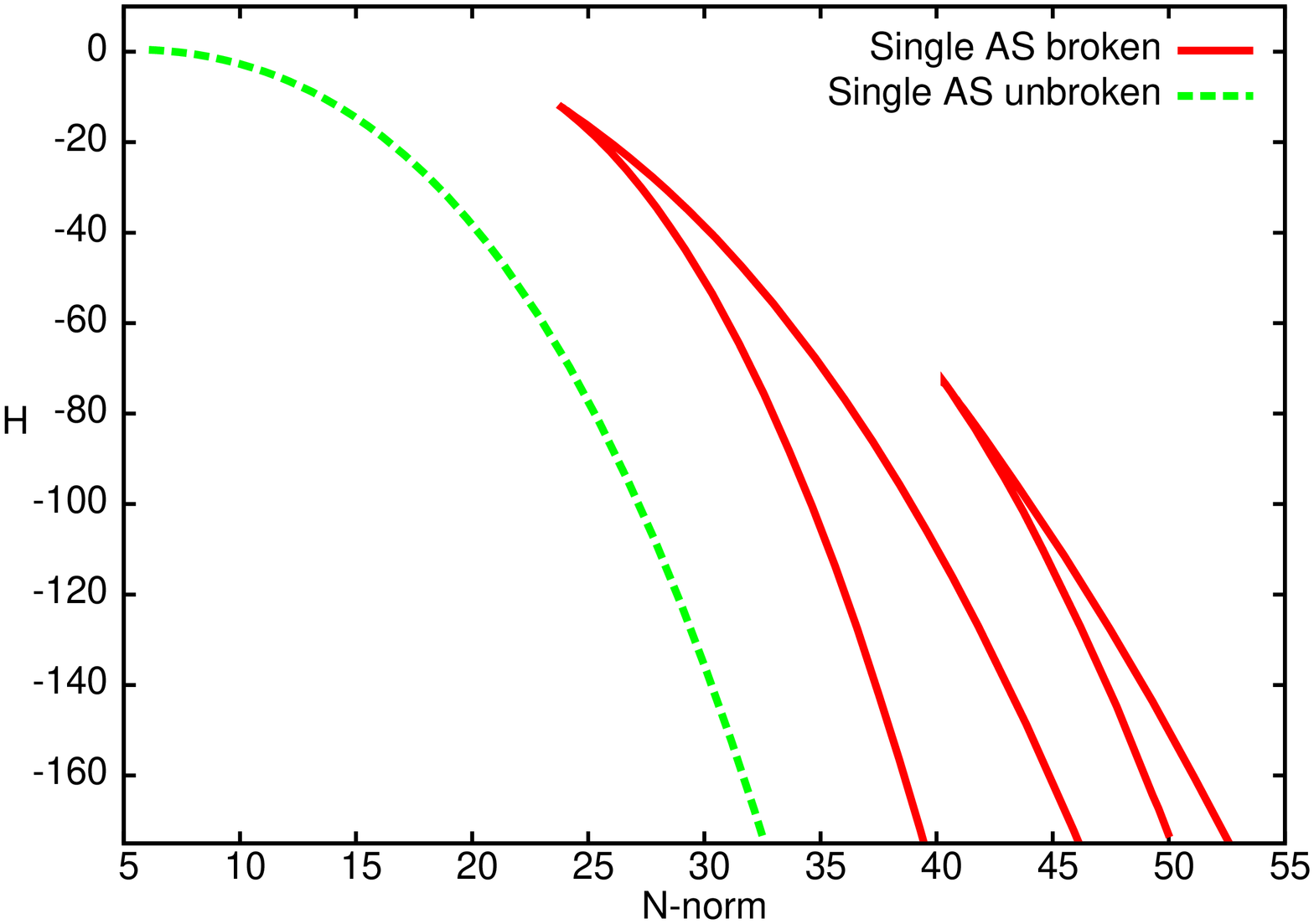}
\end{center}
\par
\vspace{-0.5cm}
\caption{(Color online) Saddle-node bifurcations which generate new pairs of
modes with broken antisymmetry in the single-component model (word ``single"
in the plots stresses this fact) for $a=0.50$~(a), $a=0.60$~(b), $a=0.70$%
~(c), and $a=0.80$~(d). }
\label{Fig6}
\end{figure}

\begin{figure}[h]
\refstepcounter{fig} \label{f-17b}
\par
\begin{center}
(a)\hspace{-0.6cm}
\includegraphics[height=.35\textheight,
angle=-90]{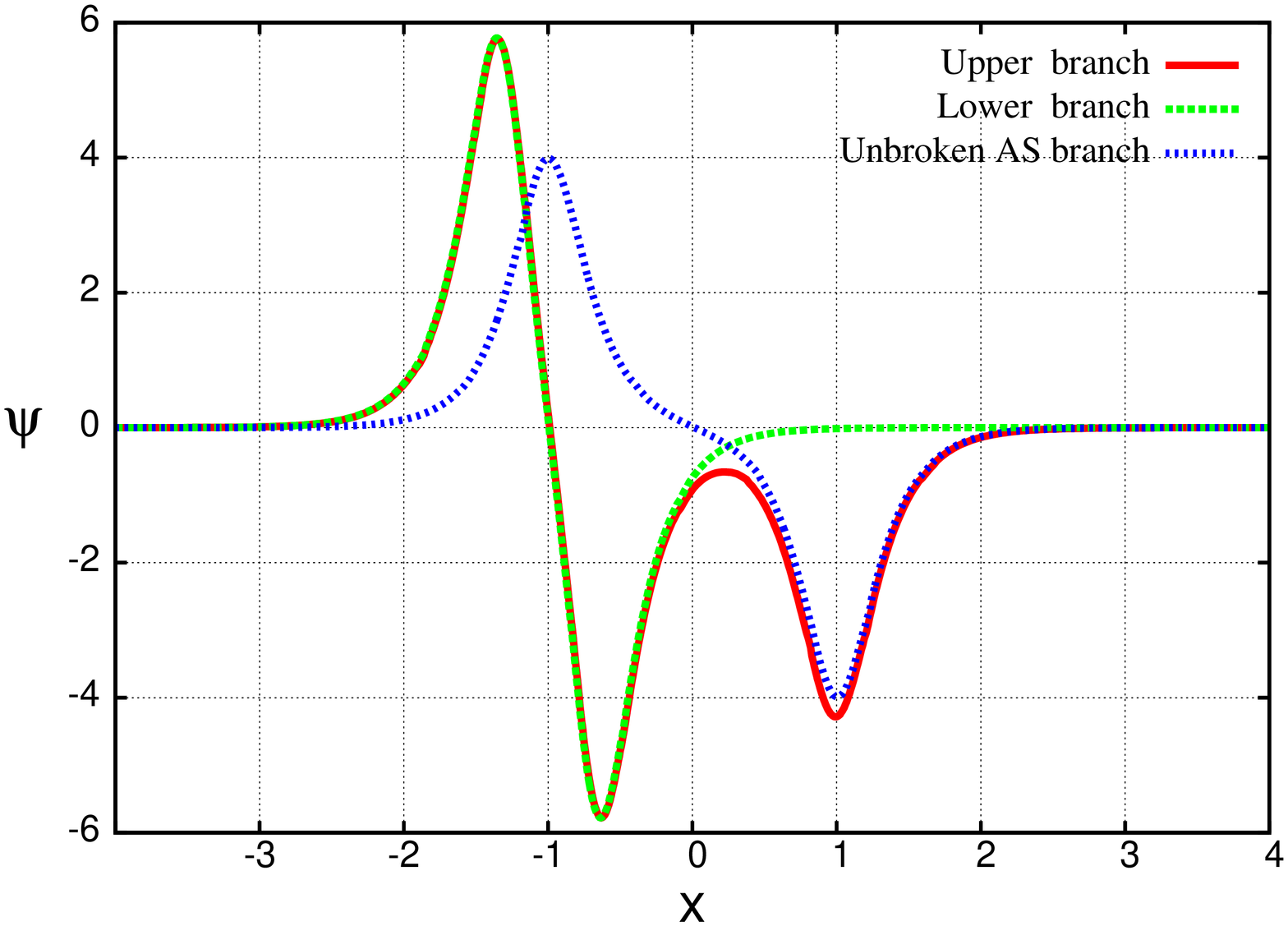}
(b)\hspace{-0.6cm}
\includegraphics[height=.35\textheight, angle
=-90]{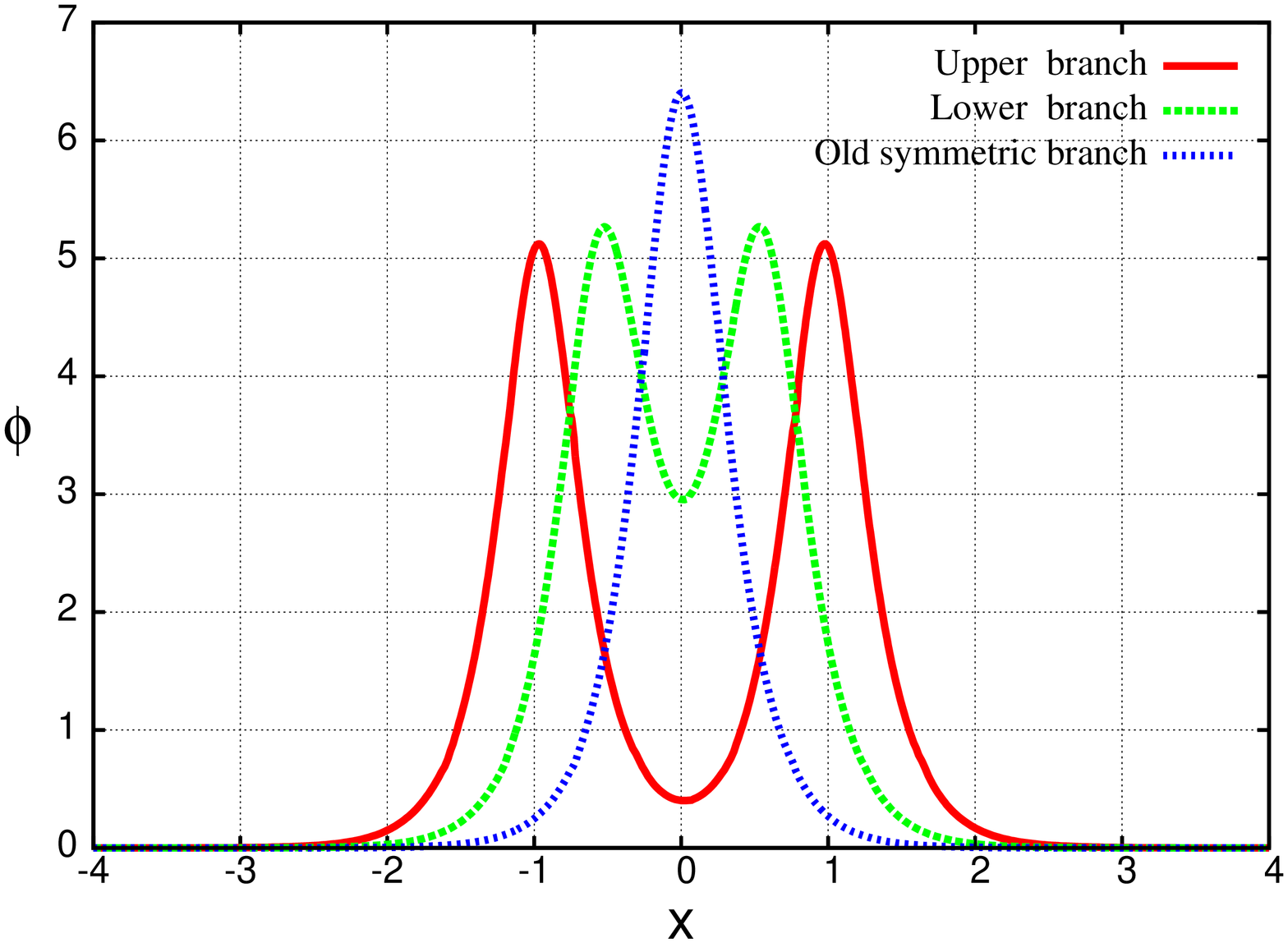}
\end{center}
\par
\vspace{-0.5cm}
\caption{(Color online) (a) A typical example of the pair of
single-component modes with broken antisymmetry, which are generated by the
saddle-node bifurcation shown in Fig. \protect\ref{Fig6}. This example is
displayed for $a=0.60$ and chemical potential $|\protect\lambda |=8.0$ of
the mode with the unbroken antisymmetry. (b) A typical example of the pair
of symmetric single-component double-peak modes which are generated by the
saddle-node bifurcation shown in Fig. \protect\ref{f-19}. This example is
displayed for $a=0.90$ and chemical potential $|\protect\lambda |=8.0$ of
the original mode with the unbroken antisymmetry. ``Upper" and ``lower"
modes, generated by the saddle-node bifurcations, are defined according to
the positions of the respective branches in Figs. \protect\ref{Fig6} and
\protect\ref{Fig8}.}
\label{Fig7}
\end{figure}

In Ref. \cite{Dong}, no bifurcations were found on the branch of
single-component antisymmetric modes, both for $a\rightarrow 0$ (where this
result was exact) and for finite $a$ (in the numerical form). Our analysis
complies with those conclusions. Nevertheless, as shown in Fig. \ref{Fig6},
our numerical studies, performed at finite $a$, reveal saddle-node
bifurcations (more than a single one, see Fig. \ref{Fig6}(c,d)), that
generate pairs of single-component modes with broken antisymmetry, which are
disjoint from the branch with the unbroken antisymmetry (for this reason,
they were not found in Ref. \cite{Dong}). An example of the pair of the
newly found solutions with the broken antisymmetry is displayed in Fig. \ref%
{Fig7}(a). An essential peculiarity of these modes, which is seen, for
instance, in the case of the lower mode displayed in Fig. \ref{Fig7}(a), is
that they tend to be trapped in a single nonlinear-potential well, instead
of being spread between both, which is the case for the mode with the
unbroken antisymmetry. It will be shown below that such modes, essentially
trapped in a single pseudopotential well, play an essential role in
bifurcation scenarios of compound modes of the AS-AS and S-AS types in the
two-component system at finite $a$.

\begin{figure}[tbh]
\refstepcounter{fig} \label{f-19}
\par
\begin{center}
(a)\hspace{-0.6cm}
\includegraphics[height=.23\textheight,
angle=-90]{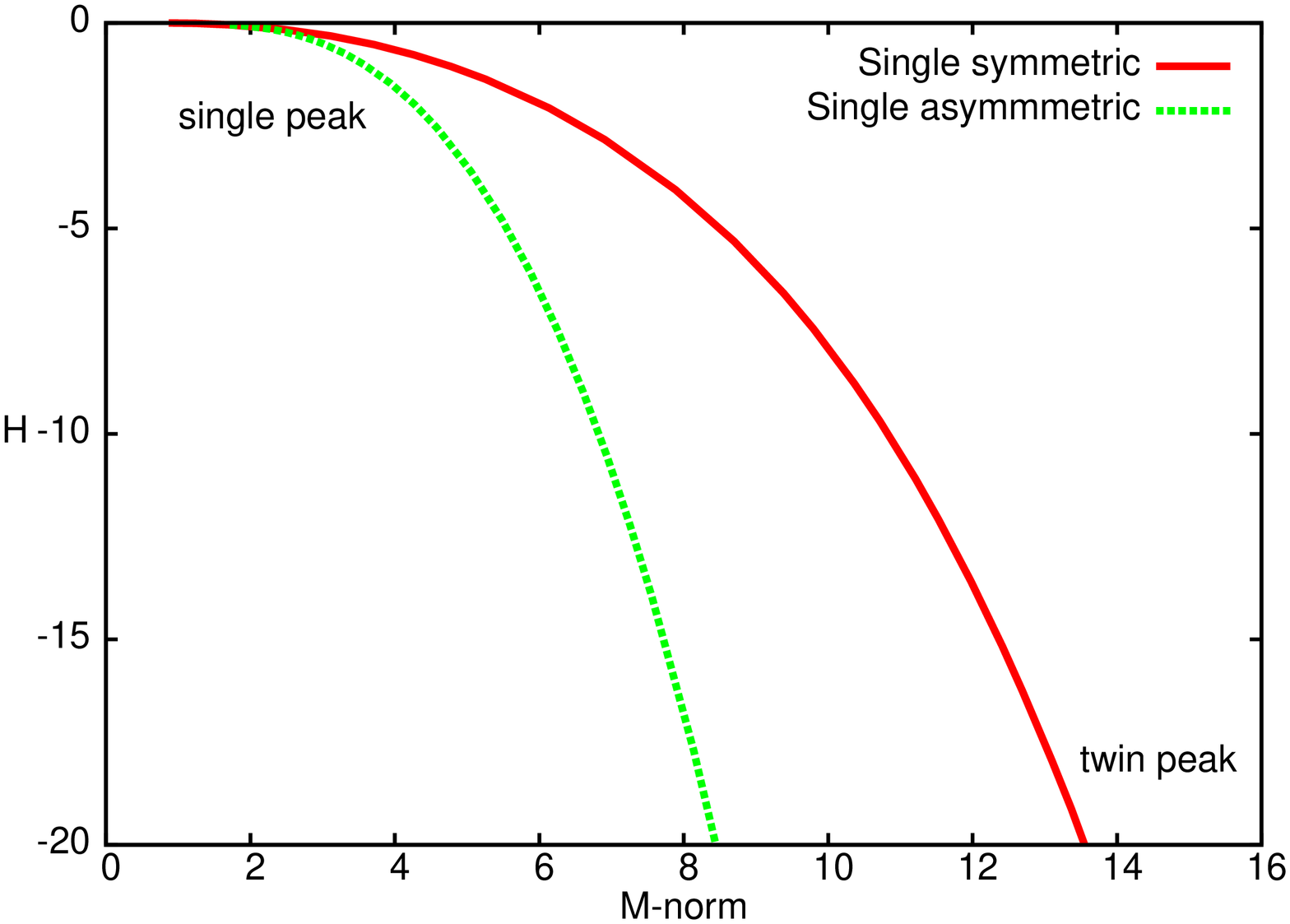} \hspace{0.5cm} (b)\hspace{-0.6cm} %
\includegraphics[height=.23\textheight, angle =-90]{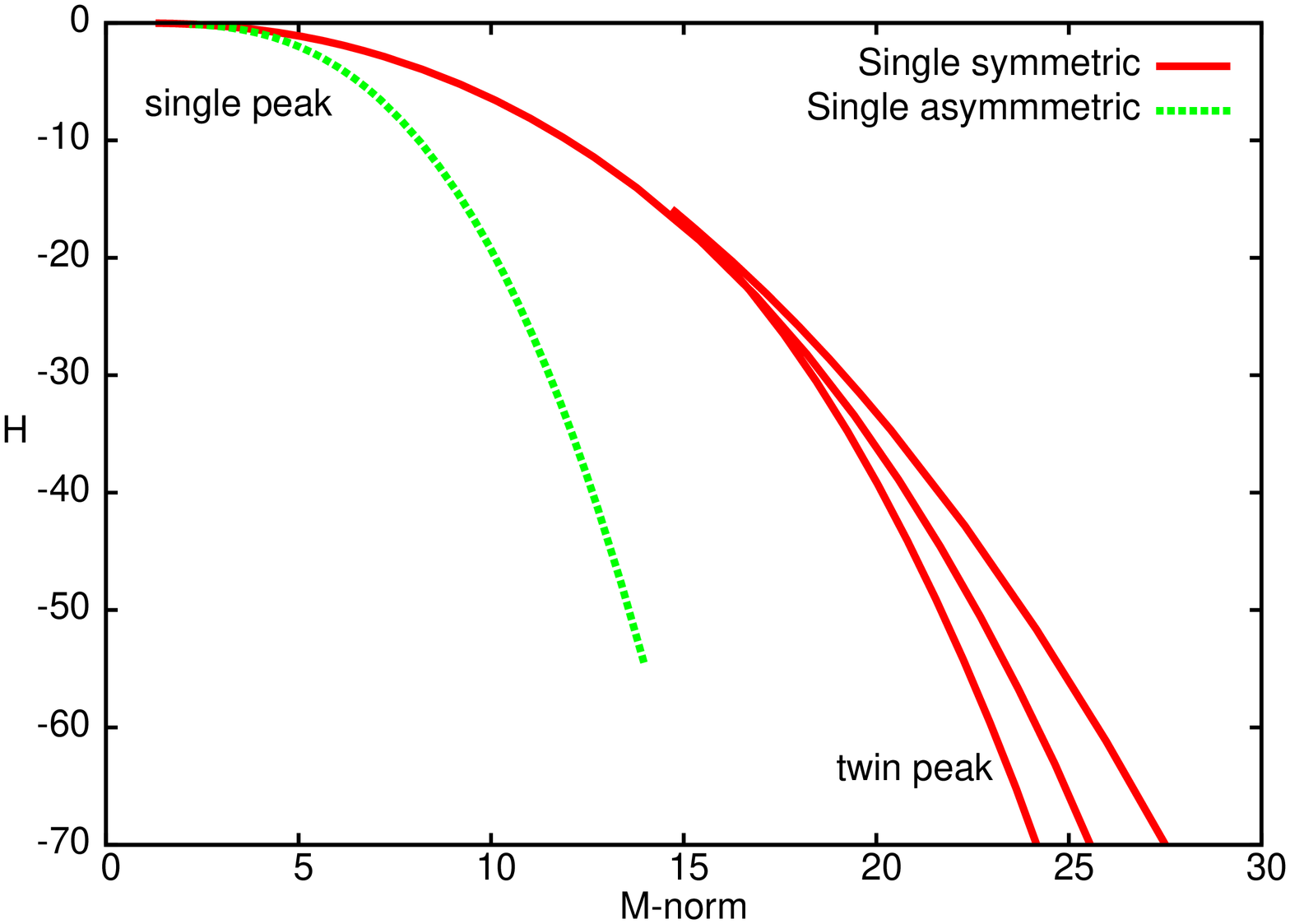}
\hspace{0.5cm} (c)\hspace{-0.6cm}
\includegraphics[height=.23\textheight,
angle =-90]{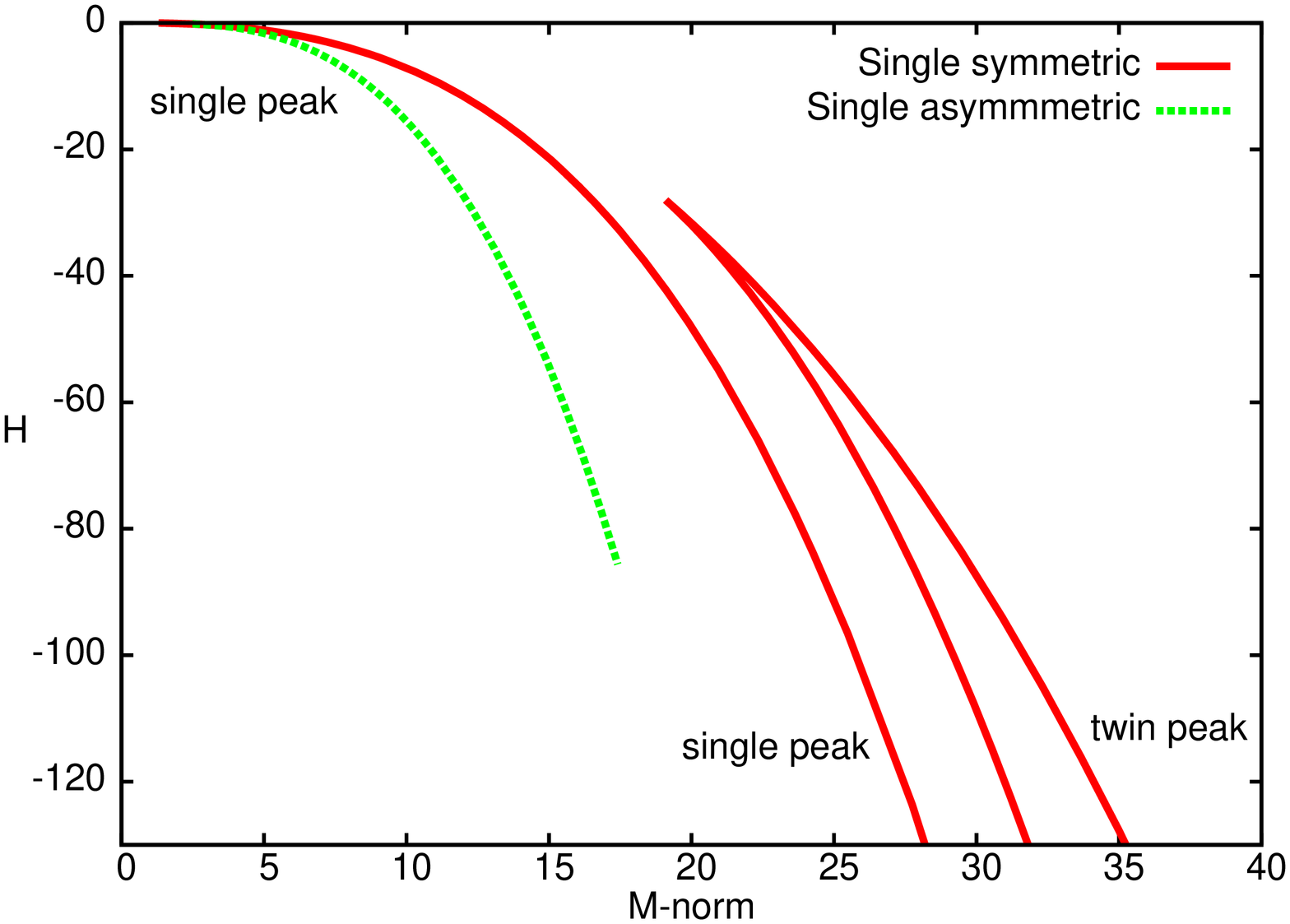}
\end{center}
\par
\vspace{-0.5cm}
\caption{(Color online) The saddle-node bifurcation which generates a new
pair of symmetric modes in the single-component model. At $a=0.60$ (prior to
the saddle-node bifurcation, panel (a)), only the ``old" branch of the
symmetric solutions is present. New symmetric branches, generated by the
bifurcation, appear as $a\geq 0.72$. The new symmetric branches are plotted,
along with the old one, for $a=0.80$ (b), and for $a=0.90$ (c). }
\label{Fig8}
\end{figure}

Furthermore, we observe similar saddle-node bifurcations of symmetric and
asymmetric solutions in the single-component model. In particular, a pair of
new symmetric-solution branches, generated by such a bifurcation, is
displayed in Fig.~\ref{Fig8}. These bifurcations take place on the ``old"
symmetric and asymmetric branches at some critical values of $a$ (in
contrast to the bifurcations displayed in Fig. \ref{Fig6}, which are
disconnected from the ``old" branch of the antisymmetric solutions). For the
asymmetric and symmetric solutions, this happens at $a=0.67$ and $a=0.72$,
respectively (the bifurcation which gives rise to the pair of the asymmetric
branches is not displayed here). As $a$ grows beyond these critical values,
the new branches move away from the parent ones. It may happen that more
pairs of branches emerge through additional saddle-node bifurcations (cf.
Fig. \ref{Fig6}, where two such bifurcations are shown for the broken
antisymmetric solutions), but we did not aim to look for them.

The difference of the newly emerging symmetric states from the original one,
which evolves continuously starting from $a=0$, is that, at relatively large
values of $a$, the original mode features a single-peak shape, while the new
solutions are shaped as twin-peak modes, see Fig. \ref{Fig7}(b).

\subsection{Bifurcations of the symmetric-symmetric (Sm-Sm) branches in the
two-component system}

To understand the evolution of the Sm-Sm branches at finite $a$ in the
two-component system, we select a characteristic value, $a=0.6$. The
analysis was performed by fixing chemical potential $\lambda $ which is
associated with component $\phi $, and scanning the configuration space of
the system by letting vary the other chemical potential, $\mu $, which is
associated with component $\psi $. At all values of $\lambda $ and $\mu $,
branches of solutions of the Sm-Sm types with unbroken and broken
symmetries, have been found, the former ones always having lower energy (\ref%
{H}), which suggests that the symmetry breaking may tend to destabilize the
modes.

The branch of the Sm-Sm solutions with the broken symmetry smoothly arises
from the corresponding \emph{single-component} asymmetric state, trapped in
one of the nonlinear potential wells. In the course of its evolution
following the variation of $\mu $, this branch evolves towards the other
well, where the \emph{other} component of the ``broken" Sm-Sm solution
approaches the respective asymmetric state of the single-component model,
while the mate component vanishes.

\begin{figure}[tbh]
\refstepcounter{fig} \label{f-30}
\par
\begin{center}
(a)\hspace{-0.6cm}
\includegraphics[height=.34\textheight, angle =-90]{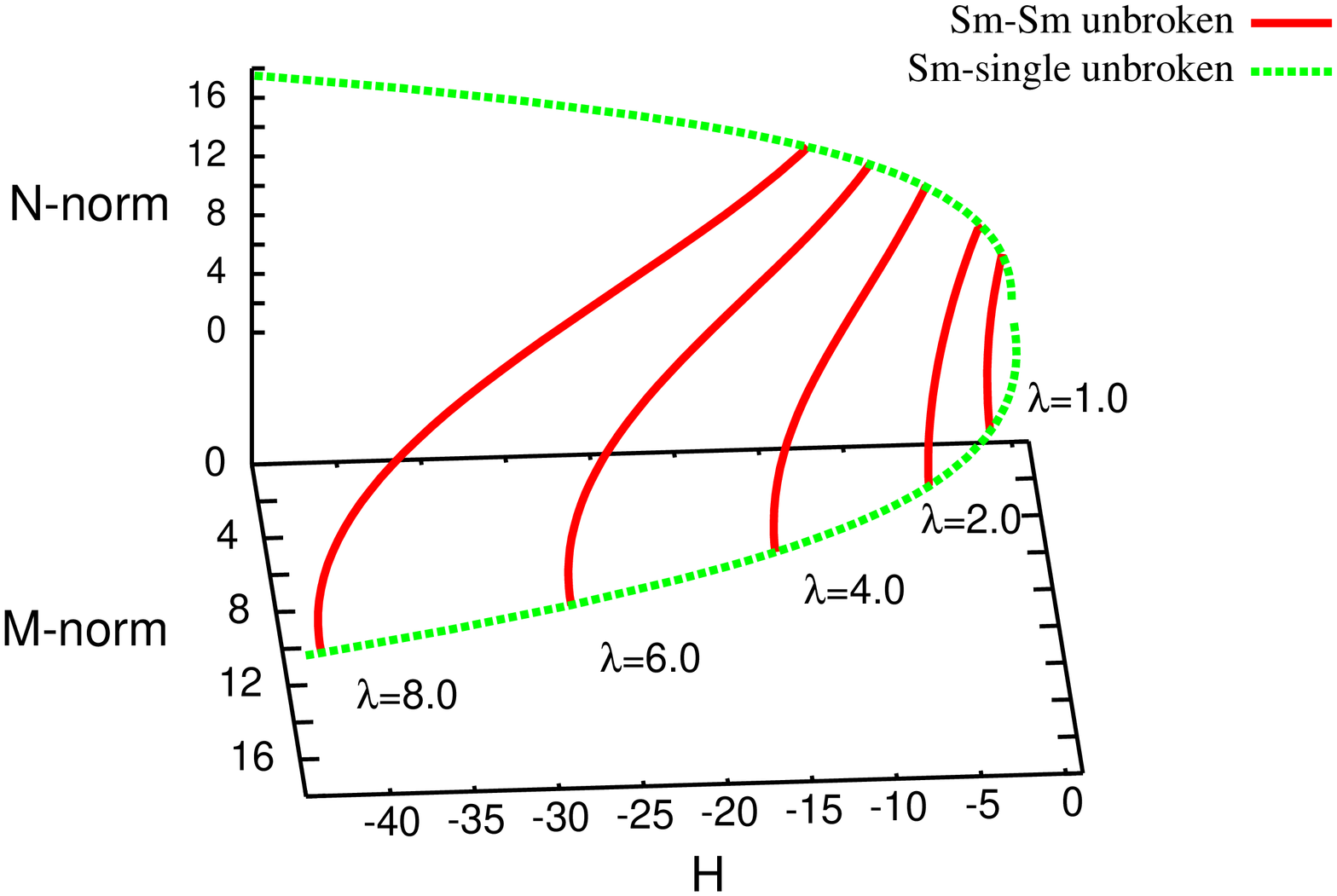} (b)\hspace{%
-0.6cm} \includegraphics[height=.34\textheight, angle =-90]{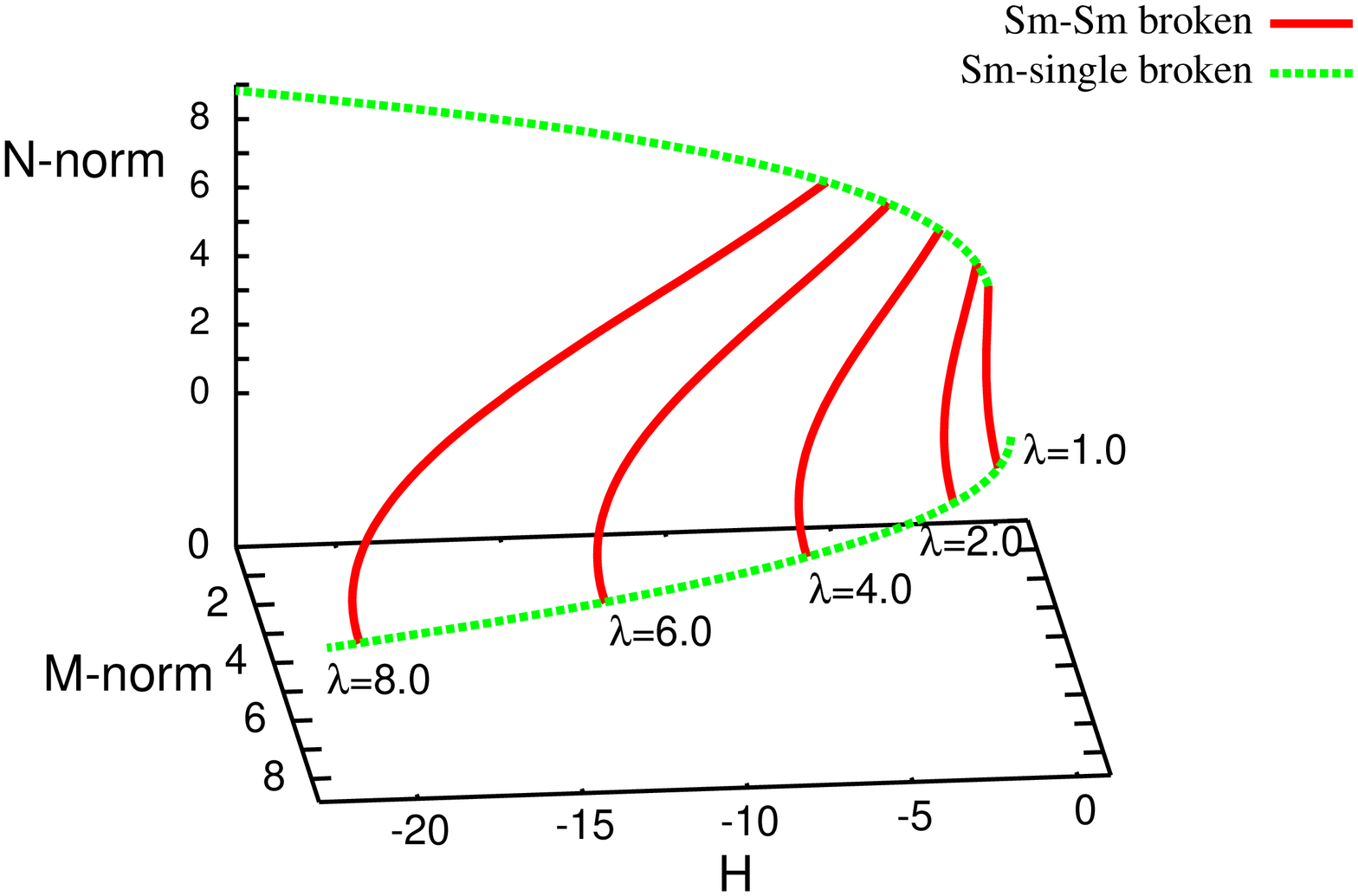}
\end{center}
\par
\vspace{-0.5cm}
\caption{(Color online) The bifurcation scenarios, at $a=0.60$, for the
Sm-Sm modes with unbroken (a) and broken (b) symmetries, displayed by means
of curves showing the evolution of energy (\protect\ref{H}) with the
variation of chemical potential $\protect\mu $, at the following fixed
values of the chemical potential of the other component: $|\protect\lambda %
|=1.0,2.0,4.0,6.0$ $8.0$. }
\label{Fig9}
\end{figure}

Similarly, the Sm-Sm branch with the unbroken symmetry arises from the
corresponding symmetric solution of the single-component model, at some
critical value of $\mu $, and eventually merges into another symmetric
state, in which, again, only a single component is present, but the other
one. Thus, in both cases of the Sm-Sm modes with broken and unbroken
symmetries, the evolution along the branch of the two-component solutions
essentially amounts to the conversion of one single-component mode (resp.,
asymmetric or symmetric) into the mode represented by the originally missing
component, while the opposite one vanishes. These bifurcation scenarios for
the Sm-Sm branches are illustrated by Fig. \ref{Fig9}. They persist in the
interval of $0.16<a<0.72$.

\begin{figure}[tbh]
\refstepcounter{fig} \label{f-31}
\par
\begin{center}
(a)\hspace{-0.6cm}
\includegraphics[height=.35\textheight,
angle=-90]{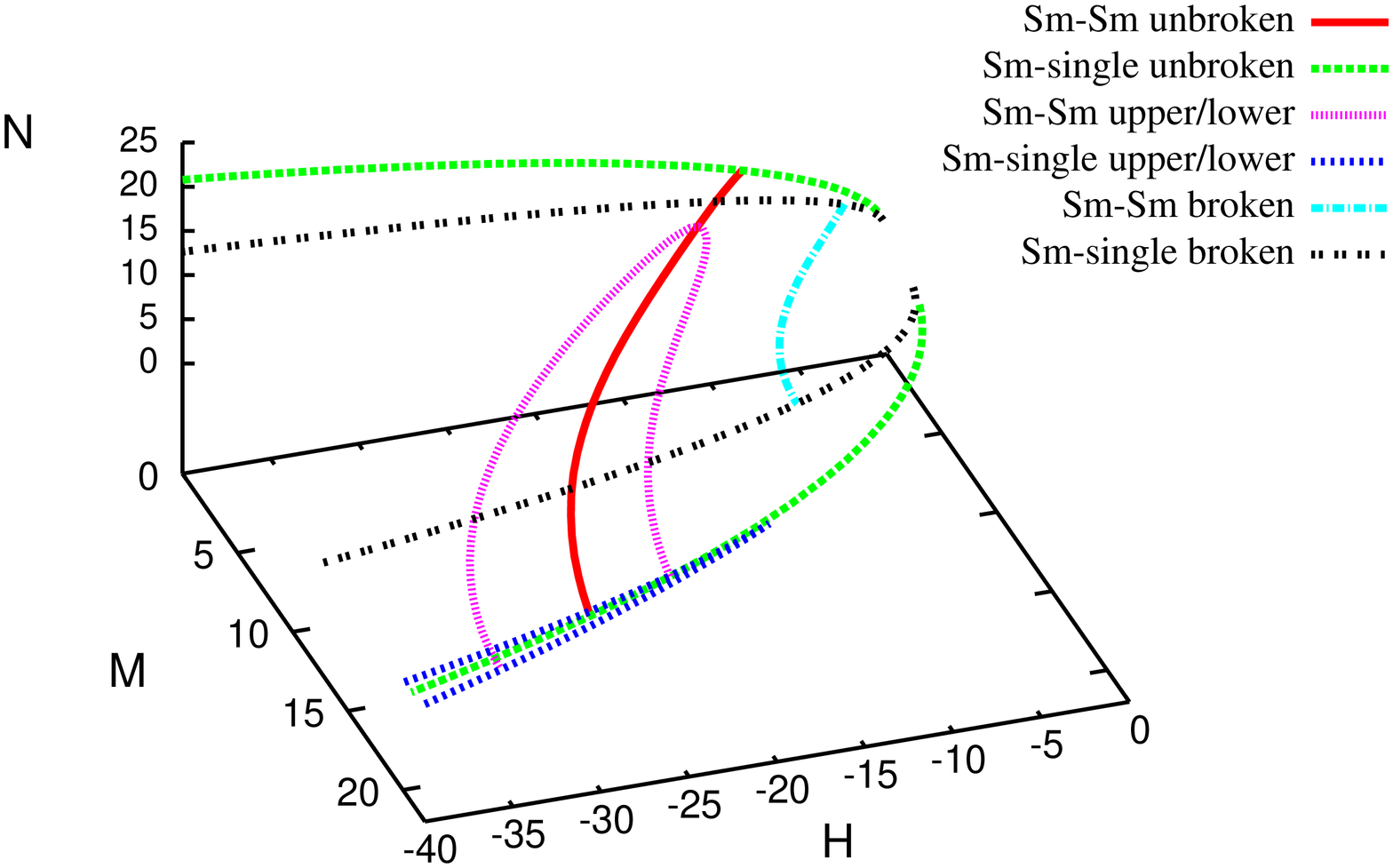} \hspace{0.5cm} (b)\hspace{-0.6cm} %
\includegraphics[height=.35\textheight, angle =-90]{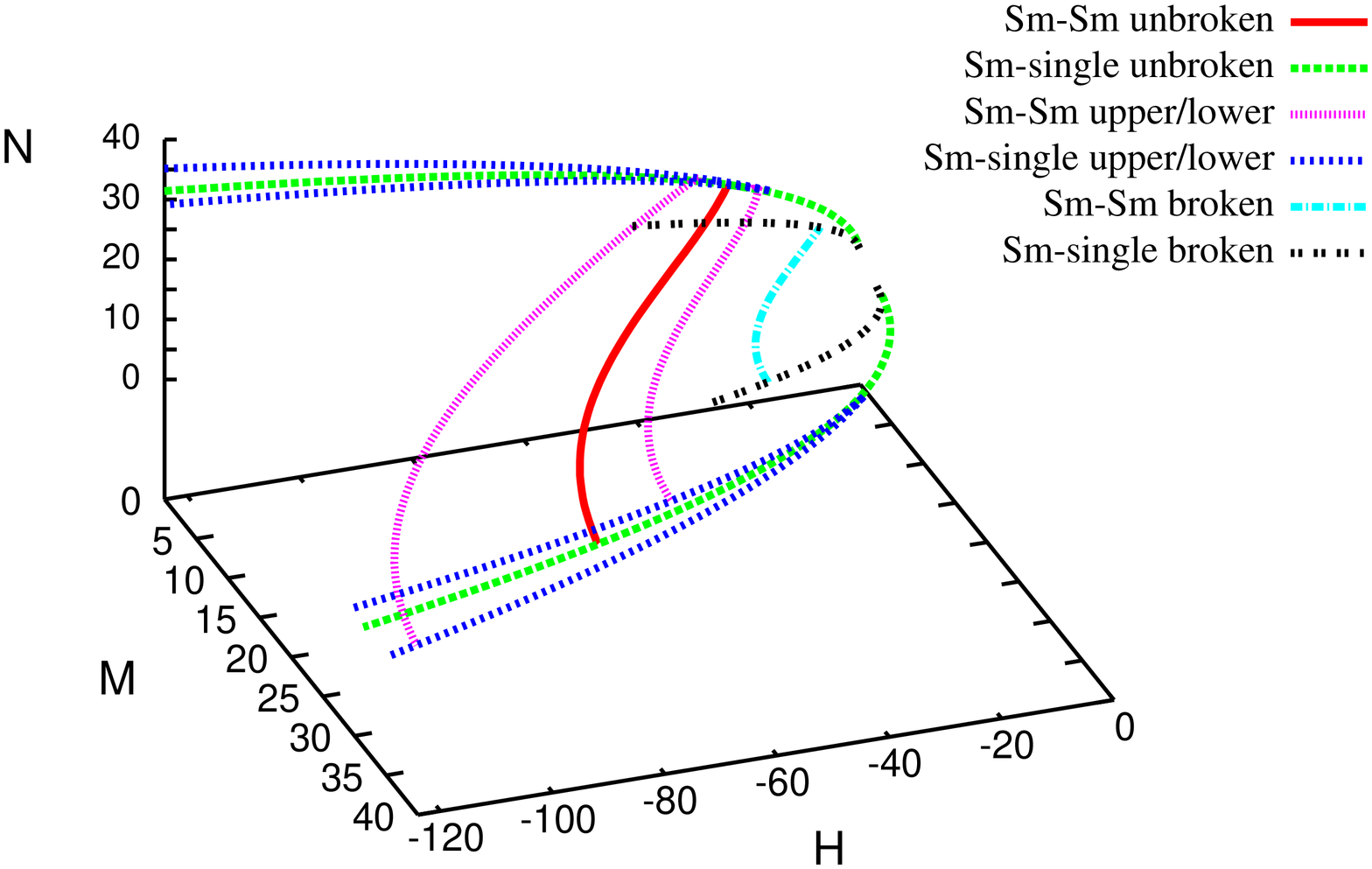} \hspace{0.5cm}
(c)\hspace{-0.6cm}
\includegraphics[height=.35\textheight,
angle=-90]{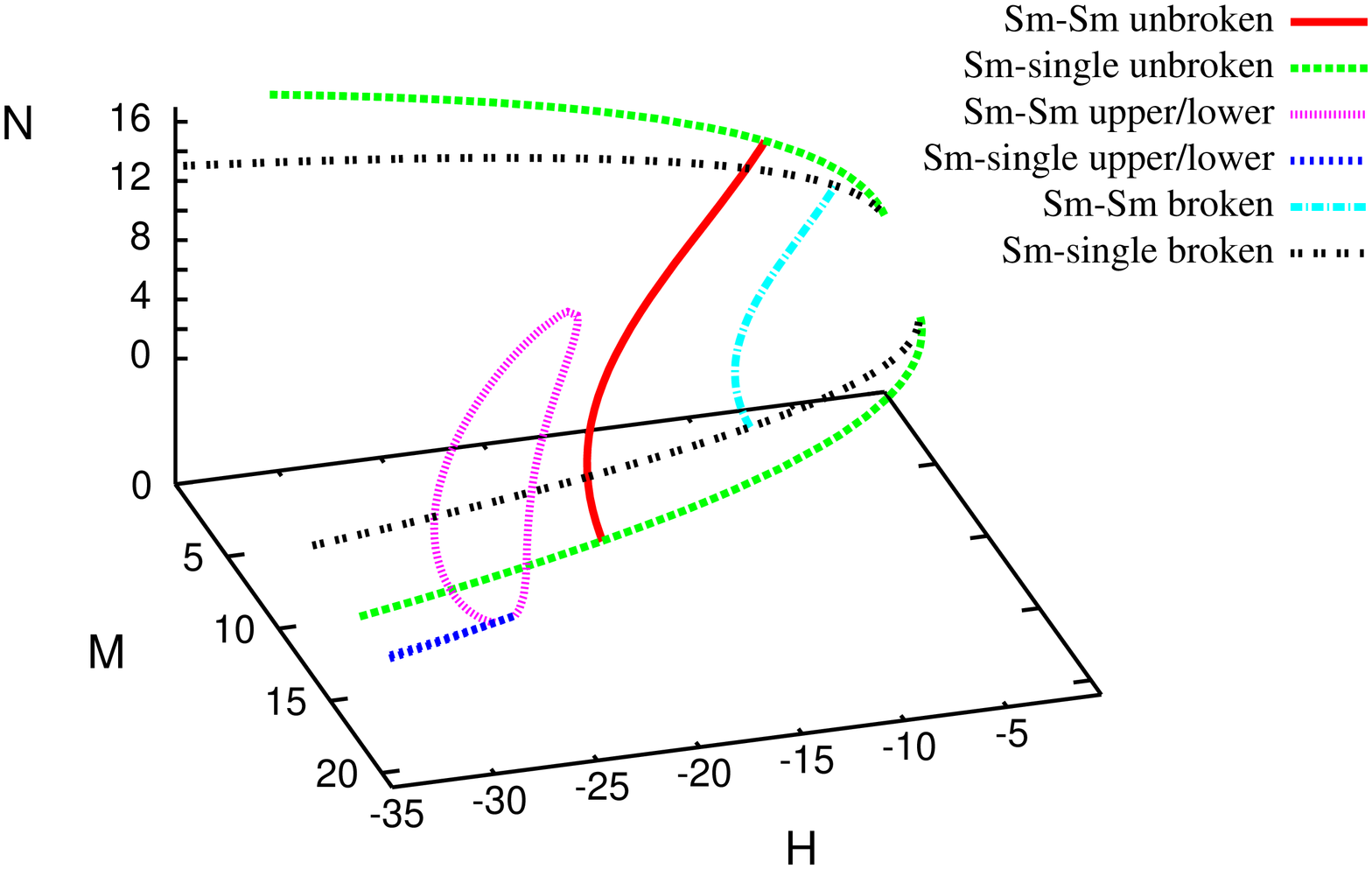} \hspace{0.5cm} (d)\hspace{-0.6cm} %
\includegraphics[height=.35\textheight, angle =-90]{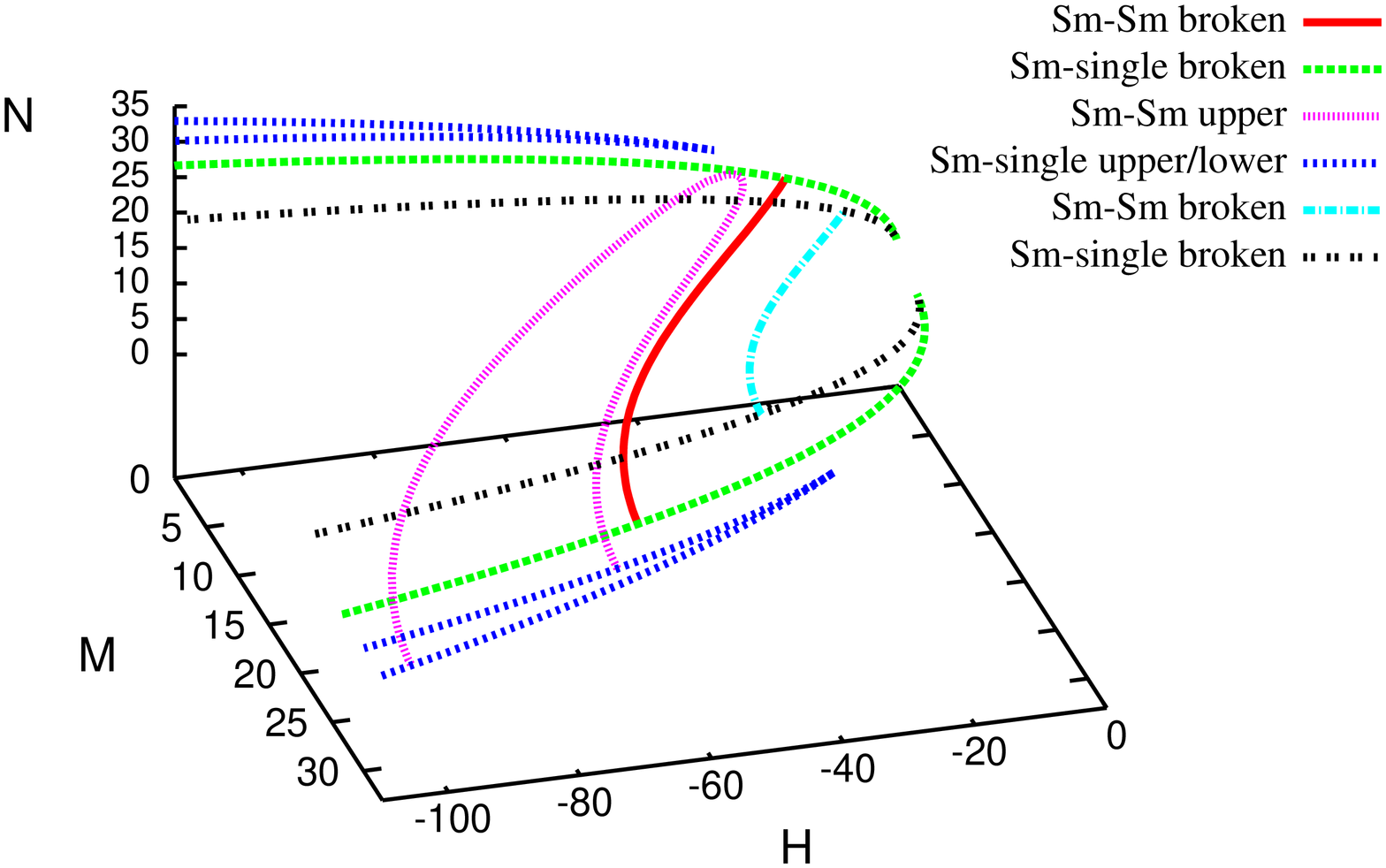} \hspace{0.5cm}
(e)\hspace{-0.6cm}
\includegraphics[height=.35\textheight,
angle=-90]{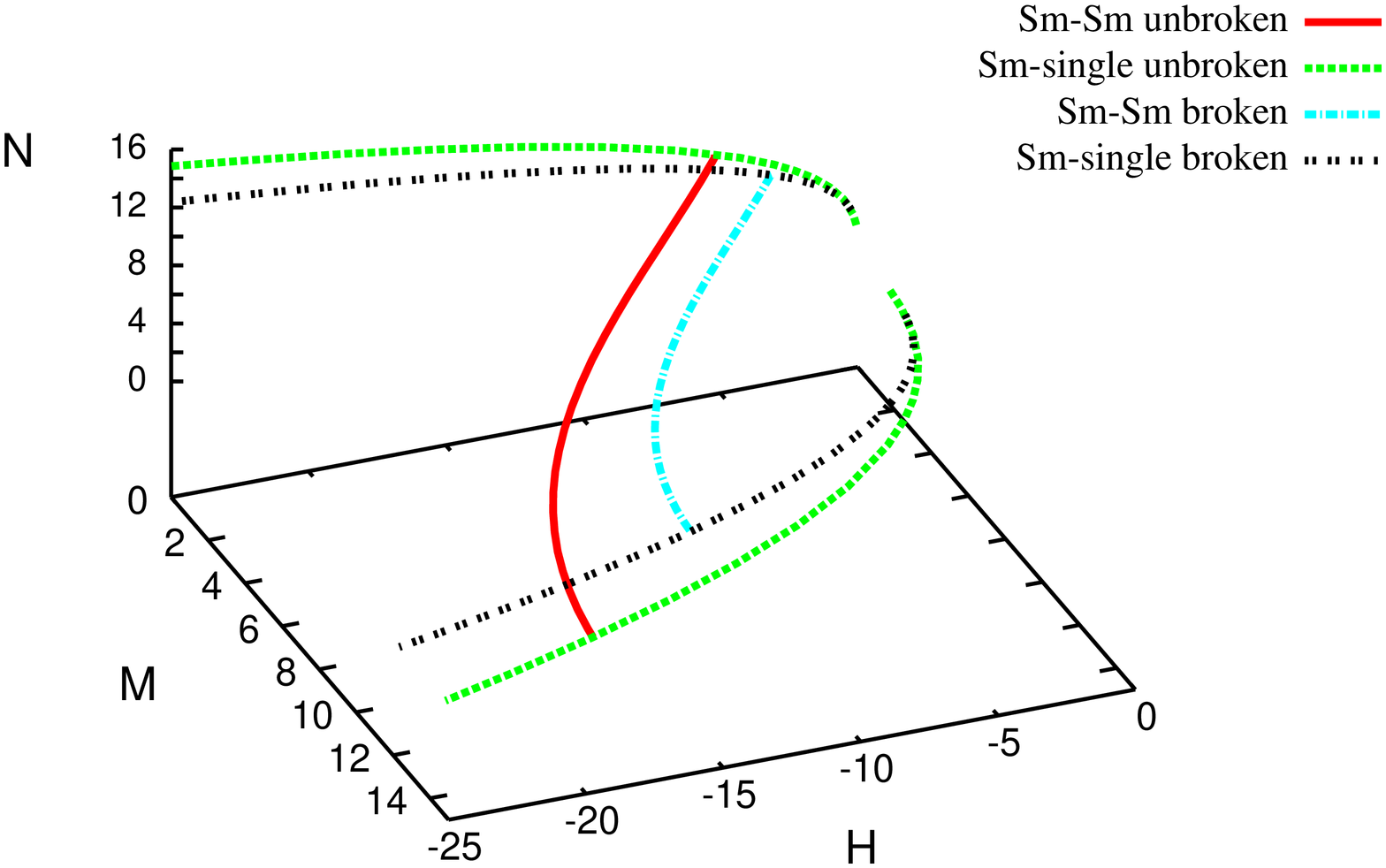} \hspace{0.5cm} (f)\hspace{-0.6cm} %
\includegraphics[height=.35\textheight, angle =-90]{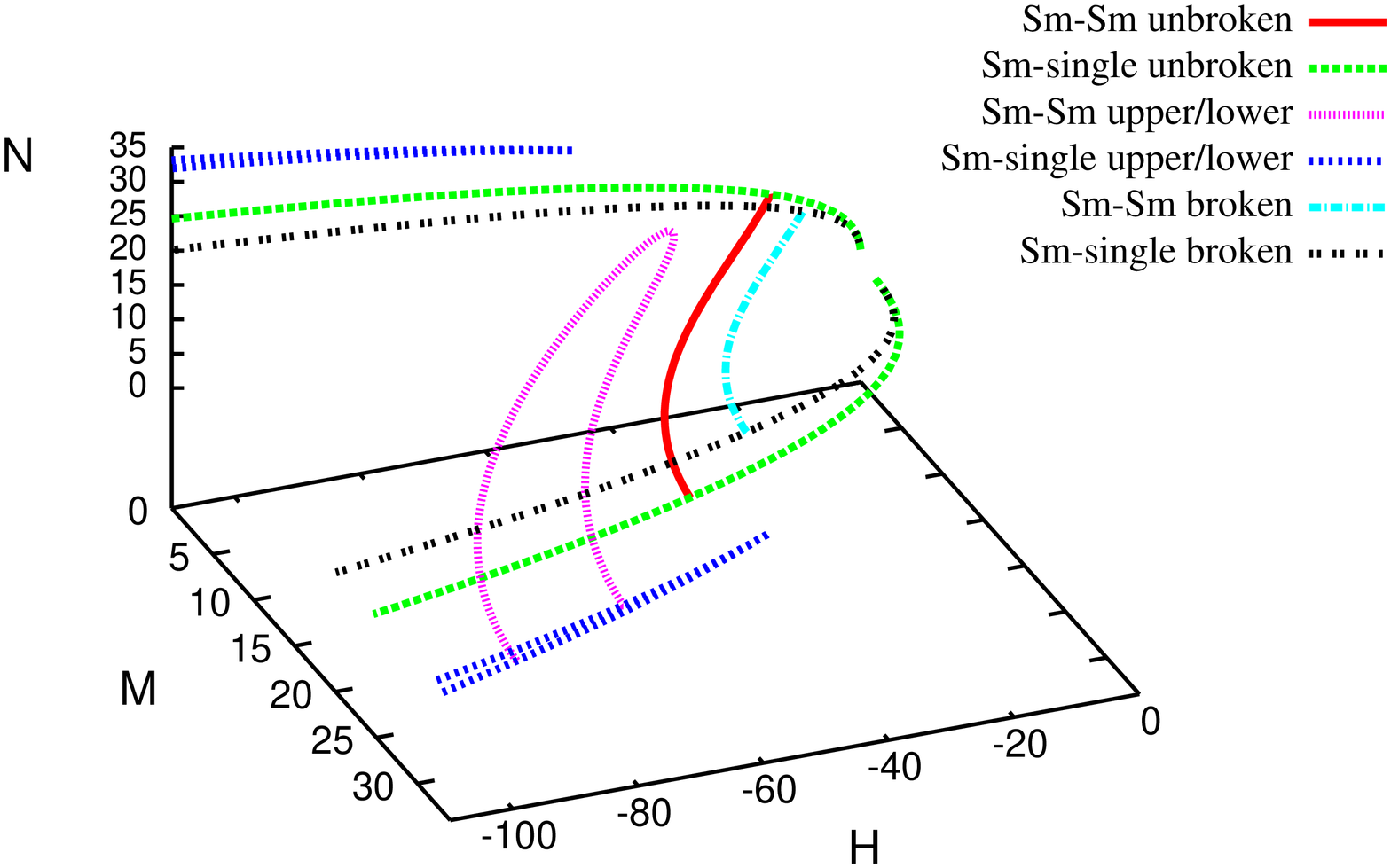}
\end{center}
\par
\vspace{-0.5cm}
\caption{(Color online) The same as Fig. \protect\ref{Fig9}, \ but for $|%
\protect\lambda |=4.0$ and $|\protect\lambda |=8.0$ (left and right
columns), for $a=0.8$, $0.9$, and $1.0$ (the first, second, and third rows,
respectively).}
\label{Fig10}
\end{figure}

At $a>0.72$, the system keeps the scenarios described above, but also gives
rise to new ones, which involve the additional symmetric modes in the
single-component model generated by the saddle-node bifurcation (see Figs. %
\ref{Fig7}(b) and \ref{Fig8}). In particular, at $a=0.8$ for fixed $|\lambda
|=4.0$, there is a new branch of solutions with the unbroken symmetry, which
smoothly arises from the lower symmetric state generated by the saddle-node
bifurcation in the single-component model. As shown in Fig. \ref{Fig10}(a),
this two-component branch makes a loop in the norm space, $\left( M,N\right)
$, and then merges into the upper symmetric state of the single-component
model, which is generated by the above-mentioned saddle-node bifurcation. In
this case, the second component of the Sm-Sm system vanishes at both ends of
the branch. For larger absolute values of the chemical potential, say $%
|\lambda |=8.0$, a different behavior is observed. Instead of coming back to
the upper state of the single-component model generated by the saddle-node
bifurcation, the first component vanishes, while the other one evolves
towards the corresponding upper-state solution of the single-component
model, as shown in Fig.~\ref{Fig10}(b).

As $a$ increases further, $|\lambda |$ must be very large to support a link
between the first and second components playing the role of the end states
of the Sm-Sm branch. For $a=0.9$, for example, both at $|\lambda |=4.0$ and $%
|\lambda |=8.0$, the Sm-Sm branch makes a loop in the norm space, emerging
from the lower-state solution of the single-component model with the second
component being absent. This branch evolves towards the upper-state
symmetric solution of the single-component model with the second component
vanishing again, as shown in Figs.~\ref{Fig10}~(c,d). The further increase
of $a$ at relatively small values of $|\lambda |$ eliminates the looped
branch; however, the branches of the two-component solutions with the broken
and unbroken symmetries, which link the corresponding single-component modes
(asymmetric or symmetric, respectively) still persist, as shown in Fig.~\ref%
{Fig10}(e). Nevertheless, at larger values of $\lambda $ the analysis
reproduces the same pattern as observed before, cf. Figs.~~\ref{Fig10}(d)
and \ref{Fig10}(f).

\subsection{Bifurcations of the antisymmetric-antisymmetric (AS-AS) branches}

The structure of the configuration space of the AS-AS solutions in the model
with finite $a$ is complex. It can also be studied by fixing, say, $a=0.6$
and one of the chemical potentials ($\lambda $, which is associated with the
$\phi $-component), and scanning the space by varying $\mu $, which is
associated with component $\psi $. In fact, the picture displayed below for $%
a=0.6$ adequately represents the situation in the interval of $0.16<a<0.76.$

Similar to the case of the Sm-Sm modes considered above, the description of
various bifurcation scenarios for the AS-AS branches, with the broken and
unbroken antisymmetry alike, essentially reduces to identifying pairs of
states of the single-component models which are linked by these branches.
First, at relatively small fixed values of $|\lambda |$ (see Figs. \ref%
{Fig11}(a,b)), we observe a branch of the AS-AS solutions with
broken antisymmetry, which arises smoothly from the ``unbroken"
AS-AS family, at some critical value of $\mu $. The asymmetry of the
states rapidly increases along this branch.

The branch extends up to a second critical value of $\mu $, where it merges
back into the ``unbroken" AS-AS branch. Shortly afterwards, the unbroken
branch itself merges into the corresponding solution of the single-component
model. This scenario implies that, typically, one of the components, e.g., $%
\phi $, gradually approaches the limit solution with the unbroken
antisymmetry, while the other component remains strongly asymmetric almost
until the end of the branch, where its asymmetry abruptly drops to zero.

\begin{figure}[tbh]
\begin{center}
(a)\hspace{-0.6cm}
\includegraphics[height=.35\textheight, angle
=-90]{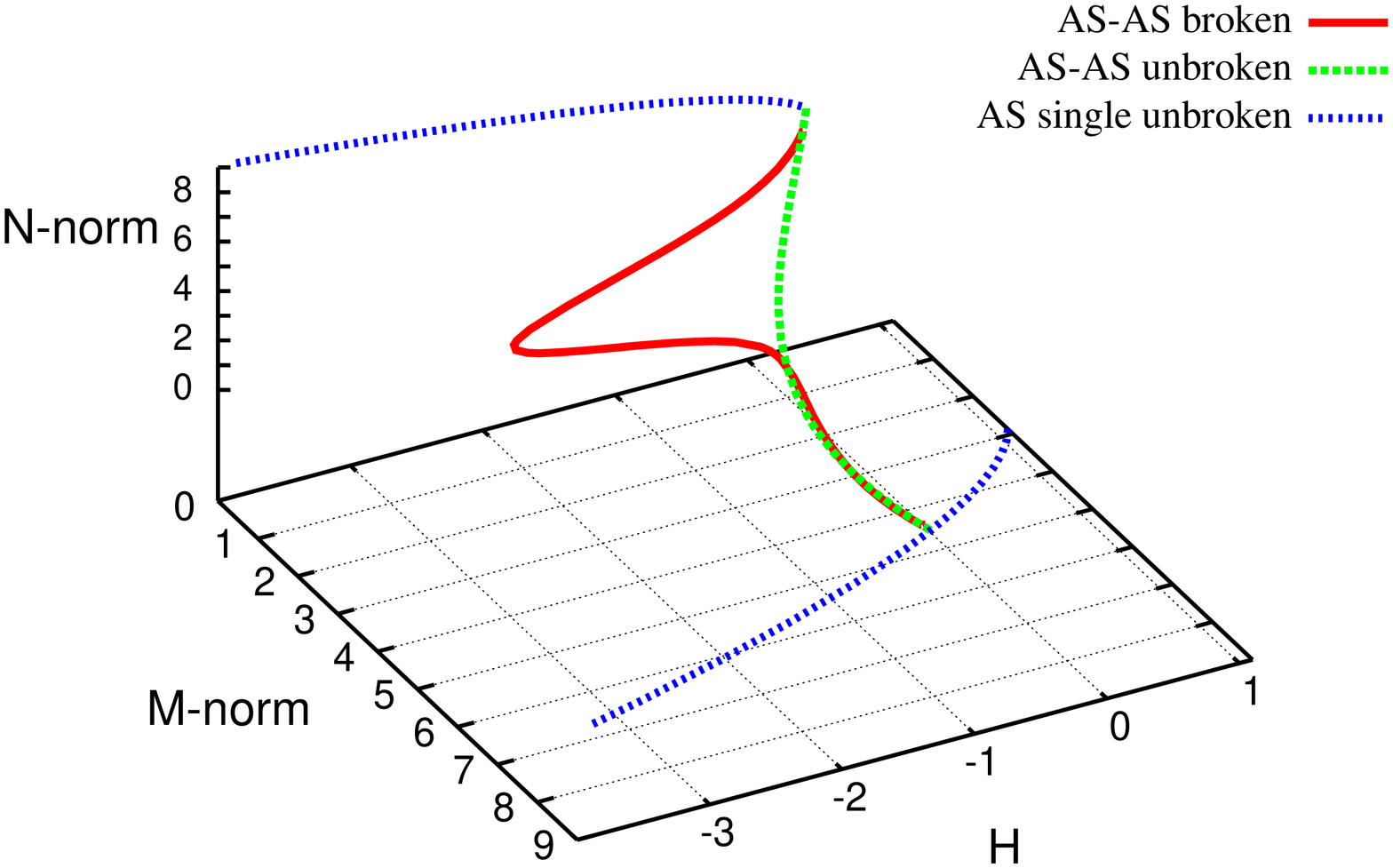} \hspace{0.5cm} (b)\hspace{-0.6cm}
\includegraphics[height=.35\textheight, angle
=-90]{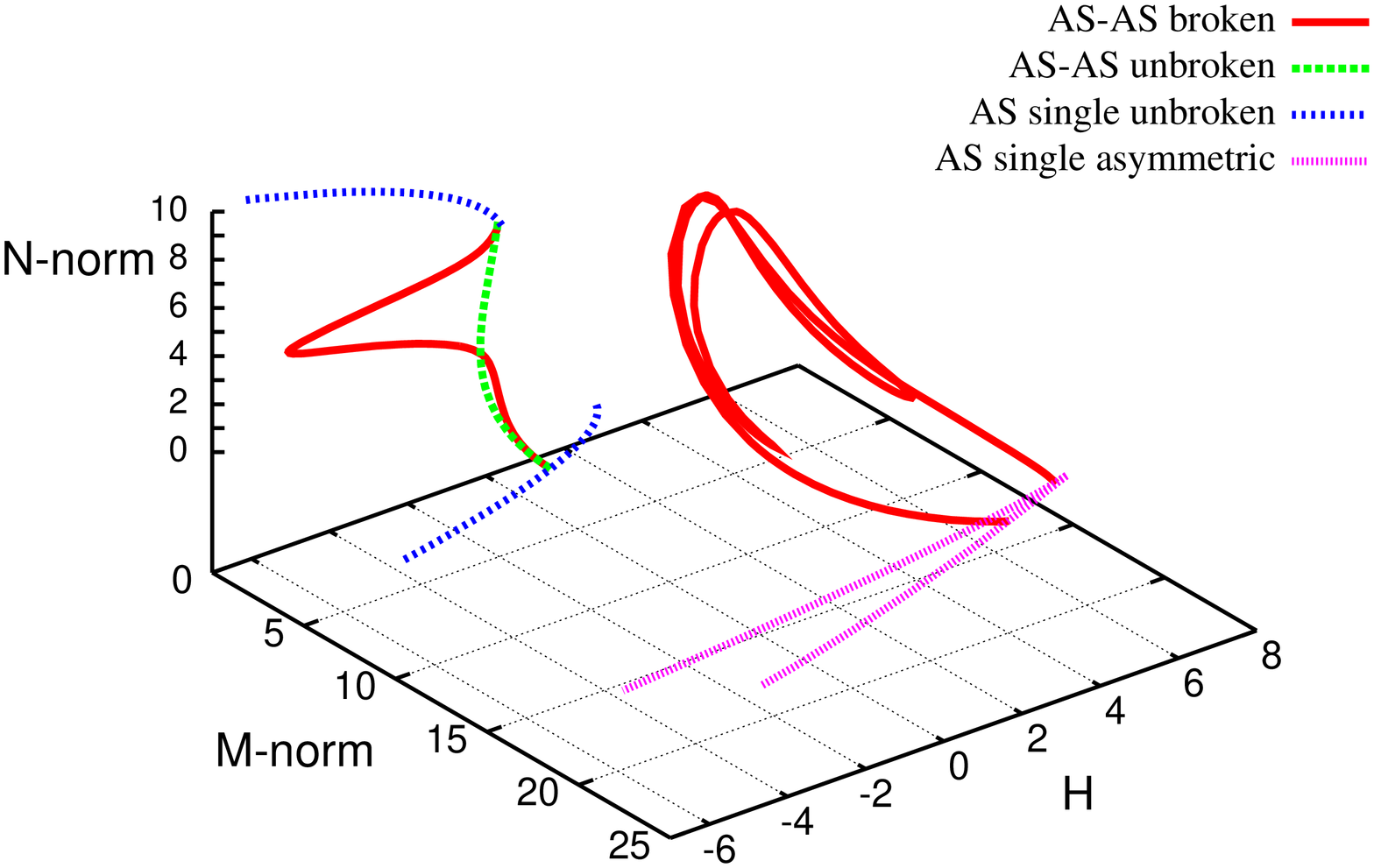} \hspace{0.5cm} (c)\hspace{-0.6cm}
\includegraphics[height=.35\textheight, angle
=-90]{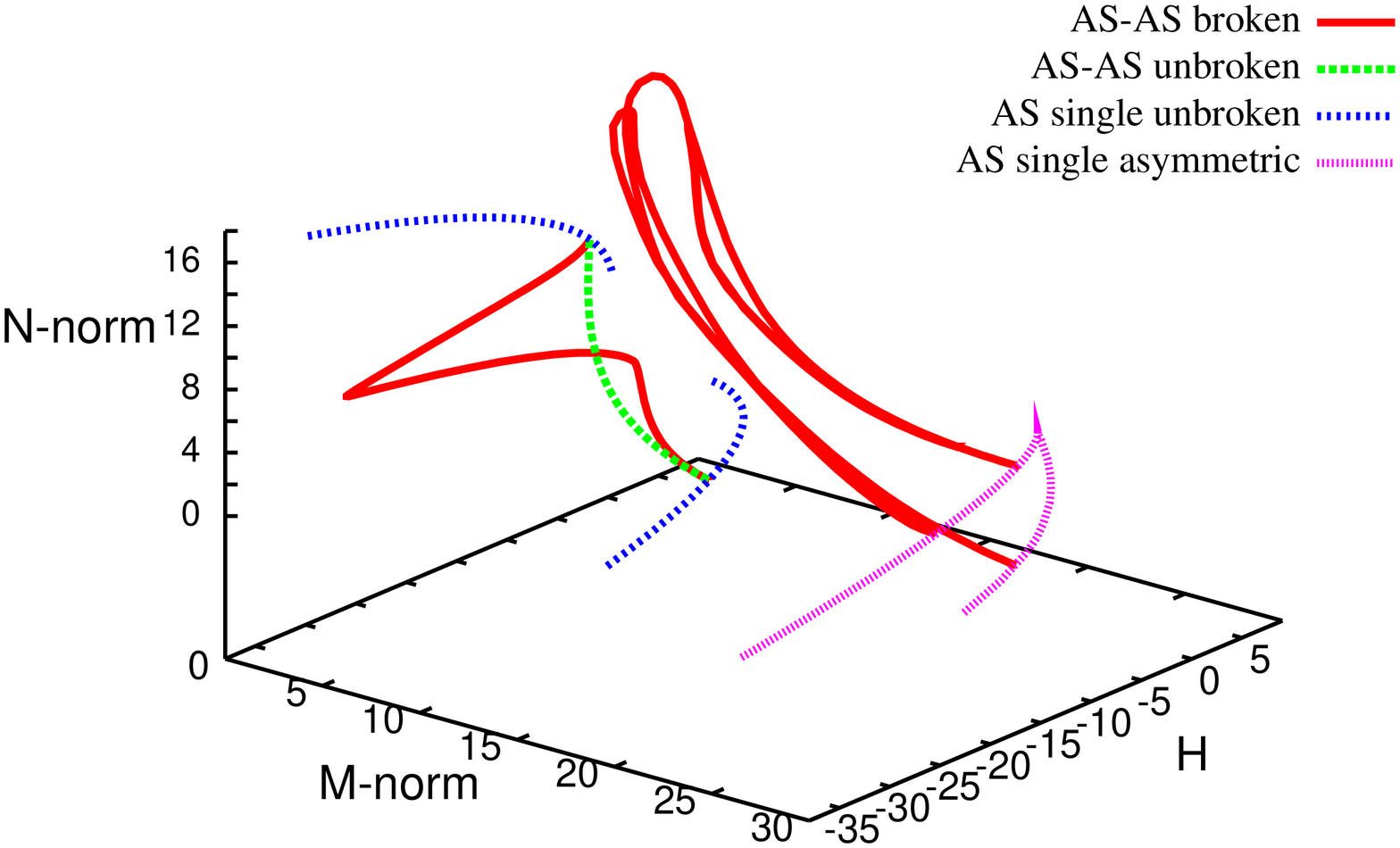} \hspace{0.5cm} (d)\hspace{-0.6cm}
\includegraphics[height=.35\textheight, angle
=-90]{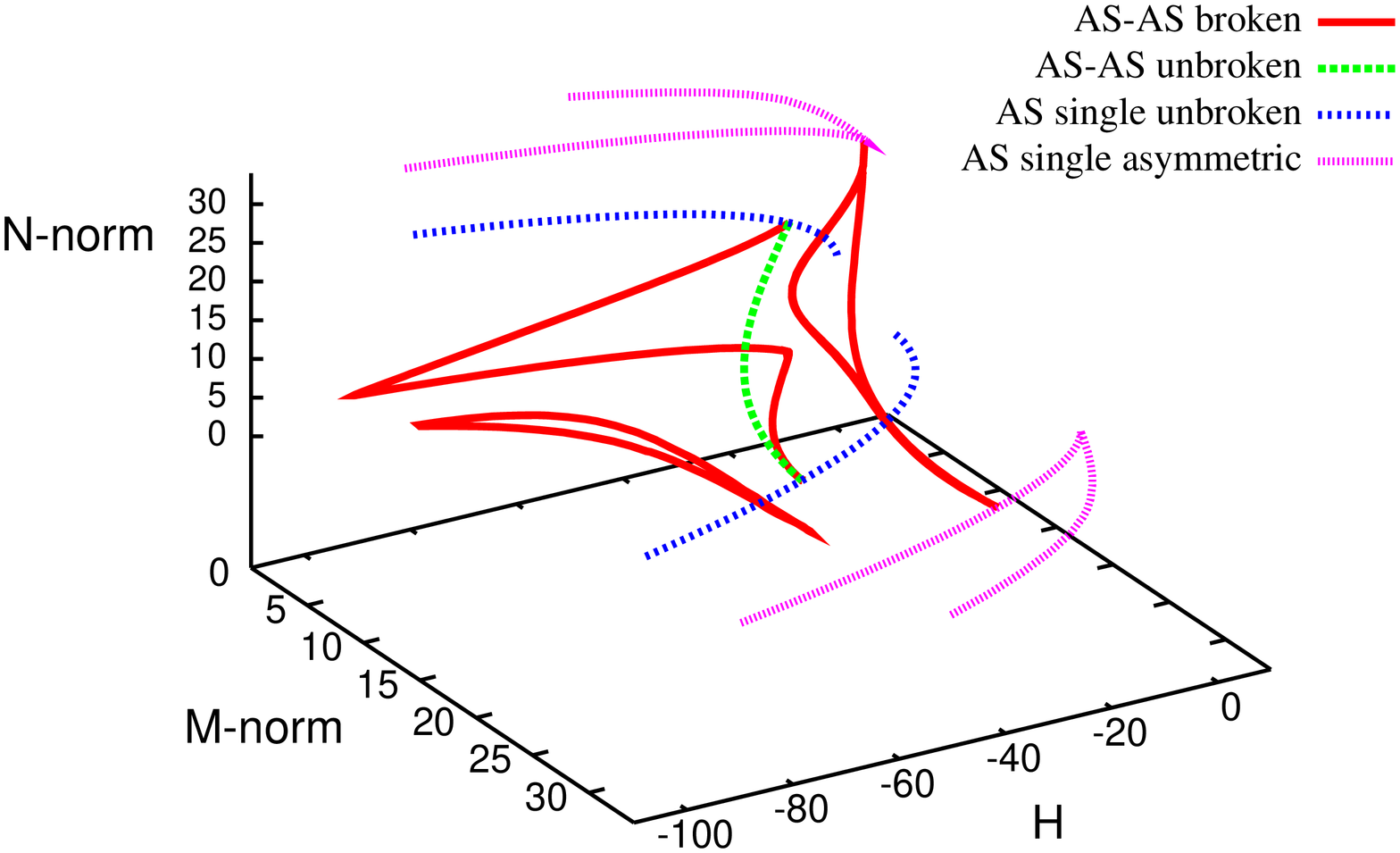}
\end{center}
\par
\vspace{-0.5cm}
\caption{(Color online) The picture of the bifurcation scenarios similar to
that in Fig. \protect\ref{Fig9}, but for AS-AS branches with $a=0.6$ and$~%
\protect\lambda =1.0~$(a), $1.5~$(b), $4.0~$(c), and $8.0$ (d). }
\label{Fig11}
\end{figure}

\begin{figure}[tbh]
\refstepcounter{fig} \label{f-26}
\par
\begin{center}
(a)\hspace{-0.6cm}
\includegraphics[height=.24\textheight, angle =-90]{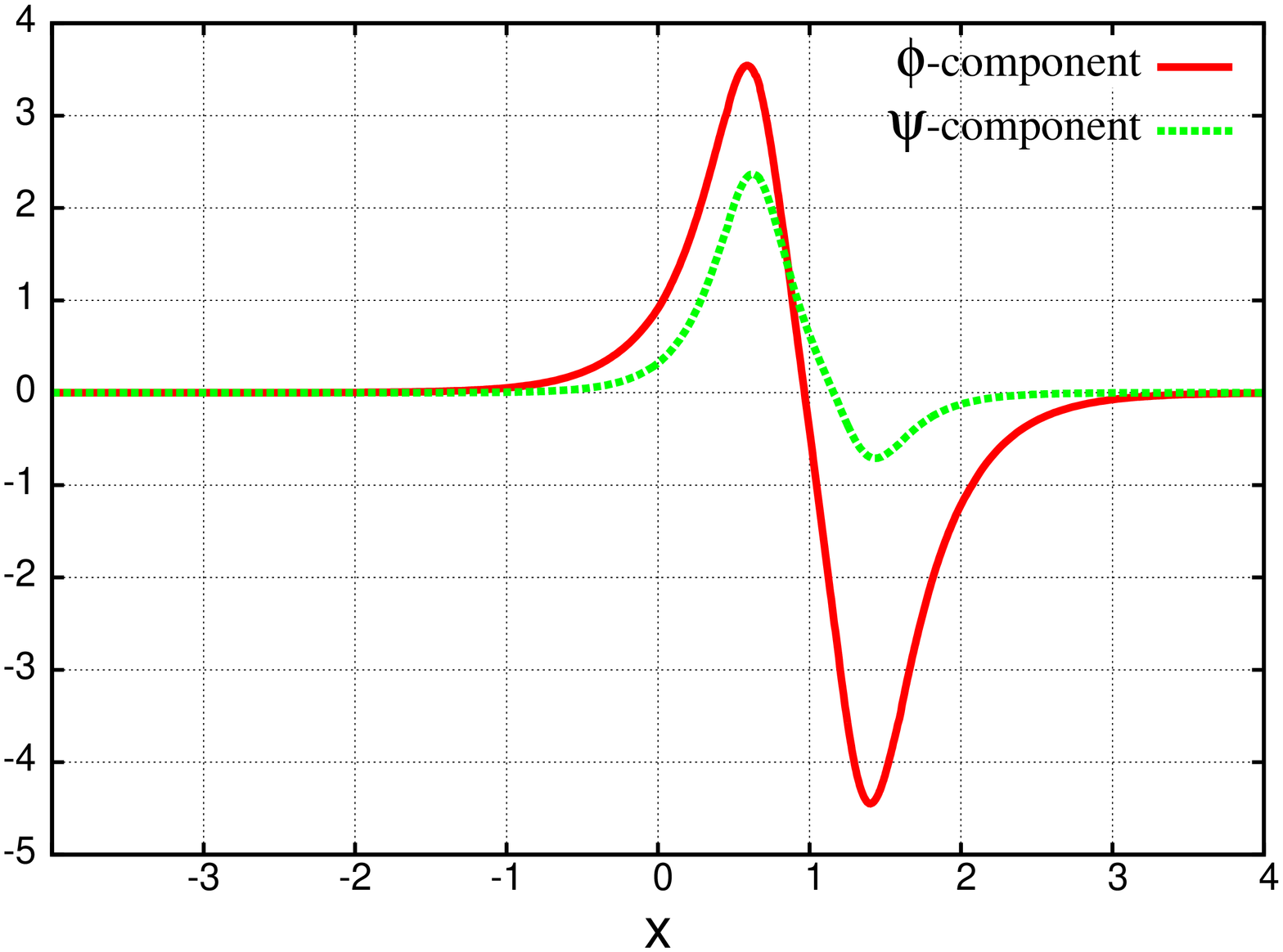} (b)\hspace{%
-0.6cm} \includegraphics[height=.24\textheight, angle =-90]{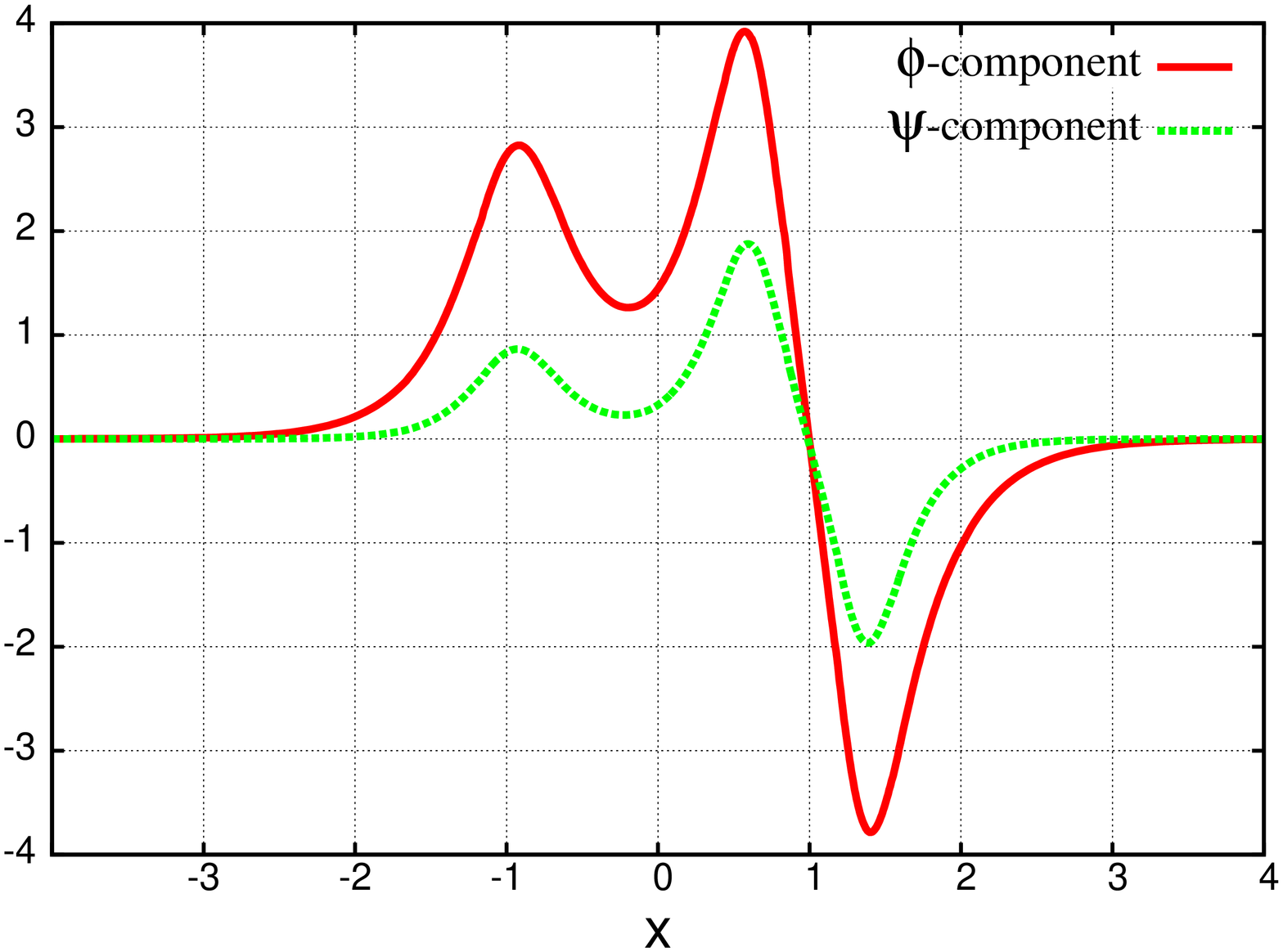}
\end{center}
\par
\vspace{-0.5cm}
\caption{(Color online) (a) and (b): Examples of AS-AS solutions belonging,
respectively, to the lower (closed loop) and upper (transition) branches in
Fig. \protect\ref{Fig11}(c), with $a=0.6$, $\protect\mu =8.0$ and $|\protect%
\lambda |=4.0$. }
\label{Fig12}
\end{figure}

As $|\lambda |$ increases beyond a critical value, the above-mentioned
broken-antisymmetry modes, generated by the saddle-node bifurcations in the
single-component model, come into the play, making the evolution of the
AS-AS branches quite involved. A new branch arises smoothly from the new
broken-antisymmetry solution of the single-component model, which is trapped
in one of the nonlinear potential wells. Fixing $|\lambda |=1.5$ and varying
$\mu $, we observe that this branch first makes a double loop in the norm
space, $\left( M,N\right) $, with both components remaining trapped in the
potential well, as shown in Figs. ~\ref{Fig11}(c,d). Surprisingly, almost
immediately another branch arises. The evolution along the new branch
amounts to a transition of one of the components from one
nonlinear-potential well to the other. One of the components on this branch
looks like an S-AS bound state, rather than a single AS state, as shown in
Fig.~\ref{Fig12}(b).

\begin{figure}[tbh]
\refstepcounter{fig} \label{f-29}
\par
\begin{center}
(a)\hspace{-0.6cm}
\includegraphics[height=.25\textheight, angle =-90]{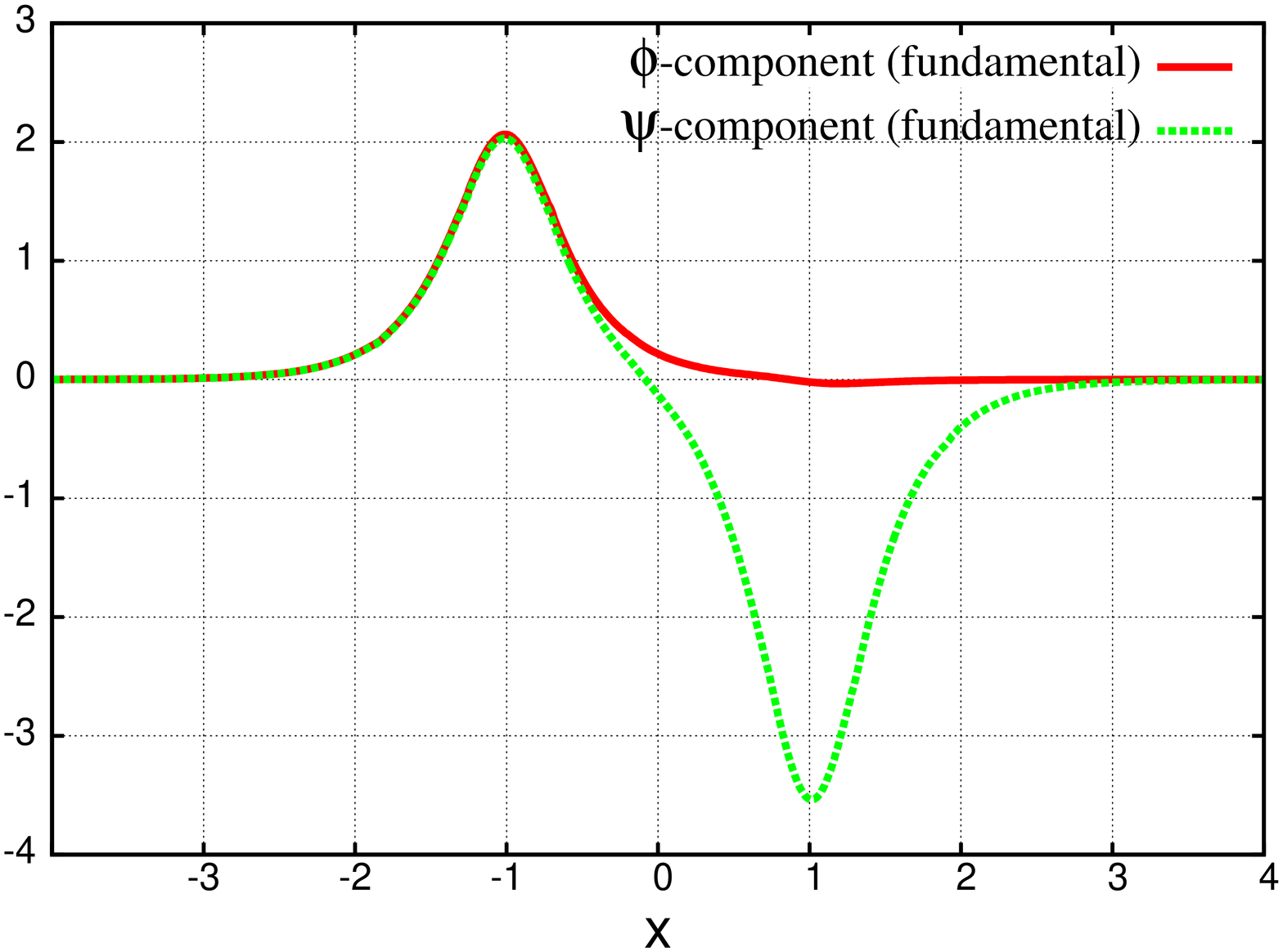} (b)\hspace{%
-0.6cm} \includegraphics[height=.25\textheight, angle =-90]{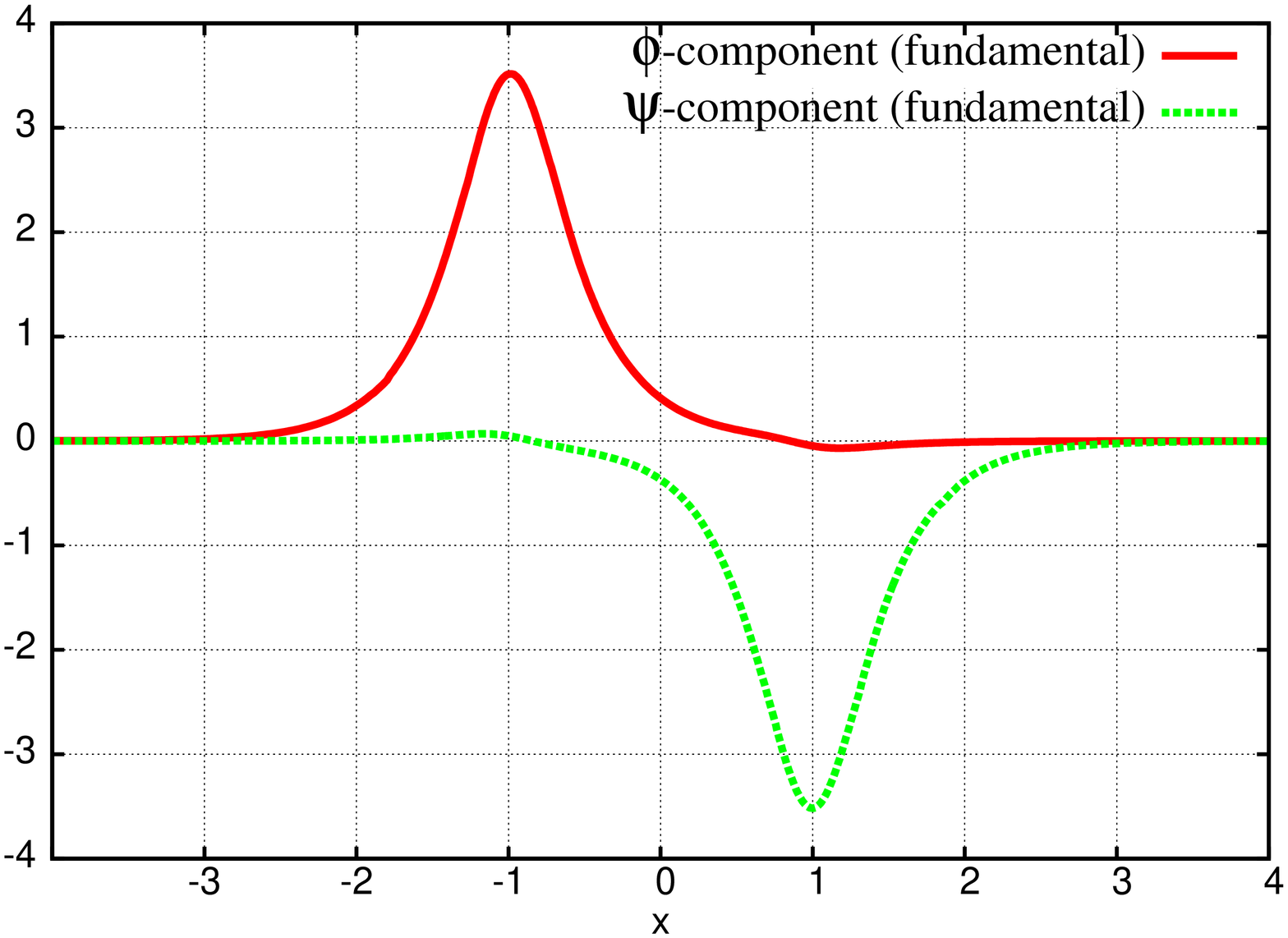} (c)%
\hspace{-0.6cm} \includegraphics[height=.25\textheight, angle=-90]{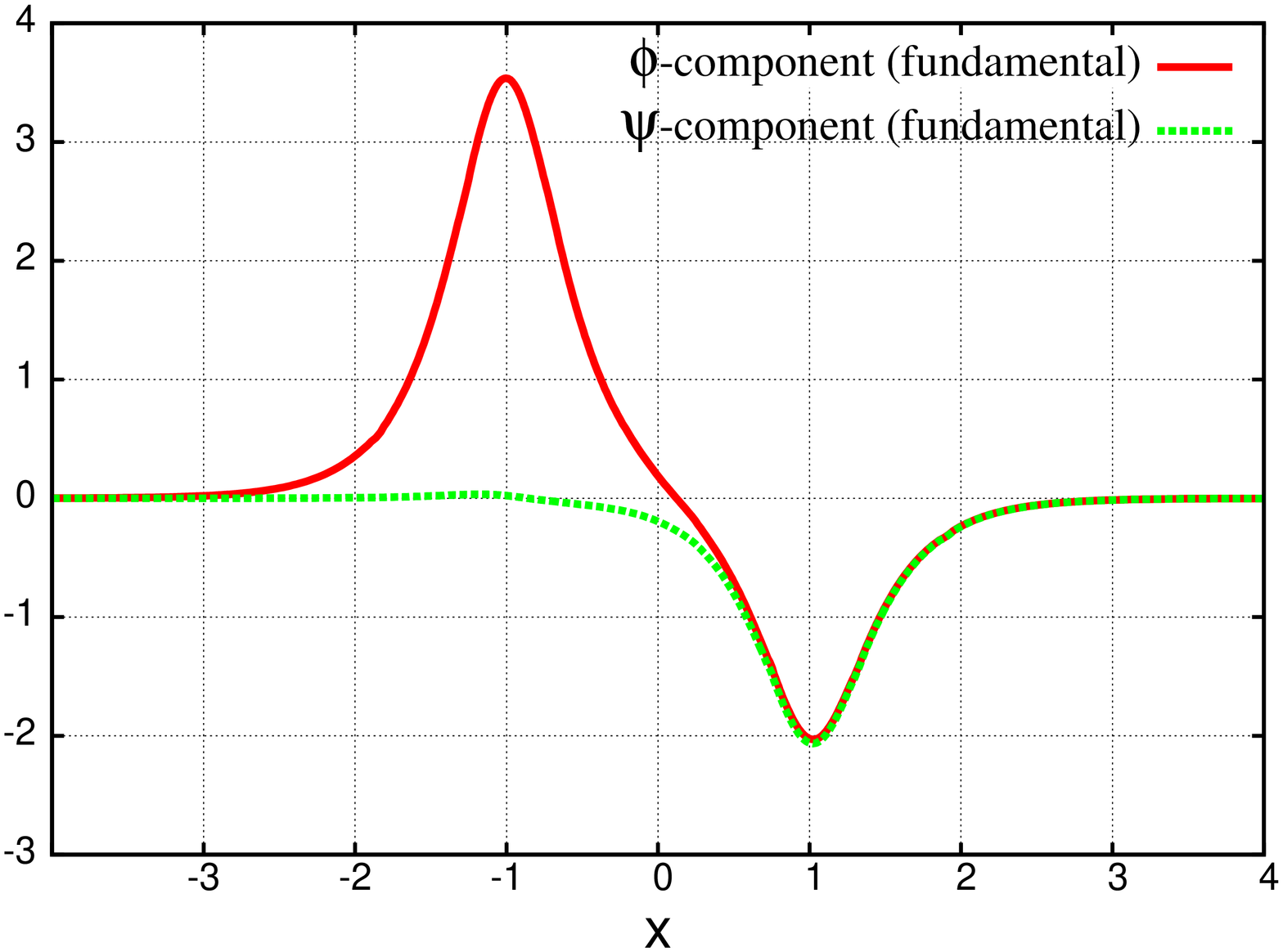}
\end{center}
\par
\vspace{-0.5cm}
\caption{(Color online) Examples of AS-AS solutions with the broken
antisymmetry at three points of the fundamental branch with $a=0.8$, $|%
\protect\mu |=4.0$ and $|\protect\lambda |=4.0$.}
\label{Fig13}
\end{figure}

To consider in further detail how the evolution of the AS-AS branches
depends on the value of $a$, we now fix chemical potential $\lambda $ and
consider a range of values of $a$. In the case of not very large $|\lambda |$%
, the AS-AS branch with the broken antisymmetry again originates from the
``unbroken" AS-AS branch, as $\mu $ increases above some critical value.
Since this branch is actually observed in the entire parameter space,
including the limit case with the $\delta $-functions ($a\rightarrow 0$), it
may be naturally called the fundamental AS-AS branch with the broken
antisymmetry. Along this branch, one of the components, say $\phi $, rapidly
grows with the increase of $|\mu |$, remaining localized in one
nonlinear-potential well [as seen in Fig.~\ref{Fig13}(a)], while the other
component, $\psi $, has two well-pronounced peaks near $x=\pm 1$. The
evolution along this branch proceeds through the decrease of the maximum
values of the second component, so that, at some point, the mode approaches
the limit case of two AS states localized around opposite minima of the
nonlinear potential, see Fig.~\ref{Fig13}(b). Finally, the system evolves
towards the mirror image of the initial configuration, where the $\phi $
component features maxima near $x=\pm 1$, while component $\psi $ is trapped
in the nonlinear potential well opposite to that where $\phi $ was
originally trapped, see Fig.~\ref{Fig13}(c). As $\mu $ approaches the second
critical value, the branch merges back into the AS-AS state with the
unbroken antisymmetry. Actually, this branch links two single-component AS
solutions with the unbroken antisymmetry.

The evolution pattern becomes still more complex at larger values of $%
|\lambda |$. In that case (for instance, at $|\lambda |=4$) , the building
blocks of the AS-AS states are again single-component AS modes with the
strongly broken antisymmetry, each localized around one minimum of the
nonlinear potential. The evolution trajectory in the configuration space may
then feature two loops. Along the first one, the system moves from the
localization in one nonlinear potential well to the other, simultaneously
swapping the two components (the one which was originally dominating and its
initially vanishing counterpart). Continuing the evolution along the second
loop, the system swaps the two components once again (not shown here in
detail).

\subsection{Bifurcations of the symmetric-antisymmetric (S-AS) branches}

The evolution of branches of the S-AS type in the system with finite $a$
also strongly differs from what was presented above for the system with the
set of the $\delta $-functions, corresponding to $a\rightarrow 0$. As well
as the solutions of the Sm-Sm and AS-AS types, in the present case the
evolution also amounts, essentially, to the switch between different states
of the single-component model, linked by the branch of the S-AS type, with
the unbroken or broken (anti)symmetries. To present the results, we again
start with $a=0.6$, fixing the chemical potentials ($\mu $) which is
associated with the symmetric component ($\phi $), and scan the
configuration space by varying the chemical potential ($\lambda $)
associated with the antisymmetric component, $\psi $.

\begin{figure}[tbh]
\refstepcounter{fig} \label{f-20}
\par
\begin{center}
(a)\hspace{-0.6cm}
\includegraphics[height=.35\textheight,
angle=-90]{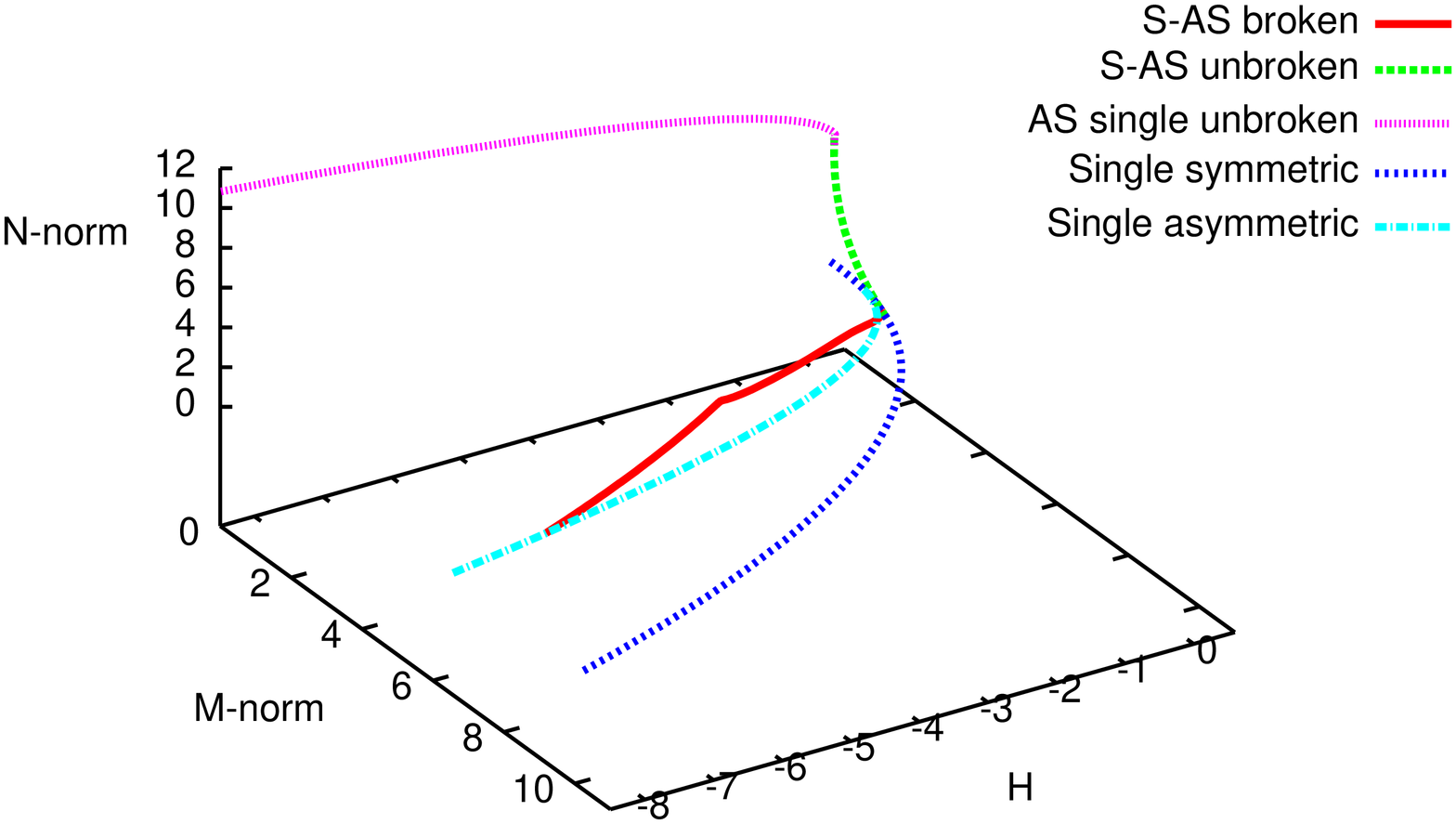} \hspace{0.5cm} (b)\hspace{-0.6cm} %
\includegraphics[height=.35\textheight, angle =-90]{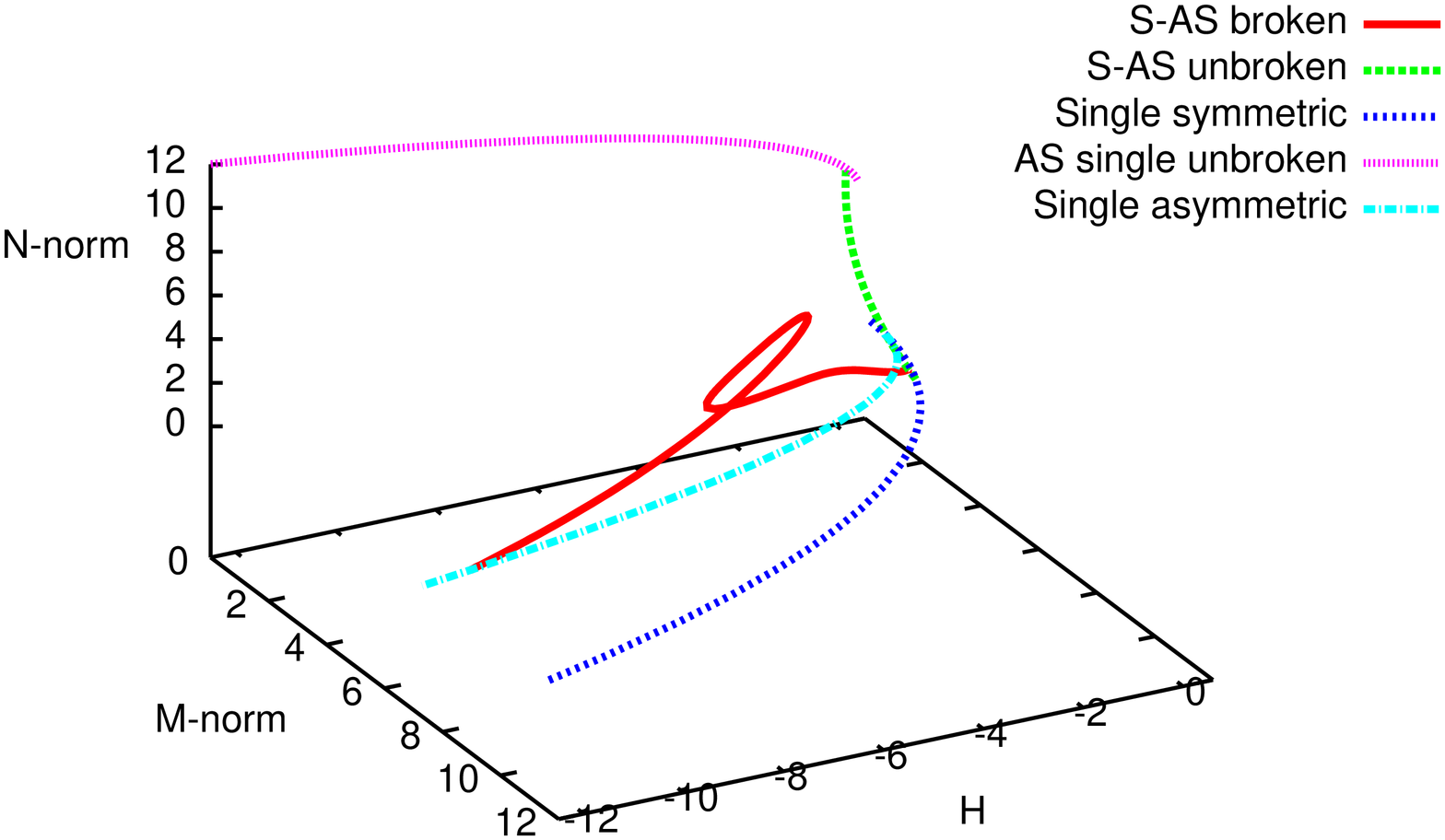}
\hspace{0.5cm} (c)\hspace{-0.6cm} \includegraphics[height=.35%
\textheight,angle =-90]{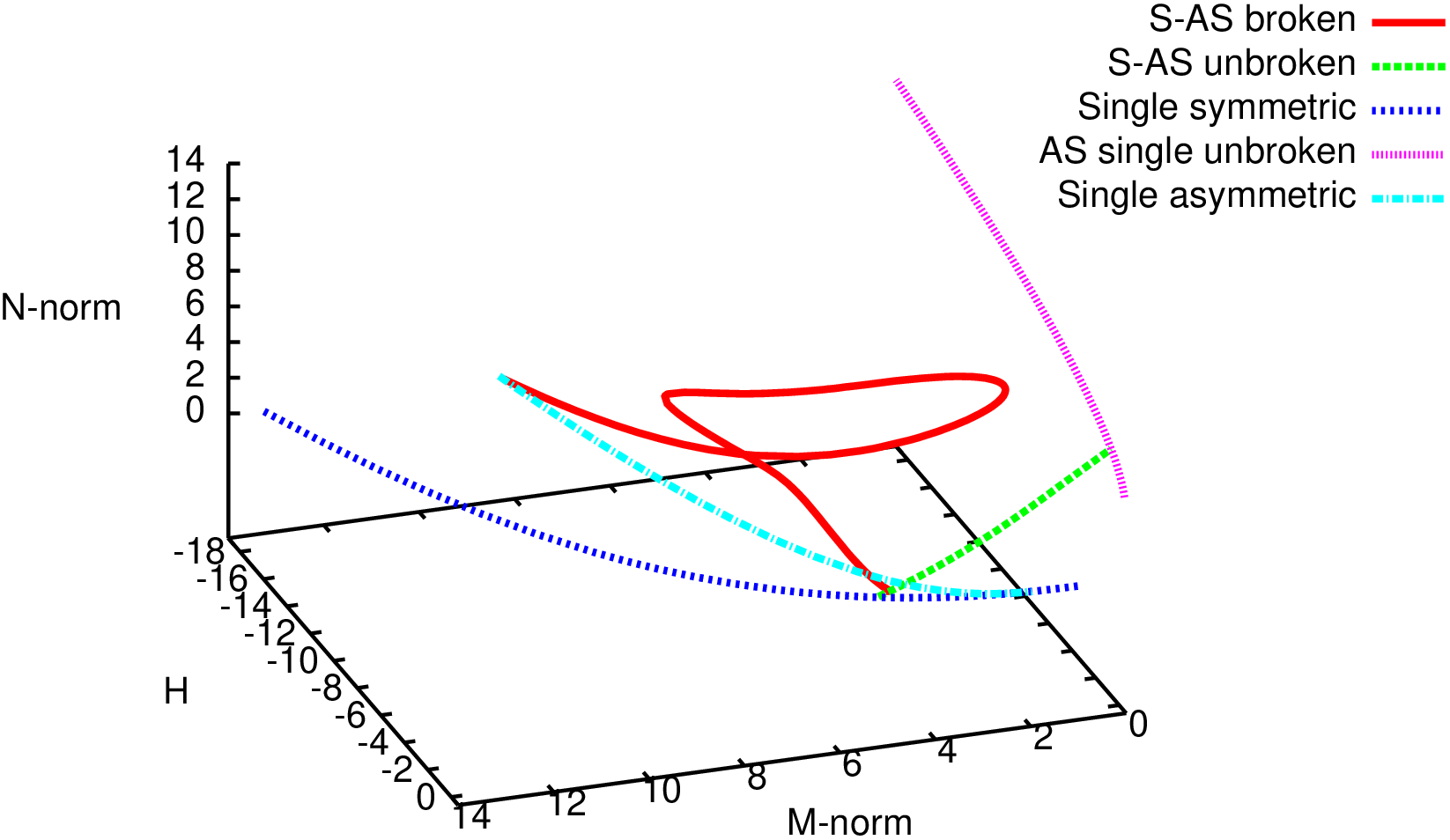} \hspace{0.5cm} (d)\hspace{%
-0.6cm}
\includegraphics[height=.35\textheight, angle
=-90]{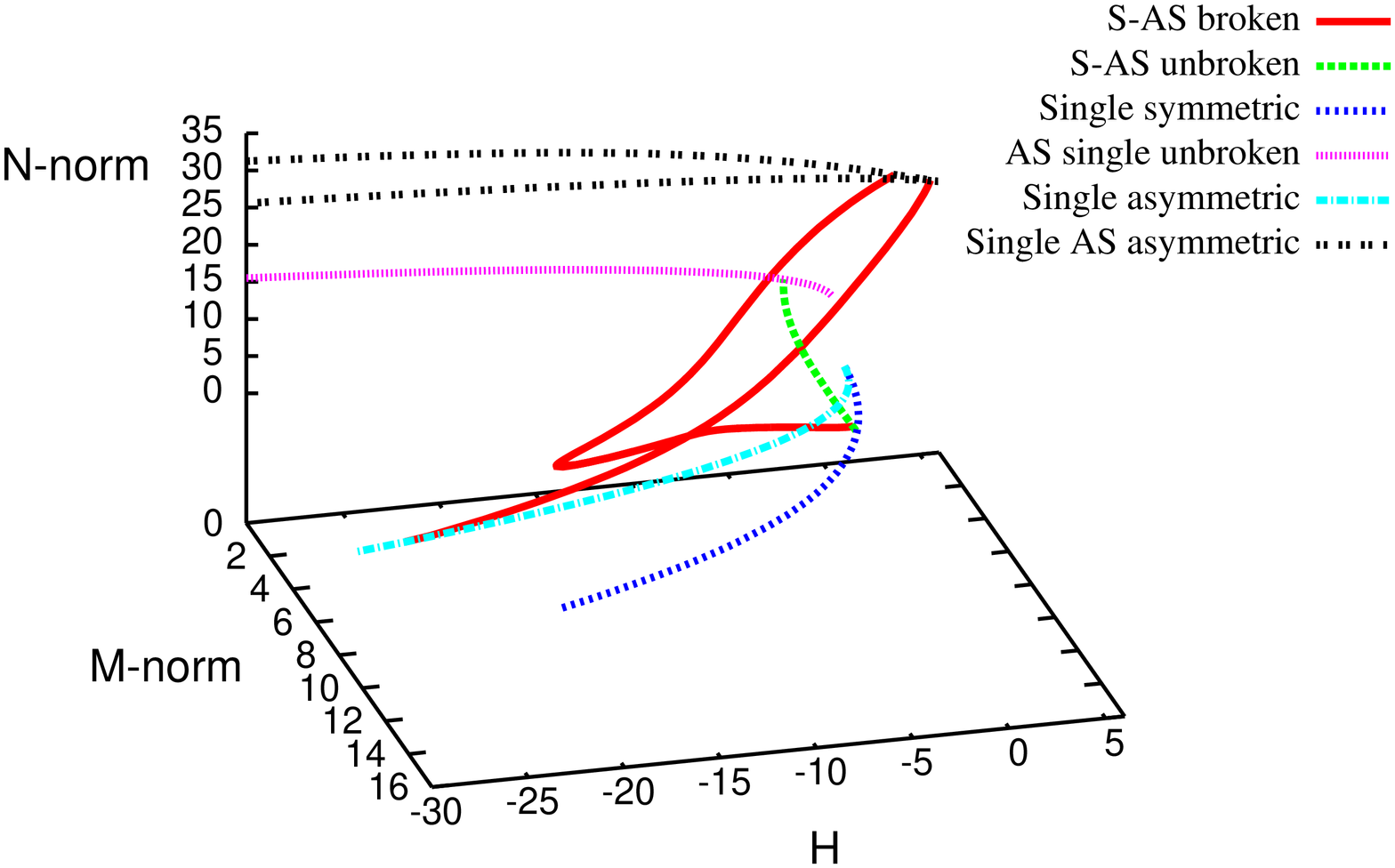}
\end{center}
\par
\vspace{-0.5cm}
\caption{(Color online) Branches of modes of the S-AS type: the evolution of
the energy (\protect\ref{H}) in the norm space, $\left( M,N\right) $, at $%
a=0.6$ and $|\protect\mu |=0.2~$(a), $|\protect\mu |=0.5~$(b), $|\protect\mu %
|=1.0~$(c), and $|\protect\mu |=1.5$ (d). }
\label{Fig14}
\end{figure}

Following the general pattern outlined above for other species of the
two-component solutions, the ``broken" S-AS branch arises from its
counterpart with the unbroken (anti)symmetries, at some critical value of $%
\lambda $. The energy of the emerging ``broken" branch rapidly grows with $%
|\lambda |$, as seen in Fig. ~\ref{Fig14}.

\begin{figure}[tbh]
\refstepcounter{fig} \label{f-13}
\par
\begin{center}
(a)\hspace{-0.6cm}
\includegraphics[height=.35\textheight,
angle=-90]
{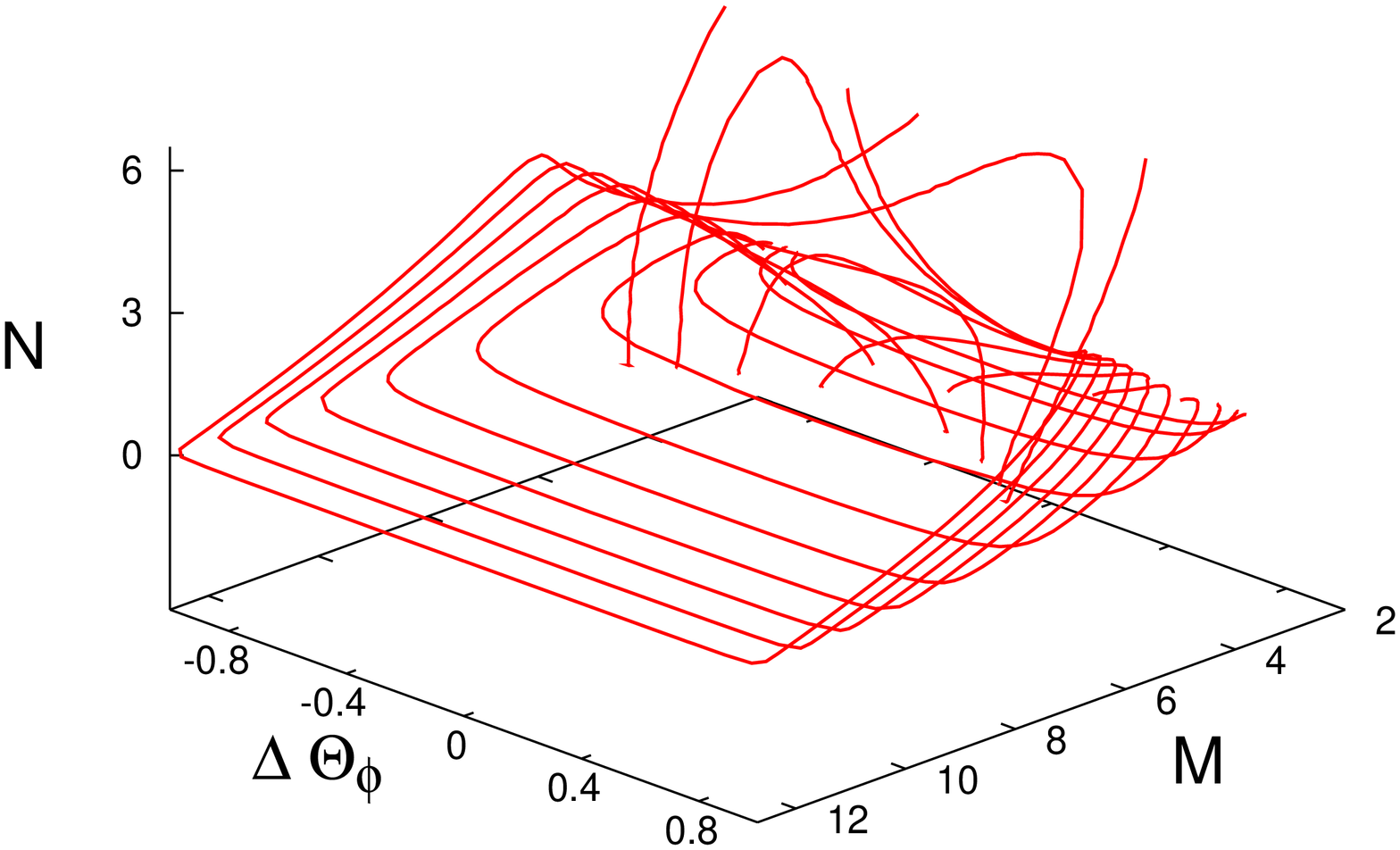}
\hspace{0.5cm} (b)\hspace{-0.6cm} %
\includegraphics[height=.35\textheight, angle =-90]
{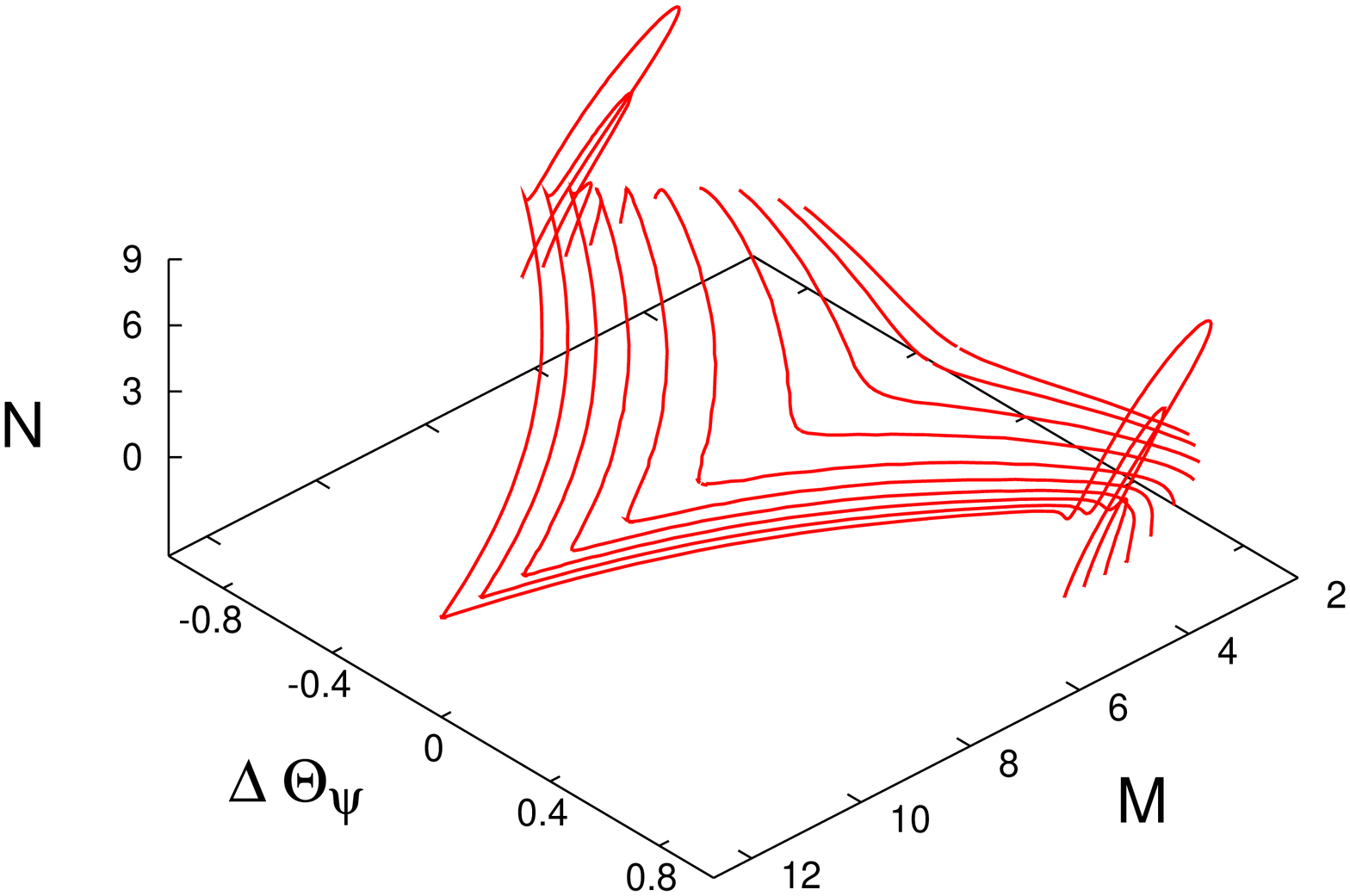}
\end{center}
\par
\vspace{-0.5cm} \caption{S-AS branches in the system with
$a=0.60$: the evolution of the asymmetry of the (originally)
symmetric field $\protect\phi $ (a) and antisymmetric field
$\protect\psi $ (b) in the norm space, $\left( M,N\right) $.}
\label{Fig15}
\end{figure}

The emergence and evolution of these S-AS branches is further illustrated by
Fig.~\ref{f-13}, which displays the pattern of the modes at $|\mu
|=1.5,1.0,0.5,0.2$. We observe that, if $|\mu |$ remains relatively small,
the increase of $|\lambda |$ leads to a bifurcation between the
``broken" and ``unbroken" S-AS modes. As $%
|\lambda |$ slightly decreases below this critical value, the norm
of the AS component in the ``unbroken" S-AS mode rapidly drops to
zero, so that the branch almost immediately merges into the branch
of symmetric solutions of the single-component model. As $|\lambda
|$ increases above the critical value, the ``broken" S-AS branch
evolves towards the limit of a single-component ``broken" AS
solution which is strongly localized in one well of the nonlinear
potential, while the ``unbroken" S-AS branch approaches the
``unbroken" AS mode of the single-component model.

Another look at the evolution of the S-AS branches is suggested by the
general circumstance, similar to that emphasized above, in the course of the
analysis of the solution families of the Sm-Sm and AS-AS types, \textit{viz}%
., that various branches of two-component solutions may be realized as links
connecting the basic modes of the three types: the two-component ones with
the unbroken (anti)symmetry, single-component
symmetric/antisymmetric/asymmetric states, and the other above-mentioned
species of single-component solutions, which are almost entirely localized
in one well of the nonlinear potential. Below, we refer to these three basic
types of the modes as A, B, and C, respectively. In particular, the picture
of the evolution of the ``broken" S-AS branches, as described above, may be
interpreted so that the broken-(anti)symmetry branch evolves from state A
towards B. For example, at small values of $|\mu |$, the ``broken" S-AS
states are weakly localized, and the increase of $|\lambda |$ drives the
system towards state B without bifurcations. At $|\mu |>0.15$, the asymmetry
of the AS component changes its sign, but there is still no bifurcation,
with energy (\ref{H}) monotonically depending on norms $M$ and $N$.

At $|\mu |>0.37$, there emerges the above-mentioned single-component mode
(C), strongly localized in one well of the nonlinear potential, with a
result that the AS state tends to be trapped near this mode, before it will
be driven to mode B. This circumstance changes results of the bifurcation,
as can be observed in terms of the asymmetries and energy in Figs.~\ref%
{Fig15} and \ref{Fig17}(a). Another interpretation is that there is an
additional state which appears at the bifurcation point, built as a
combination of a symmetric mode in one component and a new AS mode with
broken antisymmetry, see Fig ~\ref{f-14}. This means that three branches are
observed in this region, with the evolving one attracted to state C and
oscillating about it for a while, before plunging to the limit of the
vanishing AS component (state B). 
\begin{figure}[tbh]
\refstepcounter{fig} \label{f-23}
\par
\begin{center}
(a)\hspace{-0.6cm}
\includegraphics[height=.22\textheight, angle =-90]{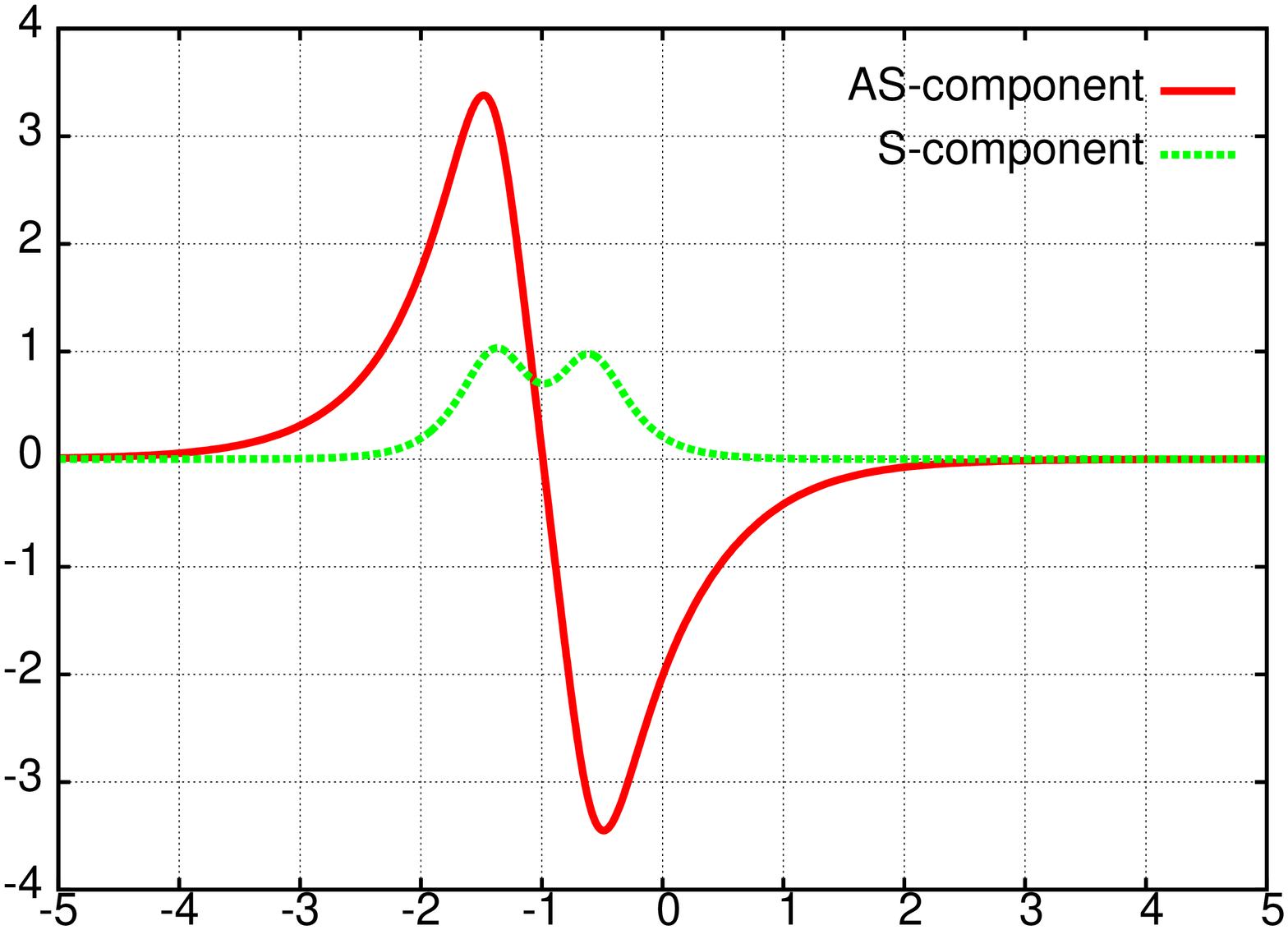
} \hspace{0.5cm} (b)\hspace{-0.6cm}
\includegraphics[height=.22\textheight, angle =-90]{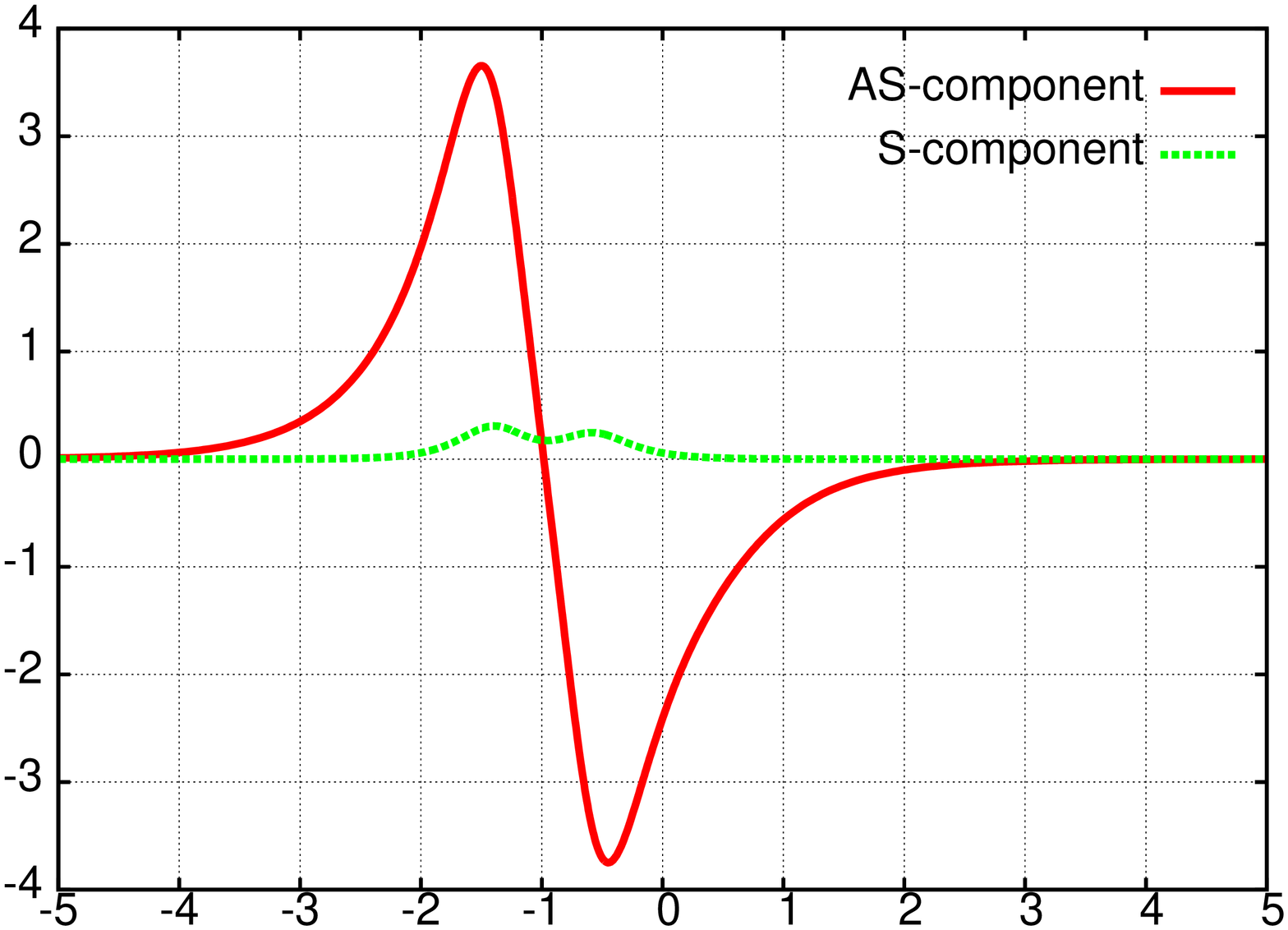
} \hspace{0.5cm} (c)\hspace{-0.6cm}
\includegraphics[height=.22\textheight, angle =-90]{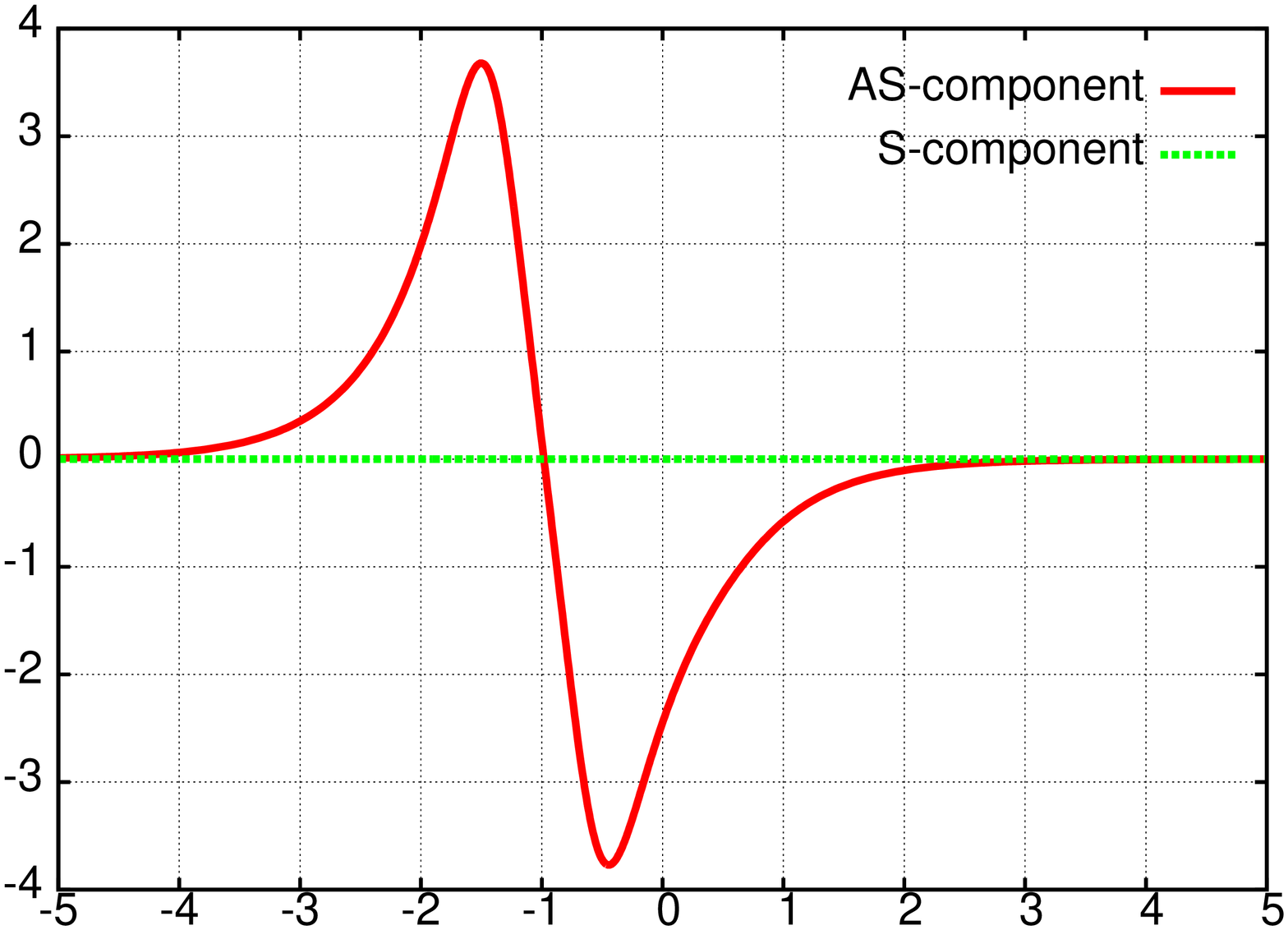
}
\end{center}
\par
\vspace{-0.5cm}
\caption{(Color online) The evolution of modes of the S-AS type with the
broken antisymmetry, approaching the effectively single-component mode (C,
as defined in the text). In terms of Fig. \protect\ref{Fig6}, C is the upper
AS state with the broken antisymmetry, generated by the saddle-node
bifurcation and trapped in one nonlinear potential well. The parameters are $%
a=0.6$, $|\protect\mu |=1.5$, and $|\protect\lambda |=6.5$~(a)$,$ $7.40~$(b)
and $\protect\lambda =7.43$~(c), respectively. }
\label{Fig16}
\end{figure}
\begin{figure}[h]
\refstepcounter{fig} \label{f-14}
\par
\begin{center}
(a)\hspace{-0.6cm}
\includegraphics[height=.35\textheight,
angle=-90]{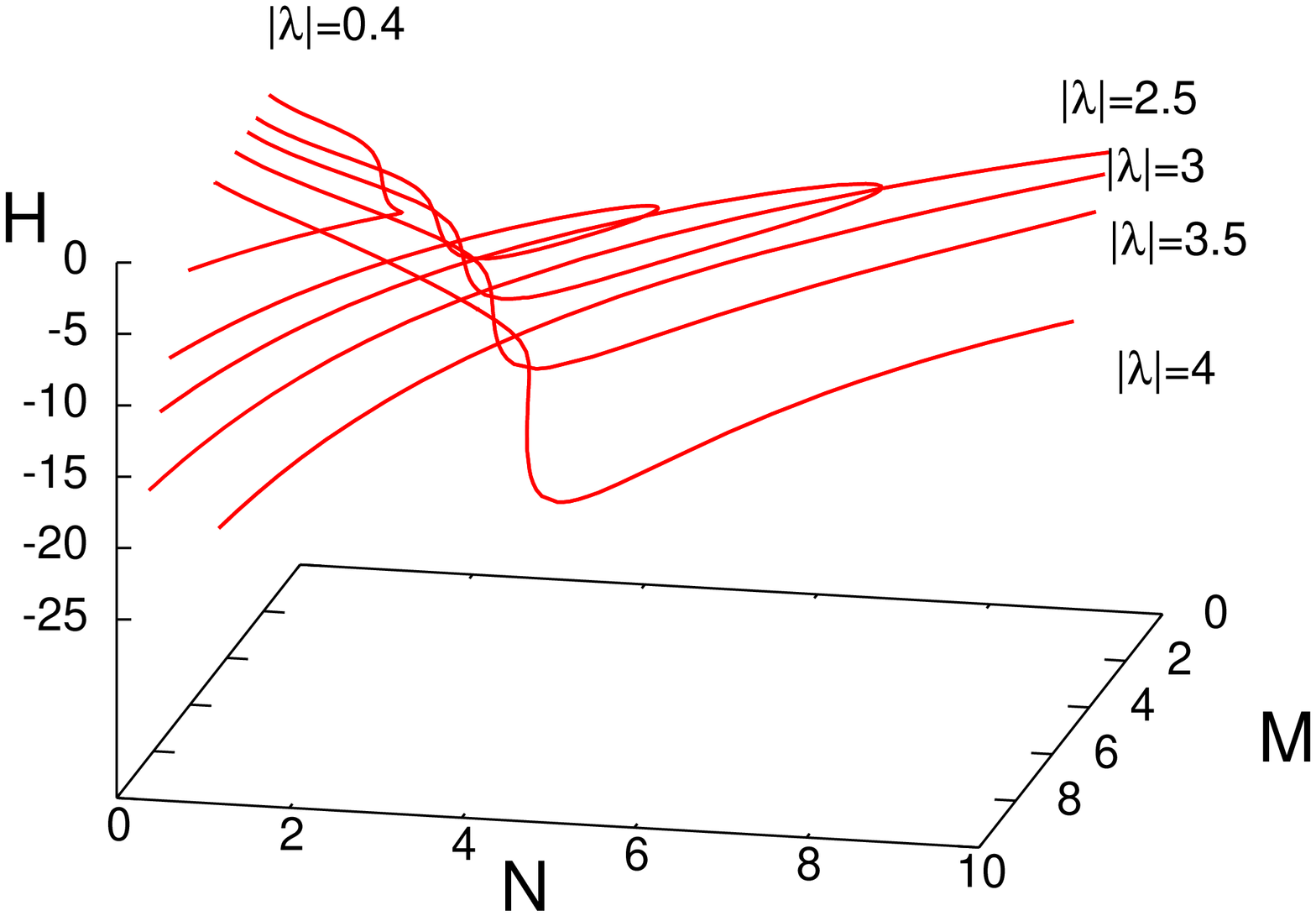}
\hspace{0.5cm} (b)\hspace{-0.6cm} %
\includegraphics[height=.35\textheight, angle =-90]{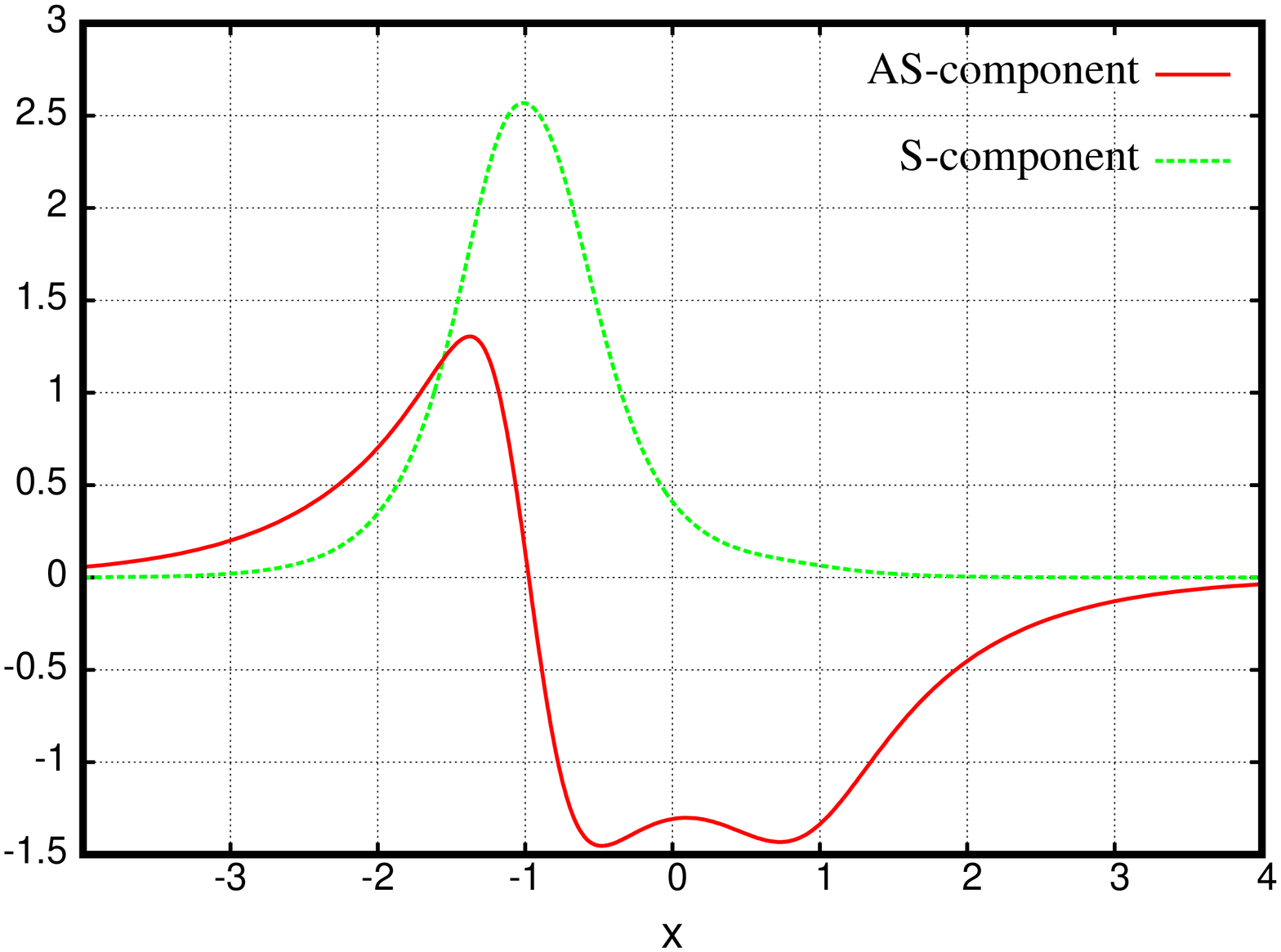}
\end{center}
\par
\vspace{-0.5cm}
\caption{(Color online) (a) The energy of the S-AS states with the broken
(anti)symmetry at $a=0.6$ and  $|\protect%
\lambda |=4.0,3.5,3.0,2.5,2.0,1.0,0.7,
0.4$, in the $\left( {M,N}\right) $ plane. (b) Profiles of the two components of the new S-AS
solution found in
a vicinity of state C (defined in the text), at $a=0.6$, and $|\protect\mu %
|=4.00,$ $|\protect\lambda |=0.791$. }
\label{Fig17}
\end{figure}

The picture of the evolution of the S-AS branches outlined above is valid in
the interval of $0.1<a<0.67$. Additional ramifications of the picture, which
are not presented here, are revealed by the analysis at larger values of $a$.

\section{Conclusions}

In this work, we have introduced the two-component one-dimensional model
with the nonlinear (pseudo)potential represented by two strongly localized
potential wells. The subject of the analysis is the evolution of various
two-component modes of the three types: Sm-Sm (symmetric-symmetric), AS-AS\
(antisymmetric-antisymmetric), and S-AS (mixed states with symmetric and
antisymmetric components), and the related SSB
(spontaneous-symmetry-breaking) effects. In the limit of the nonlinearity
modulation represented by the set of two $\delta $-functions, the solution
was obtained in the semi-analytical form. The consideration of this solution
demonstrates new features of the two-component modes with spontaneously
broken (anti)symmetries, which are qualitatively different from what was
previously reported in the single-component model. In particular, the
spontaneous breaking of the antisymmetry is possible only in the
two-component system, and, obviously, S-AS states, with the unbroken or
broken mixed symmetry exist only in the two-component system.

In the model based on the set of two nonlinear potential wells of a finite
width, the evolution and bifurcation scenarios were found to be still more
complex. First, new results for the case of finite potential wells were
reported for the single-component model. These are pairs of modes with the
broken antisymmetry, and pairs of twin-peak symmetric modes, generated by
isolated saddle-node bifurcations, in either case. Then, numerous complex
scenarios of the evolution of the Sm-Sm, AS-AS, and S-AS branches with the
broken (anti)symmetries may be interpreted in terms of routes linking three
species of basic modes: two-component states with the unbroken
(anti)symmetries, ``unbroken" or ``broken" single-component states featuring
density peaks in both nonlinear-potential wells, and, finally,
single-component states effectively trapped in a single well.

It remains to systematically analyze the stability of diverse two-component
stationary modes reported in this work. Challenging problems may also be to
extend the model to a set of several nonlinear spots, including a periodic
lattice, and, on the other hand, to develop the two-dimensional (2D) version
of the two-component system, following the recent analyses of the 2D
extensions of the single-component model \cite{Hung,new}.

\end{document}